\documentclass[apj,iop]{emulateapj}
\newif\ifemapj
\emapjtrue
\usepackage{xspace,times,perpage}
\MakePerPage{footnote}

\newcommand{\contrerasSNe}{18\xspace}
\newcommand{\contrerasPrevious}{9\xspace}
\newcommand{\fullsample}{20\xspace}

\newcommand{\kitchensinksample}{16\xspace}
\newcommand{\kitchensinksamplespell}{Sixteen\xspace}
\newcommand{\goodkitchensinksample}{14\xspace}

\newcommand{\numbergoodszgtone}{15}

\newcommand{\clusterRMS}{$0.19 \pm 0.04$\xspace}
\newcommand{\clusterzgtoneWRMS}{$0.20 \pm 0.05$\xspace}
\newcommand{\GOODSzgtoneWRMS}{$0.25 \pm 0.05$\xspace}

\newcommand{\oursamplebrightersigma}{1.5-$\sigma$}
\newcommand{\earlyhizintdisp}{$0.14 ^{+ 0.11}_{-0.08}$\xspace}
\newcommand{\earlyhizWRMS}{$0.23 \pm 0.05$\xspace}
\newcommand{\latehizintdisp}{$0.14 ^ {+ 0.06} _{-0.05}$\xspace}
\newcommand{\latehizWRMS}{$0.26 \pm 0.05$\xspace}

\newcommand{\earlyhizbrighter}{$0.18 \pm 0.09$\xspace}

\newcommand{\untargetedlow}{0.55}
\newcommand{\targetedlow}{0.13}
\newcommand{\untargetedlownoGauss}{0.50\xspace}
\newcommand{\targetedlownoGauss}{0.09\xspace}
\newcommand{\untargetedmusigma}{(\mu = 9.88, \sigma^2 = 0.92^2)}
\newcommand{\targetedmusigma}{(10.75, 0.66^2)}
\newcommand{\bestfitdelta}{-0.03}

\newcommand{\mediansigmasys}{0.15}

\newcommand{\PascalNICzpshift}{0.065}
\newcommand{\PascalNICzpAB}{23.757}
\newcommand{\PascalNICzpVega}{23.029}

\newcommand{\flatwfit}{w = -1.013^{+0.068}_{-0.073}\xspace}
\newcommand{\owCDMomegak}{\Omega _{k}=0.002 ^{+0.007}_{-0.007}}
\newcommand{\flatLCDMomegal}{\Omega_{\Lambda}={0.729}^{+0.014}_{-0.014}}
\newcommand{\wcurvedconstant}{w = -1.003^{+0.091}_{-0.095}}
\newcommand{\snealoneconstw}{$w = -1.001^{+ 0.348}_{-0.398}$}
\newcommand{\omegakoLCDM}{$\Omega_k = 0.002^{+0.005}_{-0.005}$}
\newcommand{\waflatwithsys}{$w_a = 0.14^{+0.60}_{-0.76}$}
\newcommand{\omegalambdasnealone}{0.705^{+0.040}_{-0.043}}

\newcommand{\FoMstat}{1.84}
\newcommand{\FoMsys}{1.04}

\newcommand{\Fomonesigstat}{39.3}
\newcommand{\FoMonesigsys}{22.6}

\usepackage{graphicx}
\usepackage[latin1]{inputenc}
\usepackage[english]{babel}
\usepackage[dvips]{rotating}
\usepackage{amsmath}
\usepackage{amssymb}
\usepackage{xspace}
\usepackage{natbib}
\usepackage{ifthen}
\usepackage{lscape}
\usepackage[usenames]{color}

\newcommand{\sigsamp}{\sigma_{\mathrm{sample}}}

\newcommand{\ang}{\AA\xspace}

\newcommand{\sne}{SNe\xspace}
\newcommand{\snia}{SN~Ia\xspace}
\newcommand{\sneia}{SNe~Ia\xspace}

\newcommand{\hst}{HST\xspace}

\newcommand{\acs}{ACS\xspace}

\newcommand{\zacs}{$z_{\mathrm 850}$\xspace}
\newcommand{\iacs}{$i_{\mathrm 775}$\xspace}
 
\newcommand{\om}{\mbox{$\Omega_m$}}
\newcommand{\omegam}{\mbox{$\Omega_m$}}
\newcommand{\omegak}{\mbox{$\Omega_k$}}
\newcommand{\omegal}{\mbox{$\Omega_{\Lambda}$}}

\newif\ifdraftmodep
\draftmodepfalse

\newcommand{\NOTE}[1]{\ifdraftmodep {\color {red} [{\it #1}]} \fi}

\newif\ifapjp
\apjpfalse

\usepackage{ifthen,color,graphicx}

\newcommand{\outlinestart}[1]
{
\ifthenelse{\boolean{#1}}{\begin{enumerate}}{}
}

\newcommand{\outline}[2]
{
\ifthenelse{\boolean{#1}}{\it \color {red} \item #2 \rm \color {black}}{}
}

\newcommand{\outlineend}[1]
{
\ifthenelse{\boolean{#1}}{\end{enumerate}}{}
}

\newboolean{draft}
\setboolean{draft}{false}

\begin{document}



\title{The {\it Hubble Space Telescope} 
       Cluster Supernova Survey: V. Improving the Dark Energy Constraints Above $z>1$ and 
       Building an Early-Type-Hosted Supernova Sample
\footnote{Based in part on observations made with the
  NASA/ESA {\it Hubble Space Telescope}, obtained from the data
  archive at the Space Telescope Institute. STScI is operated by the
  association of Universities for Research in Astronomy, Inc. under
  the NASA contract NAS 5-26555.  The observations are associated with
  program GO-10496.}
       }

\author{
N.~Suzuki\altaffilmark{1,2},
D.~Rubin\altaffilmark{1,2},
C.~Lidman\altaffilmark{3},
G.~Aldering\altaffilmark{1},
R.~Amanullah\altaffilmark{2,4},
K.~Barbary\altaffilmark{1,2},
L.~F.~ Barrientos\altaffilmark{5},
J.~Botyanszki\altaffilmark{2},
M.~Brodwin\altaffilmark{6,7},
N.~Connolly\altaffilmark{8},
K.~S.~Dawson\altaffilmark{9},
A.~Dey\altaffilmark{10},
M.~Doi\altaffilmark{11},
M.~Donahue\altaffilmark{12},
S.~Deustua\altaffilmark{13},
P.~Eisenhardt\altaffilmark{14},
E.~Ellingson\altaffilmark{15},
L.~Faccioli\altaffilmark{1,2},
V.~Fadeyev\altaffilmark{16},
H.~K.~Fakhouri\altaffilmark{1,2},
A.~S.~Fruchter\altaffilmark{13},
D.~G.~Gilbank\altaffilmark{17},
M.~D.~Gladders\altaffilmark{18},
G.~Goldhaber\altaffilmark{44},
A.~H.~Gonzalez\altaffilmark{19},
A.~Goobar\altaffilmark{4,20},
A.~Gude\altaffilmark{2,21},
T.~Hattori\altaffilmark{22},
H.~Hoekstra\altaffilmark{23},
E.~Hsiao\altaffilmark{1,2},
X.~Huang\altaffilmark{1,2},
Y.~Ihara\altaffilmark{11,25},
M.~J.~Jee\altaffilmark{26},
D.~Johnston\altaffilmark{12,27},
N.~Kashikawa\altaffilmark{28},
B.~Koester\altaffilmark{18,29},
K.~Konishi\altaffilmark{30},
M.~Kowalski\altaffilmark{31},
E.~V.~Linder\altaffilmark{1,2},
L.~Lubin\altaffilmark{26},
J.~Melbourne\altaffilmark{36},
J.~Meyers\altaffilmark{1,2},
T.~Morokuma\altaffilmark{11,25,28},
F.~Munshi\altaffilmark{2,32},
C.~Mullis\altaffilmark{33},
T.~Oda\altaffilmark{34},
N.~Panagia\altaffilmark{13},
S.~Perlmutter\altaffilmark{1,2}
M.~Postman\altaffilmark{13},
T.~Pritchard\altaffilmark{2,35},
J.~Rhodes\altaffilmark{14,36},
P.~Ripoche\altaffilmark{1,2},\\
P.~Rosati\altaffilmark{37},
D.~J.~Schlegel\altaffilmark{1},
A.~Spadafora\altaffilmark{1},
S.~A.~Stanford\altaffilmark{26,38},
V.~Stanishev\altaffilmark{20,39},
D.~Stern\altaffilmark{14},
M.~Strovink\altaffilmark{1,2},
N.~Takanashi\altaffilmark{28},
K.~Tokita\altaffilmark{11},
M.~Wagner\altaffilmark{40},
L.~Wang\altaffilmark{41},
N.~Yasuda\altaffilmark{42},
H.~K.~C.~Yee\altaffilmark{43},\\
(The Supernova Cosmology Project)
}

\email{nsuzuki@lbl.gov,rubind@berkeley.edu,clidman@aao.gov.au}

\altaffiltext{1}{E.O. Lawrence Berkeley National Lab, 1 Cyclotron Rd., Berkeley, CA, 94720}
\altaffiltext{2}{Department of Physics, University of California Berkeley, Berkeley, CA 94720}
\altaffiltext{3}{Australian Astronomical Observatory, PO Box 296, Epping, NSW 1710, Australia}
\altaffiltext{4}{The Oskar Klein Centre for Cosmo Particle Physics, AlbaNova, SE-106 91 Stockholm, Sweden}
\altaffiltext{5}{Universidad Catolica de Chile} 
\altaffiltext{6}{Harvard-Smithsonian Center for Astrophysics, 60 Garden Street, Cambridge, MA 02138}
\altaffiltext{7}{W. M. Keck Postdoctoral Fellow at the Harvard-Smithsonian Center for Astrophysics}
\altaffiltext{8}{Hamilton College Department of Physics, Clinton, NY 13323}
\altaffiltext{9}{Department of Physics and Astronomy, University of Utah, Salt Lake City, UT 84112}
\altaffiltext{10}{National Optical Astronomy Observatory, Tucson, AZ 85726-6732}
\altaffiltext{11}{Institute of Astronomy, Graduate School of Science, University of Tokyo 2-21-1 Osawa, Mitaka, Tokyo 181-0015, Japan}
\altaffiltext{12}{Michigan State University, Department of Physics and Astronomy, East Lansing, MI 48824}
\altaffiltext{13}{Space Telescope Science Institute, 3700 San Martin Drive, Baltimore, MD 21218}
\altaffiltext{14}{Jet Propulsion Laboratory, California Institute of Technology, Pasadena, CA, 91109}
\altaffiltext{15}{Center for Astrophysics and Space Astronomy, 389 UCB, University of Colorado, Boulder, CO 80309}
\altaffiltext{16}{Santa Cruz Institute for Particle Physics, University of California Santa Cruz, Santa Cruze, CA 94064} 
\altaffiltext{17}{Department of Physics and Astronomy, University Of Waterloo, Waterloo, Ontario, Canada N2L 3G1}
\altaffiltext{18}{Department of Astronomy and Astrophysics, University of Chicago, Chicago, IL 60637}
\altaffiltext{19}{Department of Astronomy, University of Florida, Gainesville, FL 32611}
\altaffiltext{20}{Department of Physics, Stockholm University, Albanova University Center, SE-106 91, Stockholm, Sweden}
\altaffiltext{21}{School of Physics \& Astronomy, University of Minnesota, Minneapolis, MN 55455}
\altaffiltext{22}{Subaru Telescope, National Astronomical Observatory of Japan, 650 North A'ohaku Place, Hilo, HI 96720}
\altaffiltext{23}{Leiden Observatory, Leiden University, Leiden, The Netherlands} 
\altaffiltext{25}{JSPS Fellow}
\altaffiltext{26}{Department of Physics, University of California Davis, One Shields Avenue, Davis, CA 95616}
\altaffiltext{27}{Department of Physics and Astronomy, Northwestern University, 2145 Sheridan Road, Evanston, IL 60208}
\altaffiltext{28}{National Astronomical Observatory of Japan, 2-21-1 Osawa, Mitaka, Tokyo, 181-8588, Japan}
\altaffiltext{29}{Kavli Institute for Cosmological Physics, The University of Chicago, Chicago, IL 60637} 
\altaffiltext{30}{Institute for Cosmic Ray Research, University of Tokyo, Kashiwa 277-8582, Japan}
\altaffiltext{31}{Physikalisches Institut Universit\"at Bonn, Germany}
\altaffiltext{32}{Astronomy Department, University of Washington, Seattle, WA 98195}
\altaffiltext{33}{Wachovia Corporation, NC6740, 100 N. Main Street, Winston-Salem, NC27101}
\altaffiltext{34}{Department of Astronomy, Kyoto University, Sakyo-ku, Kyoto 606-8502, Japan}
\altaffiltext{35}{Astronomy and Astrophysics, The Pennsylvania State University, University Park, PA 16802}
\altaffiltext{36}{California Institute of Technology, Pasadena, CA 91125}
\altaffiltext{37}{ESO, Karl-Schwarzschild-Strasse 2, D-85748 Garching, Germany}
\altaffiltext{38}{Institute of Geophysics and Planetary Physics, Lawrence Livermore National Laboratory, Livermore, CA 94550}
\altaffiltext{39}{CENTRA - Centro Multidisciplinar de Astrof\'isica, Instituto Superior T\'ecnico, Av. Rovisco Pais 1, 1049-001 Lisbon, Portugal}
\altaffiltext{40}{Space Sciences Lab, 7 Gauss Way, Berkeley, CA 94720}
\altaffiltext{41}{Department of Physics, Texas A \& M University, College Station, TX 77843, USA}
\altaffiltext{42}{Institute for the Physics and Mathematics of the Universe, University of Tokyo, Kashiwa, 277-8583, Japan}
\altaffiltext{43}{Department of Astronomy and Astrophysics, University of Toronto, Toronto, ON M5S 3H4, Canada}
\altaffiltext{44}{Deceased}



\begin{abstract}
We present ACS, NICMOS, and Keck AO-assisted photometry of 20 Type
Ia supernovae (\sneia) from the HST Cluster Supernova Survey. The
\sneia\ were discovered over the redshift interval $0.623 < z <
1.415$. Fourteen of these \sneia\ pass our strict selection cuts
and are used in combination with the world's sample of \sneia\ to
derive the best current constraints on dark energy. Ten of our new \sneia\
are beyond redshift $z=1$, thereby nearly doubling the statistical
weight of HST-discovered SNe Ia beyond this redshift. 
Our detailed analysis corrects for the recently identified correlation 
between \snia\ luminosity and host galaxy mass and corrects the NICMOS 
zeropoint at the count rates appropriate for very distant SNe Ia. 
Adding these supernovae improves the best combined 
constraint on dark energy density, $\rho_{\mathrm{DE}}(z)$, 
at redshifts $1.0 < z < 1.6$ by 18\% (including systematic errors).
For a flat $\Lambda$CDM universe, we find $\flatLCDMomegal$\ (68\% CL including
systematic errors). For a flat $w$CDM model, we measure a constant
dark energy equation-of-state parameter $\flatwfit$\ (68\% CL).
Curvature is constrained to $\sim 0.7$\% in the $ow$CDM model and to
$\sim 2\%$ in a model in which dark energy is allowed to vary with
parameters $w_0$ and $w_a$. Tightening further the constraints on
the time evolution of dark energy will require several improvements, including high-quality
multi-passband photometry of a sample of several dozen $z>1$ \sneia.
We describe how such a sample could be efficiently obtained by targeting cluster fields with WFC3 on HST.

The updated supernova Union2.1 compilation of 580 \sne is available 
at http://supernova.lbl.gov/Union
\end{abstract}


\maketitle


\section{Introduction}\label{sec:b}
\outline{draft}{A Decade of Dark Energy : CL, KD, KB, NS\\} 
More than a dozen years have passed since combined observations of nearby and
distant Type Ia Supernovae (\sneia) demonstrated that the expansion of
the Universe is accelerating at the current epoch
\citep{perlmutter98a,garnavich98a,schmidt98a,riess98a,perlmutter99a}. 
While the significance of the result has been boosted with the inclusion of
larger, better calibrated \snia data sets
\citep{knop03a,astier06a,wood-vasey07a,kowalski08b,hicken09c,kessler09a,guy10a} the cause of the
acceleration remains unknown. Einstein's cosmological constant, for
which $w$, the dark energy equation-of-state parameter, is exactly $-1$ and 
independent of time, is just one of several possible explanations that
is consistent with the constraints from \sneia and the constraints
from other probes, such as the Cosmic Microwave Background
\citep[CMB,][]{dunkley09a,komatsu11a} and Baryon Acoustic Oscillations
\citep[BAO,][]{eisenstein05a,percival10a}.

\outline{draft}{Why SNe at z $>$ 1 are important\\} 

\sneia\ constrain cosmological parameters through the comparison of
their apparent luminosities over a range of redshifts. 
At the highest redshifts, $z>1$, the Hubble Space Telescope (HST) 
has played and continues to play a key role, in discovering and 
confirming $z>1$ \sneia
\citep{riess04a,riess07a,kuznetsova08a,dawson09a} and in providing
high-precision optical and near-IR lightcurves.  While constant $w$
can be constrained using $z\sim0.5$ \sneia, \sneia\ at $z>1$ provide
the necessary redshift baseline to constrain time-varying $w$ and
some astrophysical systematics like intergalactic dust \citep{menard10a}.

Discovering and following distant \sneia with the HST requires substantial
amounts of telescope time because the field-of-view is quite small
compared to that of ground-based telescopes. 
Therefore, all HST \snia discovery programs have
coupled the search for \sneia and their photometric follow-up with
other scientific studies. 
The GOODS survey \citep{dickenson03a,riess04a,riess07a,kuznetsova08a}
is an example that provided a window to probe the high-$z$ universe 
for studying galaxy evolution \citep[e.g.][]{beckwith06a,bouwens06a,bundy05a}
in addition to $z > 1$ \sne.
Nevertheless, progress in building a large sample of $z>1$ \sneia
discovered with HST is slow.
In the most recent \snia compilation
\citep[Union2,][]{amanullah10a} which consists of 557 \sne after the
lightcurve quality cuts, only 16 HST-discovered $z>1$ \sneia were
available to help constrain a time-evolving $w$. 
(Well measured ground-based $z > 1$ SNe account for an additional four.)

Targeting regions that are rich in potential \snia hosts, such as
galaxy clusters, offers a more effective strategy for using HST for
\snia studies.
Some of the earliest \snia searches used this strategy when the
field-of-view of ground-based images were only a few arc minutes
across. 
The first spectroscopically confirmed high-redshift \snia was
discovered in a galaxy cluster 
\citep[SN1988U: z=0.31,][]{norgaardnielsen89a} as was 
the first high-redshift \snia observed by HST
\citep[SN1996cl: z=0.83,][]{perlmutter98a,perlmutter99a}. 
However, the advent of large-format CCDs, CCD
mosaics and imagers with wide fields of view quickly led away from
this approach in the late 80's and early 90's \citep{perlmutter91a}.

Given the large increase in the number of very distant clusters that
have been discovered over the last 10 years, the angular extent of
these clusters, and the field-of-view that is available with HST,
targeting galaxy clusters beyond $z\sim1$ is again an effective
strategy. In this paper, we discuss results from our HST Cluster SN
survey obtained using this strategy. In addition to
increasing the yield of \sneia discoveries per HST orbit by a factor
of two \citep{dawson09a}, we increase the yield of \sneia in
early-type galaxies by a factor of approximately four
(Meyers et al. 2011).

\outline{draft}{The properties of SN and their hosts\\}

\sneia hosted by early-type galaxies offer several potential
advantages over \sneia found in a broader range of host types. Stars
in early-type galaxies are considerably older and span a smaller mass
range than stars in late-type hosts. This may lead to a more uniform
progenitor population. Evidence for this can be seen in the
distribution in light curve widths. \sneia in early-type host galaxies
follow a narrower distribution than \sneia in late-type galaxies
\citep{hamuy96a,hamuy00a,riess99a,sullivan06a}.  Interestingly, the
relationship between \snia color and host galaxy type is weak
\citep{sullivan10b}. Using data from the HST Cluster SN Survey, we
confirm both of these relationships for $z>1$ \sneia in Meyers et al. (2011).

With the availability of larger, better-calibrated samples, evidence
for a correlation between host galaxy properties and \snia
luminosities after corrections for lightcurve width and \snia color is
now emerging. \citet{hicken09a} found that \sneia in early-type
galaxies (morphologically classified as E and S0 galaxies), are
$0.14\pm0.07$ mag brighter after lightcurve shape and color
corrections than \sneia in galaxies of later types. A relationship of
roughly the same significance between host galaxy mass\footnote{Host
  galaxy metallicity, specific-star-formation rate, or age are also 
  drivers, as these quantities are somewhat degenerate in current
  data.} and Hubble residuals was reported by \citet{kelly10a},
\citet{sullivan10b} and \citet{lampeitl10b}. Uncorrected, this
relationship leads to a significant systematic error in determining
cosmological parameters, as the fraction of \snia in galaxies with
high specific star-formation rates increases with increasing redshift
\citep{sullivan10b}.
We expect that the host mass correction is a proxy for more
profound physics behind the \snia explosion mechanism; \sneia in early-type
galaxies may lead to a better understanding of this correlation,
given that more accurate mass, metallicity and age can be assigned to
early-type galaxies \citep{bruzual03a,tremonti04a,maraston05a}.

An additional source of astrophysical uncertainty concerns the color
correction that is applied to \snia luminosities. There appears to be
at least two mechanisms for the redder-fainter relation: extinction
from dust in the ISM, which must play a role at some level, and an
intrinsic relation between color and luminosity due to the explosion
itself or the surrounding environment. There is no reason to believe
that the redder-fainter relationship should behave in the same way for
both mechanisms at all redshifts, but the two effects have proven 
to be hard to disentangle.

Early-type galaxies contain significantly less dust than late-type
galaxies, so separating \sneia according to early and late types
offers a way to study the intrinsic component and to perhaps estimate
the relative contribution and importance of dust in a broader sample.
An early-type only sample may also yield a Hubble diagram with smaller
statistical errors. 
Early work, based on a few dozen \sneia\ without color correction, 
suggested that \sneia\ in early-type galaxies are better standard candles 
\citep{sullivan03a}. 
The evidence from more recent works, which use larger samples and 
better data, revealed that \sneia\ exhibit intrinsic diversity in color,
but support the original findings \citep{sullivan03a} with lower 
statistical significance \citep{sullivan10b}. 

Per unit stellar mass, \sneia are far less common in passive,
early-type galaxies than in star forming, late-type galaxies
\citep{mannucci05a,sullivan06a}. Depending on the way hosts are
classified, about one in five \sneia at low redshift will be hosted by
an early-type galaxy. At higher redshifts, the fraction is expected to
decrease, due to a combination of an increase in the amount of star
formation and observational selection biases. Galaxy clusters, which
are rich in early-type galaxies, even up to $z\sim1.4$, are an
effective way of finding \sneia in early-type hosts \citep{dawson09a}.

\outline{draft}{Paper Structure\\} 
This paper is one of a series of ten papers that report supernova
results from the HST Cluster Supernova Survey (PI: Perlmutter,
GO-10496), a survey to discover and follow \sneia of very distant
clusters. Paper I \citep{dawson09a} describes the survey strategy and
discoveries. Paper II \citep{barbary10a} reports on the \snia rate in
clusters. Paper III \citep{meyers11a} addresses the properties of the
galaxies that host \sneia. Paper IV \citep{ripoche11a} introduces a
new technique to calibrate the ``zeropoint'' of the NICMOS camera at low
counts rates, which is critical for placing NICMOS-observed \sneia on
the Hubble diagram. The current work, Paper V, reports the \snia
lightcurves and cosmology from the HST Cluster SN Survey program. Paper
VI (Barbary et al in prep) will report on the volumetric field \snia
rate.  
\citet{melbourne07a}, one of several unnumbered papers in this series, 
present a Keck Adaptive Optics observation of a $z=1.31$ \snia in H-band. 
\citet{barbary09a} report the discovery of the extraordinary luminous 
supernova, SN~SCP06F6. 
\citet{morokuma10a} presents the spectroscopic follow-up
observations for \snia candidates. 
\citet{hsiao11a} develop techniques to remove problematic artifacts 
remaining after the standard STScI pipeline. 
A separate series of papers, ten to date, reports on cluster studies 
from the HST Cluster SN Survey:
\citet[][]{brodwin10a,eisenhardt08b,jee09a,hilton07b,
hilton09a,huang09a,santos09a,strazzullo10a,rosati09a}; and \citet{jee11b}.

This paper is organized as follows. In \S\ref{sec:HSTclusterSN},
we describe the HST Cluster SN Survey, the search strategy and discuss
\snia typing. In \S\ref{sec:photometry}, we describe the procedures we
used to process data and present the \snia photometry. 
In \S\ref{sec:g}, we update the Union2 sample by adding the new
\sneia from this paper, and we use the revised compilation to
constrain cosmological parameters in \S\ref{sec:cosmology}.

\section{SN Discoveries and Data}\label{sec:HSTclusterSN} \label{sec:d} \label{sec:discovery}

\outline{draft}{Overview of the Survey Motivation rates, dust, spectroscopy, details in KD}

The HST Cluster Supernova Survey targeted 25 high-redshift galaxy
clusters in the redshift range $0.9 < z < 1.5$ with the \acs camera on
\hst.  Clusters were selected from the IRAC Shallow Cluster Survey
\citep{eisenhardt08b}, the Red-Sequence Cluster Surveys 
\citep[RCS and RCS-2,][]{gladders05a,gilbank10a}, the XMM Cluster Survey
\citep{sahlen09a}, the Palomar Distant Cluster Survey
\citep{postman96a}, the XMM-Newton Distant Cluster Project
\citep{bohringer05a}, and the ROSAT Deep Cluster Survey
\citep[RDCS,][]{rosati99a}. At the time we conducted our survey, the sample
represented a significant fraction of the known $z>0.9$
clusters. Here, we summarize the SNe discovered in our survey.

\subsection{SN Sample}\label{sec:SNsample}
\label{sec:lens}

As described in \citet{dawson09a}, the survey produced a total of 39
likely SNe during the active phase of the search. In
\citet{barbary10a}, types are determined for 29 of these candidates.
(The remaining 10 do not have enough light curve information 
to determine type, since they lie outside of our fiducial search 
time window or our signal-to-noise cuts.)
Twenty SNe are classified as \sneia, with confidence
levels of secure, probable or plausible.  A secure \snia\ is one that
either has a spectrum that directly confirms it to be a \snia\ {\it or}
one that satisfies two conditions: (1) it occurred in a host whose
spectroscopic, photometric and morphological properties are consistent
with those of an early-type galaxy with no detectable signs of recent
star formation, and (2) it has a lightcurve shape consistent with
that of a \snia\ and inconsistent with all other known SN types. 
A probable \snia\ is one that does not have a secure spectrum but 
satisfies one of the two non-spectroscopic conditions that are required 
for a secure classification. 
A plausible \snia\ is one that has an indicative lightcurve but we do not
have enough data to rule out other types. 
Details of the classification scheme can be found in
\citet{barbary10a}, and details of the galaxy typing can be found in
\citet{meyers11a}.

\ifemapj 
\begin{deluxetable*}{llllrrrl}
\else 
\begin{deluxetable}{llllrrrl}
\tabletypesize{\scriptsize}
\fi

\tablewidth{0pt}
\tablecaption{Supernova from HST Cluster Supernova Survey \label{tbl:a}}

\tablehead{
\colhead{SN name} & 
\colhead{Nickname} & 
\colhead{$z$\tablenotemark{b}} & 
\colhead{$z_{\rm cluster}$\tablenotemark{c}} & 
\colhead{RA (J2000)} & 
\colhead{DEC (J2000)} & 
\colhead{E(B-V)\tablenotemark{d}} &
\colhead{Confidence}\\
}
\startdata

\cutinhead{SNe Hosted by Cluster Early-Type Galaxies}

SCP05D0\tablenotemark{a}&Frida&1.014&1.017&02:21:42.066&$-$03:21:53.12&0.025&secure\\
SCP06H5&Emma&1.231&1.241&14:34:30.140&$+$34:26:57.30&0.019&secure\\
SCP06K0&Tomo&1.415&1.414&14:38:08.366&$+$34:14:18.08&0.015&secure\\ 
SCP06K18&Alexander&1.411&1.414&14:38:10.665&$+$34:12:47.19&0.014&probable\\
SCP06R12&Jennie&1.212&1.215&02:23:00.083&$-$04:36:03.05&0.026&secure\\
SCP06U4\tablenotemark{a}&Julia&1.050&1.037&23:45:29.430&$-$36:32:45.75&0.014&secure\\

\cutinhead{SNe Hosted in the Cluster}

SCP06C1\tablenotemark{a}&Midge&0.98&0.974&12:29:33.013&$+$01:51:36.67&0.019&secure\\
SCP06F12&Caleb&1.110&1.110&14:32:28.749&$+$33:32:10.05&0.010&probable\\

\cutinhead{SNe Hosted by Early-Type Non-Cluster Members}

SCP05D6&Maggie&1.315&1.017&02:21:46.484&$-$03:22:56.18&0.025&secure\\
SCP06G4\tablenotemark{a}&Shaya&1.350&1.259&14:29:18.744&$+$34:38:37.39&0.015&secure\\
SCP06A4&Aki&1.192&1.457&22:16:01.078&$-$17:37:22.10&0.026&probable\\ 
SCP06C0&Noa&1.092&0.974&12:29:25.655&$+$01:50:56.59&0.020&secure\\ 

\cutinhead{SNe Hosted by Late Type Galaxies}

SCP06G3&Brian&0.962&1.259&14:29:28.430&$+$34:37:23.15& 0.015 & plausible \\
SCP06H3\tablenotemark{a}&Elizabeth&0.850&1.241&14:34:28.879&$+$34:27:26.62&0.019&secure\\
SCP06N33&Naima&1.188&1.026&02:20:57.699&$-$03:33:23.98&0.023&probable\\
SCP05P1&Gabe&0.926&1.1&03:37:50.352&$-$28:43:02.67&0.011&plausible\\
SCP05P9\tablenotemark{a}&Lauren&0.821&1.1&03:37:44.513&$-$28:43:54.58&0.011&secure\\
SCP06X26&Joe&1.440&1.101&09:10:37.888&$+$54:22:29.06&0.019&plausible\\
SCP06Z5\tablenotemark{a}&Adrian&0.623&1.390&22:35:24.967&$-$25:57:09.61&0.021&secure\\
\cutinhead{SNe with No Definitive Redshift Measurement}
SCP06E12&Ashley&\nodata&1.026&14:15:08.141&$+$36:12:42.93&0.009&plausible \NOTE{cluster redshift}
\enddata

\tablenotetext{a}{Spectroscopically confirmed as a SNe Ia}
\tablenotetext{b}{Redshift from SNe~Ia or host galaxy \citep[][Barbary et al. 2010, Meyers et al. 2011]{morokuma10a}}
\tablenotetext{c}{Redshift from cluster (Meyers et al. 2011, references therein)}
\tablenotetext{d}{Galactic Extinction from \citet{schlegel98a}}
\ifemapj \end{deluxetable*}
\else  \end{deluxetable}
\fi

\ifemapj \begin{figure*}[h]
\else \begin{figure}[h]
\fi

\begin{center}
\includegraphics[angle=0,width=1.0\textwidth]{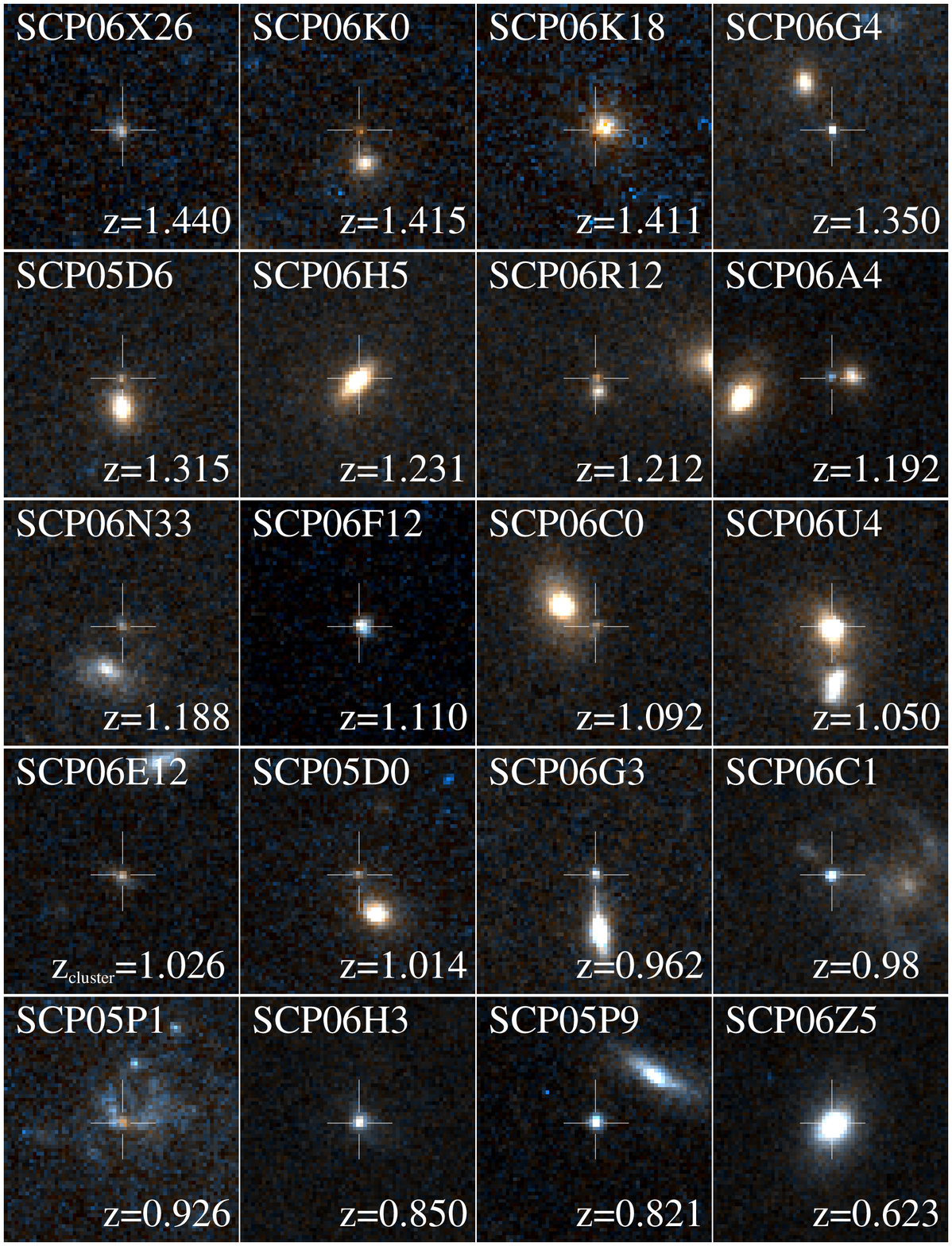}
\end{center}
\caption{\NOTE{fig:a to be revised : place holder} \label{fig:a}
Composite color (\iacs and \zacs) images of 20 \sneia\ from the HST Cluster Supernova Survey. 
Each \snia\ is shown in a box of $3.2\arcsec\times3.3\arcsec$ (North up and East left).
Note the redshift of SCP06E12 is uncertain, and we use the cluster redshift as a guide.}
\ifemapj \end{figure*}
\else  \end{figure}
\fi

\kitchensinksamplespell SNe are classified as either secure or probable. We
use these SNe in the cosmological analysis. We include the photometry
and lightcurves of an additional four plausible \sneia to illustrate
the quality of the data and the potential for a similar sample with
complete classification (and because additional host galaxy data may
later bring one of these into the larger sample). Secure, probable and
plausible \snia are listed in Table~\ref{tbl:a}, together with their position,
redshift and typing.  
Postage stamp images of the SNe and host galaxies are shown in
Figure~\ref{fig:a}.

We labeled each of our 25 clusters with 
a letter from `A' to `Z' (excluding `O' to avoid confusion with zero) and
assigned supernova names as `SCP'+[discovery year]+
[discovered cluster]+[SN ID].
The cluster IDs, coordinates, and redshifts are found in tables 
in \citet{dawson09a,barbary10a,meyers11a}.
The cluster membership is discussed in \citet{meyers11a} in detail
and summarized in Table~\ref{tbl:a} along with host type information.

Several \sneia deserve special mention:

{\it SN~SCP06C1}, a supernova discovered in 2006 in cluster C
(XMMU J2205.8$-$0159) and numbered as `1' among other transient candidates,
could not be clearly associated with a galaxy in the
cluster but was spectroscopically confirmed as a \snia\ at the cluster
redshift. It might be an example of a \snia\ that comes from a
progenitor in the intracluster stellar population \citep{sand10a}. 
SN~SCP06C1 is discussed in greater detail in \citet{barbary10a}.
As there was a bright background galaxy near the position of the supernova,
had we not obtained a spectrum of the supernova, we would have misidentified
the redshift. We note that experiments that do not obtain spectroscopy of their supernovae
and instead assume the nearest visible galaxy
is the host will have to factor cases like SN~SCP06C1 into their analyses.

{\it SN~SCP05D6, SN~SCP06G4, and SN~SCP06N33} occur behind the cluster,
and are therefore gravitationally lensed by the cluster. 
In most surveys, the lensing for \sneia\ hosted by field galaxies
averages to nearly zero. In this survey, we target regions with larger than average magnification, 
so we must make the correction for SNe behind the clusters. To estimate the amount of
magnification, we use the virial mass, $M_{200}$, from our weak lensing
measurements \citep{jee11b}. We assume a spherical Navarro-Frenk-White
profile \citep[NFW;][]{navarro97a} with the concentration parameters
determined by the $M_{200}-c$ relation in \citet{duffy08a}. The
lensing magnifications for SN~SCP05D6, SN~SCP06G4 and SN~SCP06N33 are
estimated to be $1.021^{+0.012}_{-0.008}$, $1.015^{+0.005}_{-0.004}$
and $1.066^{+0.017}_{-0.014}$ respectively. The magnification of
SN~SCP06N33 is larger than the others because the cluster is massive,
and the host galaxy is located at approximately half the distance from
cluster center as the others.  We apply these corrections when using
these \sneia\ in the cosmological fits and propagate the uncertainties
accordingly.

{\it SN~SCP06C0} is a more interesting case. The host is also behind
the lensing cluster, which means that SN~SCP06C0 will be lensed as well.
Following the methodology used to correct SN~SCP06G4,
SCP~SN05D6 and SCP~SN06N33, we find the magnification of SN~SCP06C0 to be
$1.030^{+0.007}_{-0.005}$. Upon closer inspection of the host in the
stacked ACS data and in more recent WFC3 data, a second much fainter
object, projected only 0\farcs6 from the center of the host and about
0\farcs2 from SN~SCP06C0, was detected.  The object could be a
satellite of the host, or it could be an unrelated galaxy along the
line of sight. A spectrum of SN~SCP06C0 was taken when it was about 6
days after maximum light; however, the signal-to-noise ratio was
insufficient to allow a clear detection of light from the supernova
given the observing conditions and the relative brightnesses of the
hosts and the supernova. The classification and redshift therefore
rely on correctly assigning the supernova to the brighter galaxy. In
\S ~\ref{sec:nicphottesting}, we model the surface brightness
distribution of both galaxies. At the location of the supernova, the
surface brightness of the large galaxy is four times greater than the
surface brightness of the small galaxy. Since [OII] was not detected
in spectra that were taken after the supernova had faded from view,
neither galaxy is actively forming stars. The relative supernova rate
is therefore directly related to relative surface mass density of the
the two galaxies. Using the surface brightness as a proxy for the
surface mass density, we therefore assign SN~SCP06C0 to the larger
of the two galaxies with $\sim 80$\% confidence and include
this SNe Ia in the cosmological fits.

\section{Photometry}
\label{sec:photometry} 
\label{sec:e} \NOTE{NS}
In this section, we describe the steps that were used to process the ACS
and NICMOS data after they had been processed with the standard STScI
pipelines. For the ACS data, we removed the spatially variable
background from the pipeline processed data and applied charge
transfer efficiency (CTE) and red-halo scattering corrections to the
extracted fluxes.  For the NICMOS data, we processed the data to
compensate for amplifier offsets, bright Earth persistence,
contamination from the passage of the telescope through the South
Atlantic Anomaly (SAA), residual amplifier glow and fringing, and
applied a wavelength-dependent non-linearity correction. A more detailed
description of the individual steps now follows.

\subsection{ACS Processing and Photometry} 
\outline{draft}{Postprocessing\\}
In general, the search consisted of four \zacs exposures per epoch
and a single \iacs-band exposure. All exposures are geometrically corrected (and multiple exposures are stacked) using MultiDrizzle \citep{fruchter02a,koekemoer02a}.
The photometry is performed on the
stacked \zacs-band image and the single \iacs-band drizzled image.  In addition
to the data that were taken during the search, five clusters,
CL1604+4304\NOTE{M} \citep{postman05a}, 
RDCS0910+54\NOTE{X} \citep{mei06b}, 
RDCS0848+44\NOTE{W} \citep{postman05a},
RDCS1252$-$29\NOTE{Y} \citep{blakeslee03a} and 
XMMU2235.3$-$2557 \citep{jee09a} were observed 
with ACS prior to our program (PID9290 and PID9919). 
We have included all these data and processed them and the search data in
a uniform way.  In total, 1006 ACS exposures were processed.

After processing the ACS data with the standard STScI pipeline using
the most up-to-date calibration files, we removed the spatially
variable background by masking all objects and artifacts, and
subtracting a heavily smoothed, median version of what remained. Given
the time baseline of our observations, guide stars changed between
epochs. We must therefore use objects in the images - typically 20 to
30 objects per image - to tie the relative astrometry between
epochs. The size of the residuals was typically 0.2 pixels, which is
larger than one usually expects from point sources with good
signal-to-noise ratios. We attribute this to temporal and spatial
uncertainties in the distortion correction that are applied to images
that are both temporally separated by many months and rotated with
respect to one another \citep{anderson07a}. 
The absolute astrometry is tied to the Guide Star Catalogue, version 2.3.2
\footnote{\url{http://gsss.stsci.edu/Catalogs/GSC/GSC2/gsc23/gsc23\_release\_notes.htm}},
and has an uncertainty of 0.3\arcsec.

\outline{draft}{Report Photometry Results\\}
We use apertures of 3-pixel radius to measure fluxes and report the 
\iacs and \zacs photometry for each epoch of each SN in the electronic 
version of this paper.
An example is shown in Table~\ref{tbl:b}.  
The background noise is found empirically by randomly placing apertures
within regions that are free of objects and then measuring the
dispersion in the integrated counts within these apertures. We have
compared the signal-to-noise ratios obtained for aperture and PSF
photometry and found that the difference between the two is small if
we use such small apertures. We use aperture photometry here,
as it allows for a more robust correction of CTE and red halo
scattering (described below).

The photometry is corrected for variable CTE and for flux that is
outside the aperture (referred to here as the aperture
correction). CTE depends on the position of the source, the level of
the background sky and the flux of the object that is being
corrected. It also degrades with time. We follow the formulation of
\citet{riess04b} and apply the correction factor that corresponds to
the 3-pixel radius case and the dates of our observations. On average,
we applied a 4.3\% correction to the \snia flux. When calculating CTE
corrections, we include the local background that had been previously
subtracted.

\outline{draft}{ACS z-band color dependent aperture correction\\}

We also apply a color-dependent aperture correction on the ACS 
\zacs-band measurements.  In ACS \zacs-band data, long-wavelength photons
scatter off the back-side of the CCD, causing a degradation in the
PSF. The effect is commonly known as red-halo scattering
\citep{sirianni98a}.
The
TinyTim\footnote{\url{http://www.stsci.edu/software/tinytim/tinytim.html}}
PSF does not account for this effect, hence it does not
reproduce the observed \zacs-band PSF.
We studied how the PSF changes with wavelength using standard
stars taken with a series of narrow-band filter observations from HST
calibration programs PID9020 and PID10720.  We measured the red-halo
scattering correction factor as a function of aperture radius and
wavelength. For a given aperture, we then can treat the correction as
a modification of the F850LP throughput and zeropoint. We discuss the
details of this procedure in the Appendix.

\outline{draft}{zero points\\}

We have used the updated STScI ACS Vega zeropoints for light curve
fitting. The latest zeropoints are from the STScI 
web site\footnote{\url{http://www.stsci.edu/hst/acs/analysis/zeropoints}}
which were posted on May 19th, 2009.  The ACS zeropoints changed on
July 4th, 2006 due to a change in the detector temperature.  The STScI
definition assigns Vega ($\alpha$~Lyr) a magnitude of $0.0$ in every 
filter.
However, in the Landolt system, which the \snia photometry in the
literature refers to, Vega is not zero.  To be consistent with
literature \sneia, we introduce this non-zero Vega magnitude correction
to our zeropoints.  For both the \iacs and \zacs-bands, the correction
is 0.024 mag \citep{fukugita96a}. We therefore adjust the STScI
zeropoints by this amount \citep{astier06a}, and use Vega zero points of 
\iacs$=25.291$(STScI)$+0.024=25.315$ and \zacs$=24.347$(STScI)$+0.024=24.371$ 
for data taken before July 2006 
(corresponding to a detector temperature of $-77$C), and 
\iacs$=25.277$(STScI)$+0.024=25.301$ and \zacs$=24.323$(STScI)$+0.024=24.347$ 
for data taken after July 2006 
(corresponding to a detector temperature of $-81$C).

\ifemapj \begin{deluxetable*}{lllrrrcrrrr}
\else \begin{deluxetable}{lllrrrcrrrr}
\fi

\tablewidth{0pt}
\tablecaption{Photometry Data \label{tbl:b}}
\tablehead{
\colhead{SN name} & 
\colhead{Instrument} & 
\colhead{Filter} & 
\colhead{MJD} & 
\colhead{Flux\tablenotemark{a}} & 
\colhead{Flux Error\tablenotemark{a}} &
\colhead{Vega Zeropoint}\tablenotemark{d} &
\colhead{Exptime}&
\colhead{Nexp\tablenotemark{b}} &
\colhead{Raw Flux\tablenotemark{c}} & 
\colhead{Raw Flux Error\tablenotemark{c}}\\
\colhead{} & 
\colhead{} & 
\colhead{} & 
\colhead{} & 
\colhead{(counts/s)} & 
\colhead{(counts/s)} & 
\colhead{} &
\colhead{(s)} &
\colhead{} &
\colhead{(counts/s)} & 
\colhead{(counts/s)} 
}
\startdata
SCP05D0&ACS&F850LP&53564.098&-0.0283&0.0547&24.371&2000&4&-0.0188&0.0365 \\
SCP05D0&ACS&F850LP&53589.117&0.7733&0.0573&24.371&2000&4&0.5153&0.0381 \\
SCP05D0&NICMOS&F110W&53604.074&0.5092&0.0230&23.029&2560&2&\nodata&\nodata \\
SCP05D0&ACS&F850LP&53610.836&1.5084&0.0649&24.371&2000&4&1.0040&0.0432 \\
SCP05D0&ACS&F850LP&53633.184&0.7290&0.0704&24.371&1500&4&0.4764&0.0460 \\
SCP05D0&ACS&F775W&53633.215&0.7989&0.2229&25.315&375&1&\nodata&\nodata \\
SCP05D0&ACS&F850LP&53654.434&0.2388&0.0694&24.371&1500&4&0.1520&0.0442 \\
SCP05D0&ACS&F775W&53654.469&0.3908&0.1725&25.315&375&1&\nodata&\nodata \\
SCP05D0&ACS&F850LP&53679.258&0.1792&0.0708&24.371&1500&4&0.1137&0.0450 \\
SCP05D0&ACS&F775W&53679.273&-0.1432&0.1823&25.315&375&1&\nodata&\nodata \\
SCP05D0&ACS&F850LP&53704.266&0.0279&0.0715&24.371&1500&4&0.0174&0.0446 \\
SCP05D0&ACS&F775W&53704.305&0.0624&0.2072&25.315&375&1&\nodata&\nodata \\
SCP05D0&ACS&F850LP&53965.207&0.0356&0.0738&24.347&1360&4&0.0216&0.0448 \\
SCP05D0&ACS&F775W&53965.246&-0.0034&0.1981&25.301&515&1&\nodata&\nodata 
\enddata
\tablecomments{The complete set of \sneia\ photometry data is available
in its entirety in the electronic edition of the {\it Astrophysical Journal}.
A portion is shown here for guidance regarding its form and content.}
\tablenotetext{a}{Flux, corrected for CTE and color-dependent aperture correction (for 
ACS) and the count rate non-linearity (for NICMOS). 
For the ACS F850LP filter, \zacs, this is the result of the iterative Method I in Appendix \ref{sec:k}}
\tablenotetext{b}{The number of exposures.}
\tablenotetext{c}{CTE corrected but not aperture-corrected flux
for the ACS F850LP data.  
This flux column is used as an input for the modified filter Method II in Appendix \ref{sec:k}.
Note we use the modified filter response curve and shifted zeropoint as described in
Appendix \ref{sec:k}}
\tablenotetext{d}{The zeropoint has changed slightly after July 4th, 2006 (MJD=53920)
due to the change in detector temperature.}
\ifemapj \end{deluxetable*}
\else  \end{deluxetable}
\fi

\subsection{NICMOS Processing and Photometry} \label{sec:f}
\outline{draft}{Postprocessing NICMOS : refers to Eric's paper \NOTE{NS} \\}

All NICMOS science frames were processed with the latest {\tt CALNICA}
pipeline \citep[version 4.4.1][]{dahlen08a} and then corrected for
three well-known anomalies: 
the offset between amplifiers, which
affects all NICMOS exposures and is removed using the STSDAS {\tt
PyRAF} task {\tt PEDSKY}; persistence after passage of the telescope
through the South Atlantic Anomaly (SAA); and persistence after
exposing the detectors to the limb of the Earth. Nine exposures are
affected by the SAA, which leaves persistent signals from SAA cosmic
rays.  We applied the STSDAS {\tt PyRAF} task {\tt SAACLEAN}
\citep{barker07a} to remove SAA persistence effects from the
images. When a NICMOS observation is immediately preceded by an ACS
data dump, the process could delay the NICMOS placement of the filter
blank, subjecting the detectors to 
the bright limb of the Earth,
which imprints a persistent pattern on subsequent
exposures \citep{riess08a}. Four exposures were affected in this way
and were corrected using the STSDAS software {\tt NIC\_REM\_PERSIST}. 
At this point, the mode of the flux distribution
in each image is measured and recorded. These values are used as the
sky levels for the count-rate non-linearity correction.

Even after correcting NICMOS data for these well-studied anomalies,
significant large-scale background non-uniformities remain. We developed
methods to extract and remove the background structures; these are
detailed in \citet{hsiao11a}. Briefly, the models for
the background structures are studied and characterized using
approximately 600 NICMOS exposures observed through the F110W filter and
processed with the procedures described above. Principal component
analysis applied on these images revealed that the intensity of the
residual corner amplifier glow depends on the exposure sequence.
The amount of residual glow decays exponentially and resets
every orbit. With exposure times on the order of 1000 seconds, the
exposures can be separated into two glow groups, each with
approximately constant intensity. This makes it possible to extract
the residual glow algebraically. The structured background is modeled
as a combination of a constant component and a component that scales
with the sky level and exposure time. The models are derived from the
algebraic manipulation of stacked images for each glow group. The
resulting constant component of the model is dominated by residual
amplifier glow at the corners and residual persistence structure at
the center. The model component which scales with sky level and
exposure time displays a curious fringe pattern whose origin is
unknown. The model components are fit to individual exposures via
scale parameters to create the customized background models to be
subtracted from the individual exposures. In a final step, the bias
offsets apparent in the middle column and middle row are
removed. Additional details can be found in \citet{hsiao11a}.

\outline{draft}{NICMOS non-linearity correction 1\\}

\subsubsection{NICMOS Count-rate non-linearity}\label{sec:nicmosZP}

The NICMOS data are critically important for measuring the color of
$z>1$ \sneia. Any uncertainty in the NICMOS calibration severely
limits the usefulness of \sneia observed with NICMOS. In particular,
the NIC2 detector exhibits a count-rate dependent non-linearity
\citep{bohlin05a}, the severity of which is a function of wavelength.

This non-linearity previously has only been studied at count rates
three orders of magnitude higher \citep{dejong06a} than the count rate
of a typical \snia at $z=1$, meaning that correcting the flux of \snia
at $z=1$ requires significant extrapolation and has a level of
uncertainty that is difficult to quantify. For example, comparisons
between ground-based near-IR data and a different NICMOS camera (NIC3)
showed that little or no correction is required for that camera
\citep{mobasher05a}. It is difficult to reconcile this finding with
the findings of \citet{bohlin06a} and \citet{dejong06a}. A simple test
at the flux levels relevant for the supernovae in this paper shows a
difference of 12\% between NIC2 and NIC3 when the non-linearity
corrections are made, revealing significant problems with these
extrapolations.

For the NIC2/F110W filter, the degree of count-rate 
non-linearity is $\sim~0.06$ mag per factor of 10 change 
in count rate \citep{bohlin06a,dejong06a}. 
The count rates from stars that are used
to determine the NICMOS zero points are five orders of magnitude
higher than the count rate of a typical \snia at $z=1$. This
corresponds to a $\sim~0.3$ mag correction for the NIC2/F110W filter!

\outline{draft}{NICMOS non-linearity 3\\}
Since this is so important to the cosmological results, we have
developed a method to address this count-rate non-linearity
calibration directly (Ripoche et al. 2011). We analyze ACS, NICMOS,
and ground-based near-IR observations of early-type galaxies from
clusters RCS J0221.6$-$0347 ($z=1.02$), RDCS J1252.9$-$2927 ($z=1.24$), 
and XMMU J2235.3$-$2557 ($z=1.39$). The space and
ground-based data are used to constrain the spectral energy
distributions (SED) of these galaxies, which are then numerically
integrated through the F110W filter transmission curve and compared to
the counts measured with NICMOS. The principle advantage of the
technique is that the count rate from early type galaxies at this
redshift is similar to that measured for \sneia, i.e. about 0.03
counts/second/pixel (the contribution from amplifier glow is 
comparable). 
We applied this technique using three galaxy clusters
that have deep ground based near-IR imaging data from the VLT and deep
images with the ACS and NICMOS camera. 
All three clusters are at different redshifts and produced 
consistent results. 
At the low count rates that are applicable to high-redshift \sneia,
we find that the prescription of \citet{bohlin07a} and 
\citet{dejong06a} over-predicts the zeropoint correction for the NIC2
camera with the F110W filter by \PascalNICzpshift\ mag. 
We therefore use our zeropoint of \PascalNICzpVega\ (Vega magnitude ) or 
\PascalNICzpAB\ (AB magnitudes).  Additional details can
be found in \citet{ripoche11a}.

At high count rates, the count-rate non-linearity size has a strong
dependence with wavelength across filters \citep{dejong06a}, being
considerably stronger in bluer filters. The SED of an early-type
galaxy at $z\sim 1.2$, is a good match to a \snia\ about 20 rest-frame
days after maximum, but is redder than a supernova SED at maximum
(though this is compensated somewhat by the fact that the background
level is about 1/3 of the source flux and is blue in the F110W bandpass). 
The size of the count-rate non-linearity
correction will thus also depend weakly on the phase and redshift,
varying from 0.02 magnitudes at maximum to no additional correction 20
rest-frame days after maximum. Since the wavelength-dependence of the
non-linearity may not be even this strong at low count rates, we apply
half the correction applicable at each phase, and add (in quadrature)
an additional 0.01 magnitudes to the F110W zeropoint error to account
for this uncertainty. When added to the 0.006 mag statistical error,
and 0.021 mag systematic error \citep{ripoche11a}, 
this gives a total uncertainty on the
zeropoint of 0.024 magnitudes. For the GOODS supernovae with NICMOS
observations, we start with the original flux given by \citet{riess07a}
(after converting the magnitude measurements to fluxes using the given zeropoint of 22.92), but
increase the flux by 0.01 magnitudes, representing half the correction
for the (possible) wavelength-dependence of the count-rate non-linearity.

\outline{draft}{PSF fitting with Galaxy Models \NOTE{by LF and DR} \\} 

\subsubsection{Galaxy Models}
After the postprocessing described above \citep{hsiao11a}, we
measure fluxes from the eight \sneia\ with NICMOS observations by
performing PSF photometry on the images. In all cases, the \sneia are
not separated enough from their hosts to allow us to fit for the
supernova flux alone; rather we fit a model of the host galaxy
as well. By performing PSF photometry using analytic galaxy models, we
avoid resampling the images (the better PSF sampling for the ACS data
negates this advantage of PSF photometry), and extract the maximum possible
signal-to-noise from our observations. We fit an analytic model of
the host galaxy even when we have reference images, as this gives
higher signal-to-noise, and nearly uncorrelated photometry between epochs\footnote{Had we subtracted the flux in the reference images at the location of the supernova, the errors from this flux would have to be propagated as a covariance for all the other epochs. The errors on the galaxy model at the position of the supernova are typically much smaller.}.

Model PSFs for the supernovae are obtained with the TinyTim software
using supernova SED templates from \citet{hsiao07a} redshifted to the
supernova redshift and warped as a function of wavelength to match the
photometry. After the lightcurve fitting is complete, new PSFs are
generated from the SEDs based on this photometry and the process is
repeated. Model PSFs for the galaxies are obtained with TinyTim by
appropriately redshifting a galaxy spectrum from \citet{bruzual03a}
with an age of $2.5~\mathrm{Gyr}$ and a solar metallicity; the
exact shape of the galaxy spectrum does not greatly affect the
results. The PSFs used are $3\arcsec$ in diameter, comparable to the
patch fit in each NICMOS image.

Although there is virtually no information at scales smaller than
about half a pixel, all PSFs are seven times oversampled. This
oversampling is necessary because the PSF is made slightly wider by
the convolution with the subsampled pixels, increasing the flux of the
derived photometry. In order for this effect to be negligible, seven
times oversampling must be used. Finally, a correction is made to
match the photometry from the $3\arcsec$ TinyTim PSFs to the
$30\arcsec$ TinyTim PSFs used in Ripoche et al. (2011). These
differently-sized PSFs show different structure far in the wings, but
the flux in the core changes by $3.5\%$, with negligible variation.

We generally model the host galaxies as ellipsoids, with radial
profiles given by second degree polynomial splines. These splines have
ten nodes, with spacing that asymptotically approaches an exponential
away from the core. The higher node density near the core provides
more freedom to model the host where the flux changes quickly with
position. In the few pixels closest to the core, where the spline
changes rapidly, we numerically integrate over each subpixel before
convolving with the PSF. On the basis of our tests (see
\S\ref{sec:nicphottesting}), the hosts of some supernovae were modeled
with modifications to this basic scheme, as discussed in the following
section.

\subsubsection{Photometry Testing}\label{sec:nicphottesting}
Three ingredients all have to be correct in order to achieve
photometry with low bias and variance: the PSF model, the galaxy
model, and the supernova centroid. Deriving a PSF from a field star
(details in the SN~SCP06C0 discussion below) and comparing against
TinyTim gives photometry consistent to a few mmags, so we do not
believe this is a major contribution to our errors.

Testing the host galaxy model and supernova centroiding is more
involved. For each observation, we subtract the best-fit supernova
light, and place simulated supernovae (at the same flux level) in the
images. The only place one cannot do this test is at the location of
the actual supernova, as putting a simulated PSF in this location
yields a measurement that will be highly correlated with the
measurement of the supernova. We therefore do not place any simulated
supernovae closer than two pixels to the best-fit location of the
supernova. By examining the bias and variance of the extracted fluxes
from a large number of simulations ($\sim100$), we can choose the
galaxy model which gives the most precise and accurate\footnote{We
found that precision and accuracy correlated in our simulations.}
fluxes for each particular supernova. We emphasize that the results of
these simulations were the only metric used in choosing the detailed
model. In particular, there was no feedback from the shape of the
lightcurve or the Hubble diagram since these would have undercut the
principles of ``blind'' analysis we tried to maintain (see
\S\ref{sec:g}). The same basic galaxy model (discussed above) was used
for the NICMOS photometry of each supernova, with the following
exceptions.

\begin{itemize}

\item SN~SCP06C0: As mentioned in \S\ref{sec:SNsample}, there is
a small galaxy about $0.6\arcsec$ from the likely host of SN~SCP06C0,
and just $0.2\arcsec$ from SN~SCP06C0 itself. We note that the surface
brightness of the small galaxy is one fourth of that host at the
location of the supernova. The host also has some azimuthal asymmetry
visible, indicating a possible merger. The cluster XMM1229+01 was also
observed as part of a program to cross-calibrate NICMOS (Ripoche et
al. in preparation) and deep, well-dithered images were obtained in
the WFC3 F110W filter, allowing a more-flexible background model to
subtract both galaxies. We modeled the galaxies with a 2D second-order
spline, with nodes placed in a grid every $0.076 \arcsec$ (the natural
pixel scale of NICMOS). The WFC3 F110W PSF was modeled as a
combination of the elliptical galaxy model and a 2D spline (with a
spacing of $0.1 \arcsec$) using dithered images of a field star. (This
is the same empirical PSF model used for testing TinyTim for NICMOS,
although there the 2D spline nodes are spaced at the natural pixel
scale of NICMOS.) Our testing indicates that this method achieves the
same signal-to-noise ratio as the other supernovae that have simpler
galaxy subtractions.

\item SN~SCP06A4: We found a small amount of azimuthal asymmetry
in the host. Adding a second-order 2D spline to the galaxy model, with
a node spacing of $0.38 \arcsec$ (five times the natural pixel scale
of NICMOS 2) successfully modeled this asymmetry, without adding
additional measurement uncertainty to the supernova flux.

\item SN~SCP05D6: The host galaxy requires two elliptical components to
 be fitted well. These components are forced to have the same
 centroid, but are allowed different orientations, ellipticities, and
 radial profiles. One component forms a bulge, while the other one
 forms a disk. In one epoch
  contaminated by the SAA, aperture photometry with a one-pixel radius
  aperture on the galaxy-model-subtracted images gave better
  signal-to-noise than PSF photometry, so we used this instead.

\item SN~SCP06U4: This supernova was on the core of a galaxy that
  appears to be merging with another galaxy. Similarly to SN~SCP05D6,
  a second elliptical component was needed to model the host, (in this
  case, a third, detached component was used to model the fainter
  companion). Our simulated supernovae revealed that, rather than using
  one host galaxy model to extract photometry, even more precise results were obtained averaging photometry results derived using
  the elliptical model and the 2D spline model (discussed above for SN~SCP06C0).
  Using this procedure results in a change in flux well inside the error bar.

\item SN~SCP06H5: The one NICMOS observation of this supernova was
 our most challenging extraction. The observation of the supernova was
 11 rest-frame days after maximum, and it is only $\sim 0.1 \arcsec$
 from the core. As with SN~SCP05D6, the host galaxy requires two elliptical components to
 be fitted well.(Comparing to the 2D spline model discussed above, we obtain
 photometry that is the same to within a small fraction of the error bar.)

 The signal-to-noise ratio of this measurement is low, likely implying
 some amount of bias due to centroiding error. However, this is the
 only measurement with a signal-to-noise this low, so no correlation
 is introduced with any other measurement.

\end{itemize}

\subsection{Keck AO Photometry}

The photometry of $z > 1.2$ \sne has been almost exclusively measured
from HST images \citep{knop03a,riess04a,riess07a,amanullah10a}.  At these
redshifts, the rest frame $B$-band is redshifted beyond 9000 \ang, and
falls in the near-IR (NIR). NIR observations are typically much easier
from space, not only because of the higher spatial resolution of HST
compared to ground-based seeing-limited systems, but also because of
the lower NIR sky noise in space.  However, we show here that adaptive
optics on large ground-based telescopes can overcome these limitations
and allow high-$z$ SNe to be studied from the ground.

We observed the z=1.315 SN~SCP05D6 with the Keck Laser Guide Adaptive
Optics (LGS AO) system. These observations were made in the $H$-band
(1.6 $\mu$m), which corresponds to the rest-frame $R$-band. The
diffraction-limited resolution of the Keck AO images was
$\sim0.05\arcsec$, or a factor of three better than the spatial
resolution of HST at these wavelengths.  The high spatial
resolution meant a much better separation of the SN from the galaxy
core compared to HST. 
It also allows greater contrast between the SN and the sky background.
We obtain a photometric precision of 0.14~mag at H$\sim$24~mag in a 
one hour exposure with Keck AO, showing the potential of AO in 
\snia cosmology.

\citet{melbourne07a} reports the details of the Keck AO
photometry, here we briefly summarize the observations.
The Keck LGS AO observing runs for the Center for Adaptive 
Optics Treasury Survey  \citep{melbourne05a} coincided with 
the HST Cluster SN Survey program, and we successfully observed
SN~SCP05D6 at three epochs, before, near and after
the lightcurve maximum.

A 14~mag star, $25\arcsec$ away was chosen to
provide AO tip-tilt correction. The $\sim~11$~mag sodium LGS was
pointed at the galaxy to provide higher-order AO corrections. The
observations were sampled with a $0.01\arcsec$ pixel-scale, 
allowing the diffraction-limited Keck PSF to be fully resolved.
Individual exposures of 60s were taken with five dithered
pattern positions. 
The sequence was repeated until sufficient depth was reached. 
Total exposure times varied from 30 min to 1 hour per epoch.

We were also fortunate to have a 17.9 magnitude natural PSF star only
4$\arcsec$ away from SN~SCP05D6, so the PSF near the location of the
SN was well-determined. 
From the star, we measured a FWHM of $0.055\arcsec$
while we had a mean $H$-band seeing of $0.4\arcsec$.  
Using the observed PSF, the
host galaxy was modeled by GALFIT \citep{peng02a} and subtracted from
the image, providing a clean measurement of the SN diffraction-limited
core.  Relative photometry with respect to the nearby PSF star was
performed at each epoch, and calibrated by the UKIRT standard star FS6
(the photometry is reported in Table~\ref{tbl:b}).  The photometric
uncertainty was estimated by simulations of model PSFs embedded into
the AO image at the same galacto-centric radius as the actual SN.

We fit the SN-lightcurve with the photometric data from HST/ACS
F775W, F850LP, HST/NICMOS F110W, and this Keck/AO observation. We
found that the Keck AO data was consistent with the HST observations
(Figure \ref{fig:b}). Although the uncertainty in the AO measurement 
was larger than that of HST, including it reduces the color 
uncertainty by more than 10\%, and reduces the sensitivity of 
the fit to the NICMOS zeropoint uncertainty by more than a third.


\section{Augmenting the Union2 Supernova Compilation: Union2.1}
\label{sec:g}

\sneia are an excellent probe of dark energy, as they measure the
magnitude-redshift relation with very good precision over a wide range
of redshifts, from $z=0$ up to $z\sim1.5$ and possibly beyond.  While
some individual sets of \sneia are now, by themselves, large enough to
provide constraints on some cosmological parameters
\citep{guy10a,kessler09a}, they do not yet constrain the properties of
dark energy as well as analyses that combine individual data-sets to
create a compilation of \sneia that covers a broader range of
redshifts. In \citet{kowalski08b}, we developed a systematic
methodology for combining the many available datasets into one
compilation, called the ``Union'' compilation.

There are many positive features behind the philosophy adopted by the
Union analysis. It includes all \snia data-sets on an equal footing,
with the same lightcurve fitting, cuts, and outlier rejection.
Estimates of the systematic error are entered into a covariance
matrix, which can be used for fitting any cosmological model.  Choices
about how to do the analysis and what cuts to apply are done with the
cosmological results hidden. This type of ``blind'' analysis mitigates
biases that arise from inadvertently scrutinizing some data more than
others. In \citet{amanullah10a}, we adopted this strategy to create
the Union2 compilation. This paper also revised and improved the Union
analysis in several significant ways. Firstly, it augmented the Union
sample with new \snia data-sets from the literature, including 102
low-redshft \sneia from the CfA3 survey \citep{hicken09b}, 129
intermediate-redshift \sneia from the SDSS SN survey
\citep{holtzman08a}, five intermediate-redshift \sneia discovered from La
Palma \citep{amanullah08a}, and six 
new high-redshift \sneia. 
The paper revised the analysis by replacing the SALT
lightcurve fitter with SALT2 \citep{guy05a,guy07a}, and handled many
systematic errors on a supernova-by-supernova basis in a covariance
matrix.

In the current paper, we use the analysis procedure that was
used for the Union2 compilation with only one significant change: a
correction for the host-mass \snia-luminosity relation, described below. The
HST calibration and the associated errors have also been updated, as
described in Section \ref{sec:diagnostics}. We refer to this new
compilation as ``Union2.1.''

\subsection{Host Mass Correction to \snia Luminosities}\label{sec:masscorrection}

There is evidence that \snia luminosity correlates with the mass of
the host galaxy, even after the corrections for color and light curve
width have been applied \citep{kelly10a,sullivan10b,
  lampeitl10b}. Since low-redshift \sneia are predominantly from
surveys that target catalogued galaxies, the host galaxies of \sneia in
these surveys are, on average, more massive than the host galaxies of
distant \sneia from untargeted surveys. \sneia from low-redshift
samples therefore have brighter absolute magnitudes. Left uncorrected, the
correlation biases cosmological results \citep{sullivan10b}.

\newcommand{\msobs}{m^{\mathrm{obs}}_{\star}}
\newcommand{\mstrue}{m^{\mathrm{true}}_{\star}}
\newcommand{\msthresh}{m^{\mathrm{threshold}}_{\star}}
\newcommand{\loms}{\mstrue < \msthresh}

\citet{sullivan10b} find that the correlation can be corrected by fitting a
step in absolute magnitude at $\msthresh =\ 10^{10} m_{\sun}$.
There are two complications with making this correction: most of the
SNe in the Union2 compilation do not have host mass data available in
the literature, and \snia\ hosts with masses close to the cutoff may
scatter across, decreasing the fitted size of the step. To address
these problems, we adopt a probabilistic approach to determining the proper
host mass correction to apply to each supernova, correcting each
supernova by the probability that it belongs in the low-host-mass
category. (The low-host-mass category was chosen because most of the 
low-redshift supernovae are from high-mass galaxies, so correcting 
the low-host-mass supernovae minimizes the correlation between 
$M_B$ and the correction coefficient.)

Suppose we have a
mass measurement $\msobs$ and we would like to estimate the
probability that the true mass $\mstrue$ is less than the mass threshold.
We begin by
noting that
\begin{eqnarray}
&&P(\msobs, \mstrue) = \\
&&P(\msobs | \mstrue) P(\mstrue) \;. \nonumber
\end{eqnarray}
We can then integrate this probability over all true host masses less
than the threshold:

\begin{eqnarray}
&&P(\loms | \msobs) = \\
&&\int_{\mstrue = 0}^{\msthresh} P(\msobs | \mstrue) P( \mstrue)\nonumber
\end{eqnarray}
up to a normalization constant found by requiring the integral to be
unity when integrating over all possible true masses. $P(\mstrue)$ is
estimated from the observed distribution for each type of
survey. The SNLS \citep{sullivan10b} and SDSS \citep{lampeitl10b} host
masses were assumed to be representative of untargeted surveys, while
the mass distribution in \cite{kelly10a} was assumed typical of nearby
targeted surveys. As these distributions are approximately log-normal, we use this model for $P(\mstrue)$
using the mean and RMS from the log of the host masses from these surveys (with the average
measurement errors subtracted in quadrature), giving
$\log_{10} P(\mstrue) = \mathcal{N}\untargetedmusigma$ for untargeted
surveys and $\log_{10} P(\mstrue) = \mathcal{N}\targetedmusigma$
for targeted surveys. When host mass measurements are available,
$P(\msobs |\mstrue)$ is also modeled as a log-normal; when no
measurement is available, a flat distribution is used.

For a supernova from an untargeted survey with no host mass
measurement (including supernovae presented in this paper which are
not in a cluster), $P(\loms)$ is the integral of $P(\mstrue)$ up to the threshold mass: \untargetedlow. Similarly, nearby supernovae
from targeted surveys without host galaxy mass measurements are given
a $P(\loms)$ of \targetedlow. (Very similar numbers of \untargetedlownoGauss and \targetedlownoGauss are derived from the observed distribution, without using the log-normal approximation.) We must make the correction for
supernovae in clusters, as these are from a targeted survey. We
take advantage of the simpler SEDs of early-type galaxies to precisely
measure these masses\footnote{C-001 and F-012 are in clusters, but
  are not hosted by early-type hosts. We use the untargeted value for
  their host-mass--luminosity relation correction.}.

The best-fit mass-correction coefficient, $\delta$, is much smaller in magnitude
($\bestfitdelta$) than that found in other studies ($\approx
-0.08$). This may be due to the small value for $\delta$ from the
first-year SNLS data, as shown in Table \ref{tb:subsets}. We include
the difference in $\delta$s as a systematic, as discussed in
\S\ref{sec:systematics}.
For this analysis, we assumed the host-mass correction does not evolve with 
redshift.

\subsection{Light-Curve Fitting}
\label{sec:lcfitting}

\outline{draft}{Why SALT2 : Brief Description of SALT2\\} Following
\citet{amanullah10a}, we use SALT2 \citep{guy07a} to fit supernova
lightcurves. The SALT2 model fits three parameters to each SNe: an
overall normalization, $x_0$, to the time dependent spectral energy
distribution (SED) of a \snia, the deviation, $x_1$, from the average
lightcurve shape, and the deviation, $c$, from the mean \snia $B-V$
color. The three parameters, $x_1$, $c$, and integrated $B$-band flux
of the model SALT2 SED at maximum light, $m_B^\mathrm{max}$, are then
combined with the host mass to form the distance modulus

\begin{equation}
  \mu_B = m_B^\mathrm{max} + \alpha\cdot
  x_1 - \beta\cdot c + \delta \cdot P(\loms) - M_B\, ,\label{eq:magcor}
\end{equation}
where $M_B$ is the absolute $B$-band magnitude of a \snia with
$x_{1}=0$, $c=0$ and $P(\loms)=0$. The parameters
$\alpha$, $\beta$, $\delta$ and $M_B$ are nuisance parameters that
are fitted simultaneously with the cosmological parameters. The
\snia\ photometry data and SALT2 light curve fits are shown in Figure
\ref{fig:b}. The fitted SALT2 parameters are listed in Table \ref{tbl:c}
as well as the host galaxy host stellar mass and lensing magnification
factor.

\ifemapj \begin{figure*}[]
\else \begin{figure}[]
\fi

\begin{center}
\includegraphics[angle=270,scale=.33,width=0.68\columnwidth]{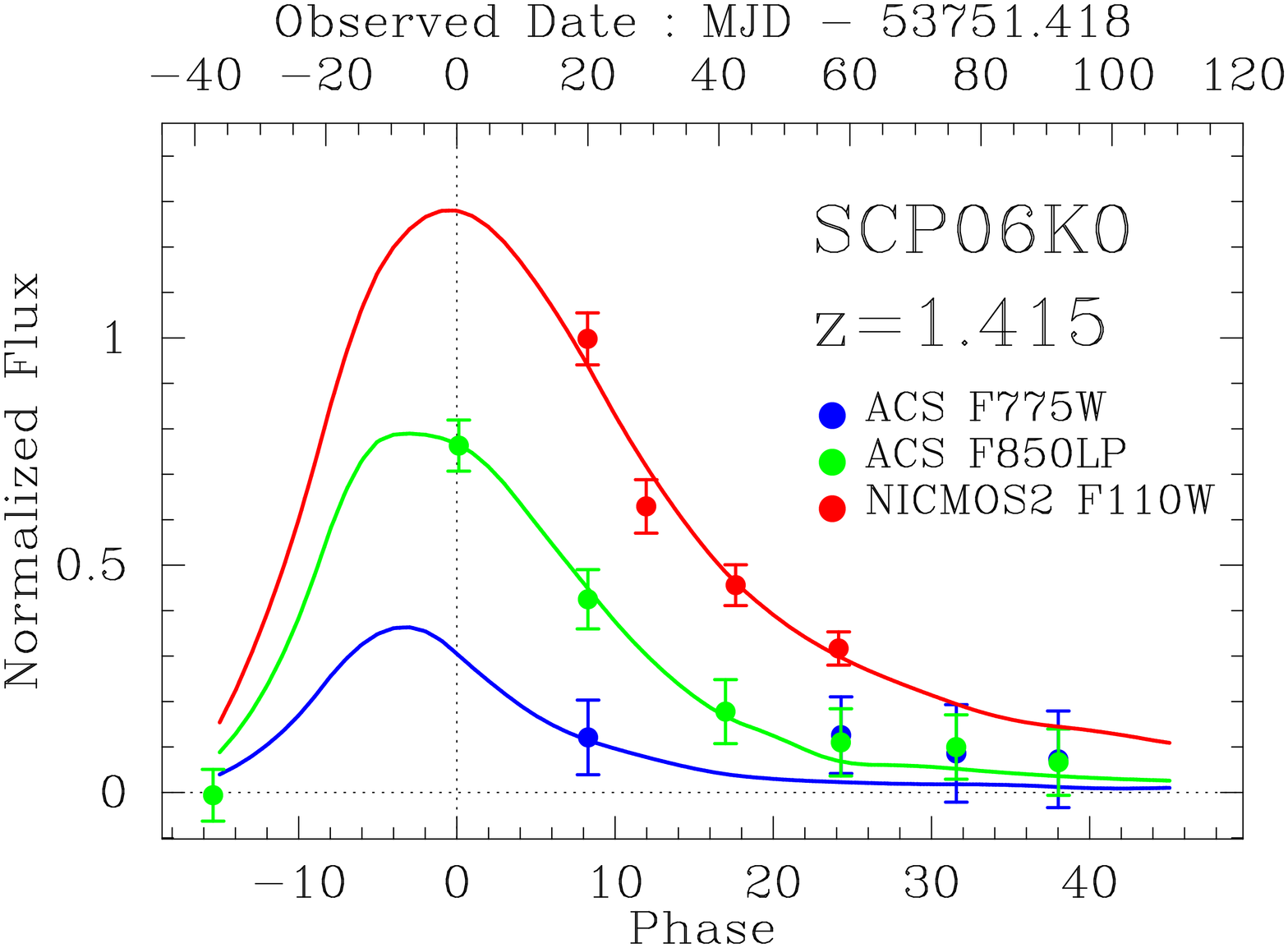}
\includegraphics[angle=270,scale=.33,width=0.68\columnwidth]{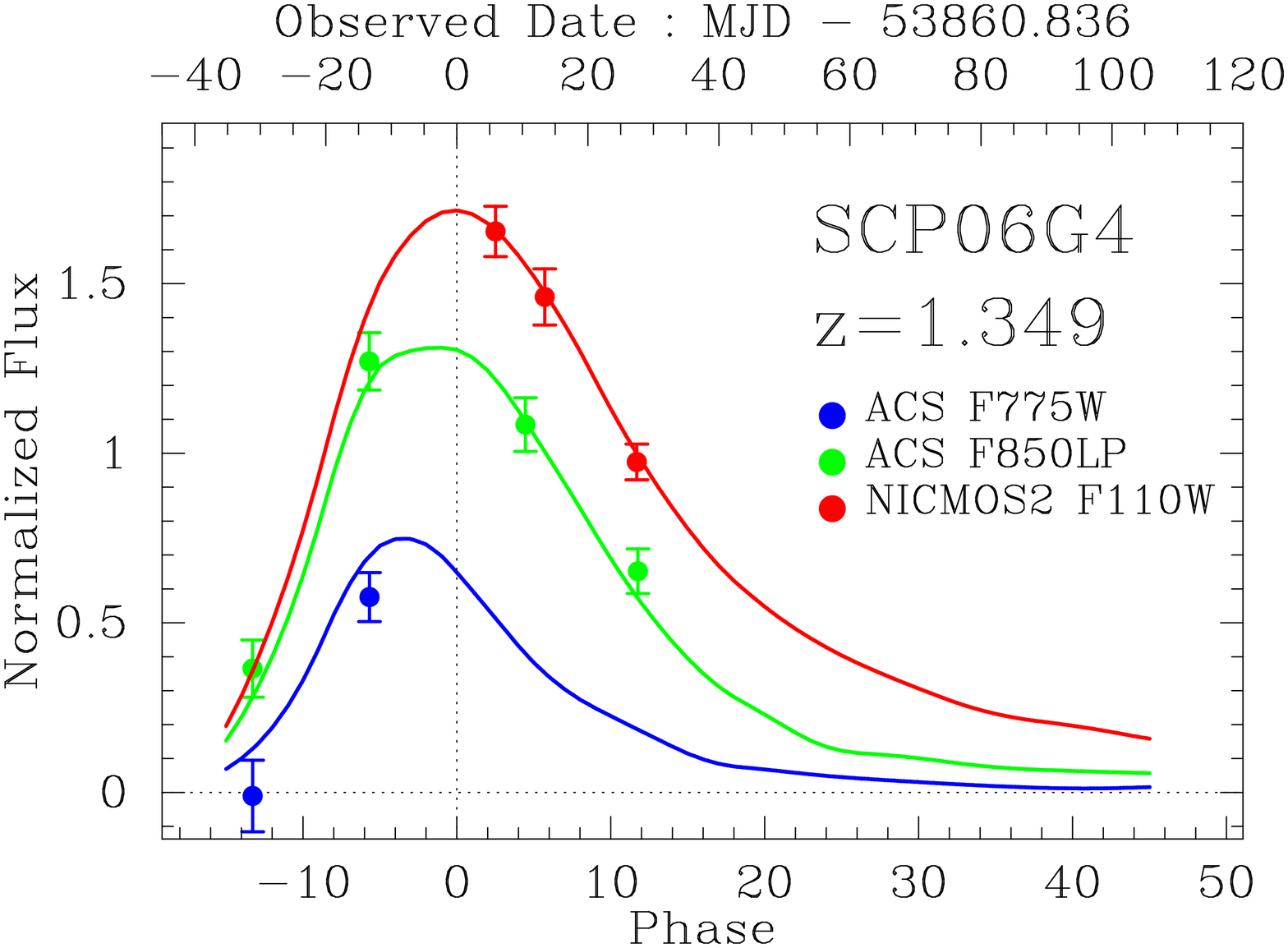}
\includegraphics[angle=270,scale=.33,width=0.68\columnwidth]{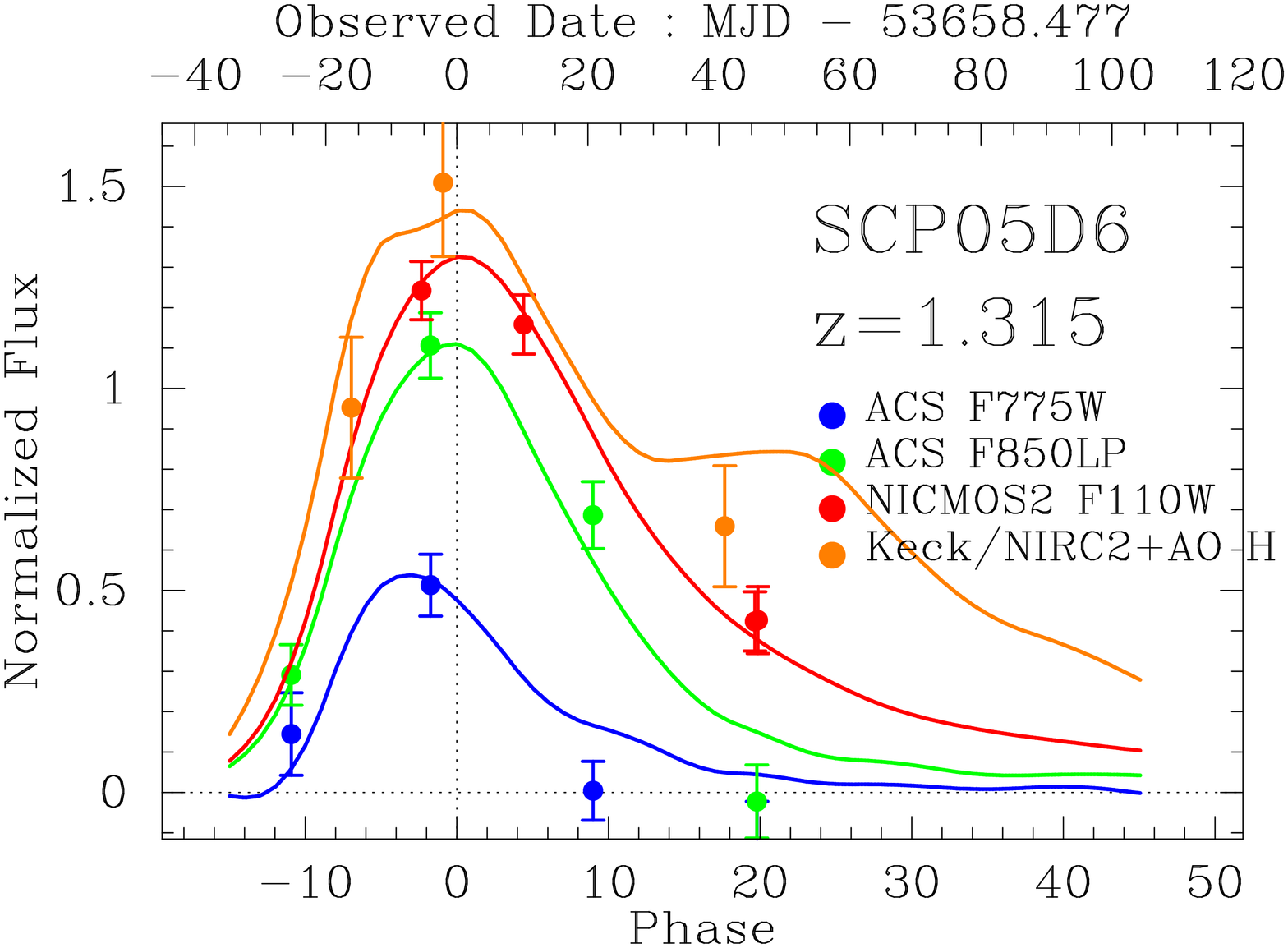}
\vspace{2mm}
\includegraphics[angle=270,scale=.33,width=0.68\columnwidth]{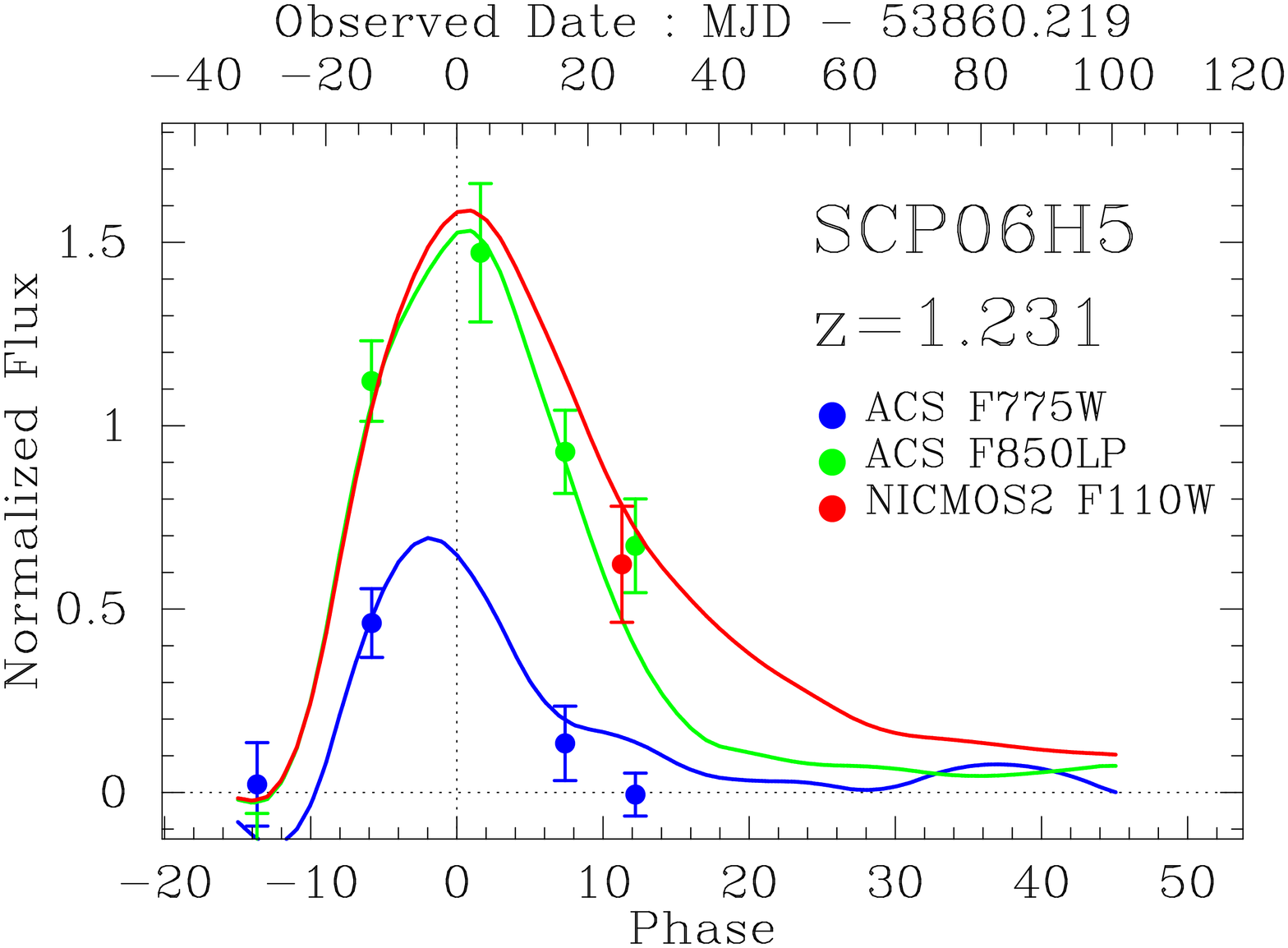}
\includegraphics[angle=270,scale=.33,width=0.68\columnwidth]{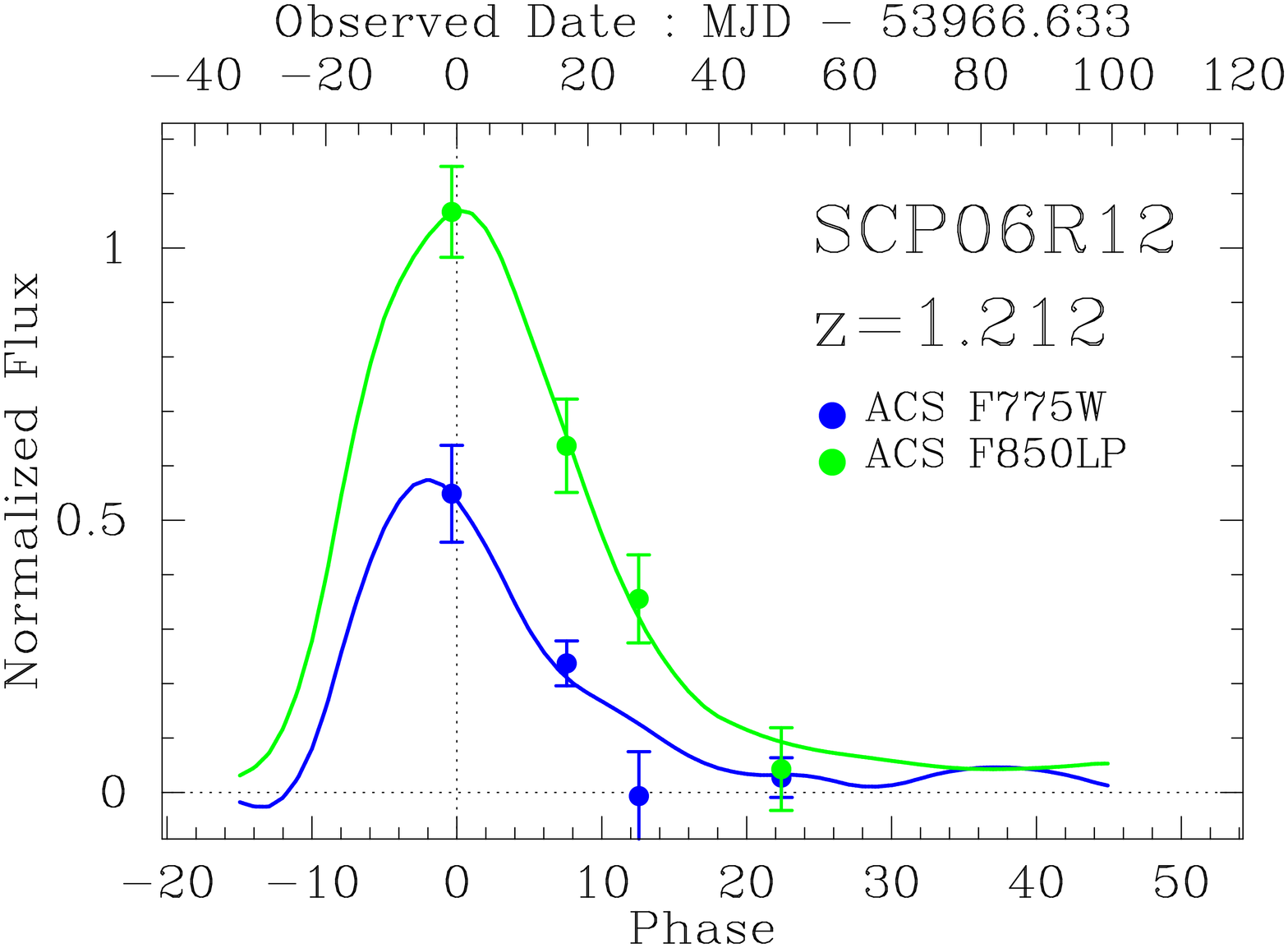}
\includegraphics[angle=270,scale=.33,width=0.68\columnwidth]{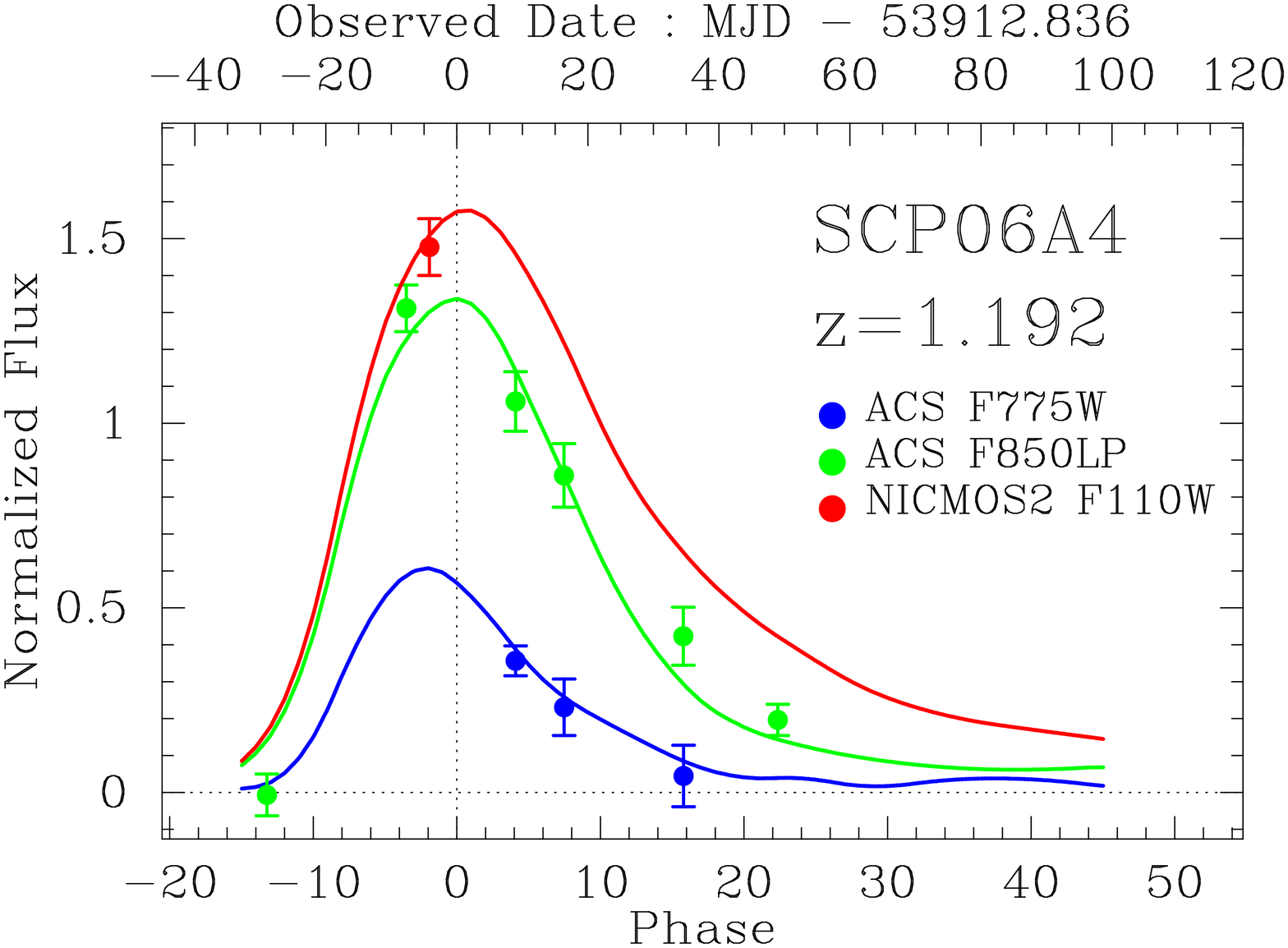}
\vspace{2mm}
\includegraphics[angle=270,scale=.33,width=0.68\columnwidth]{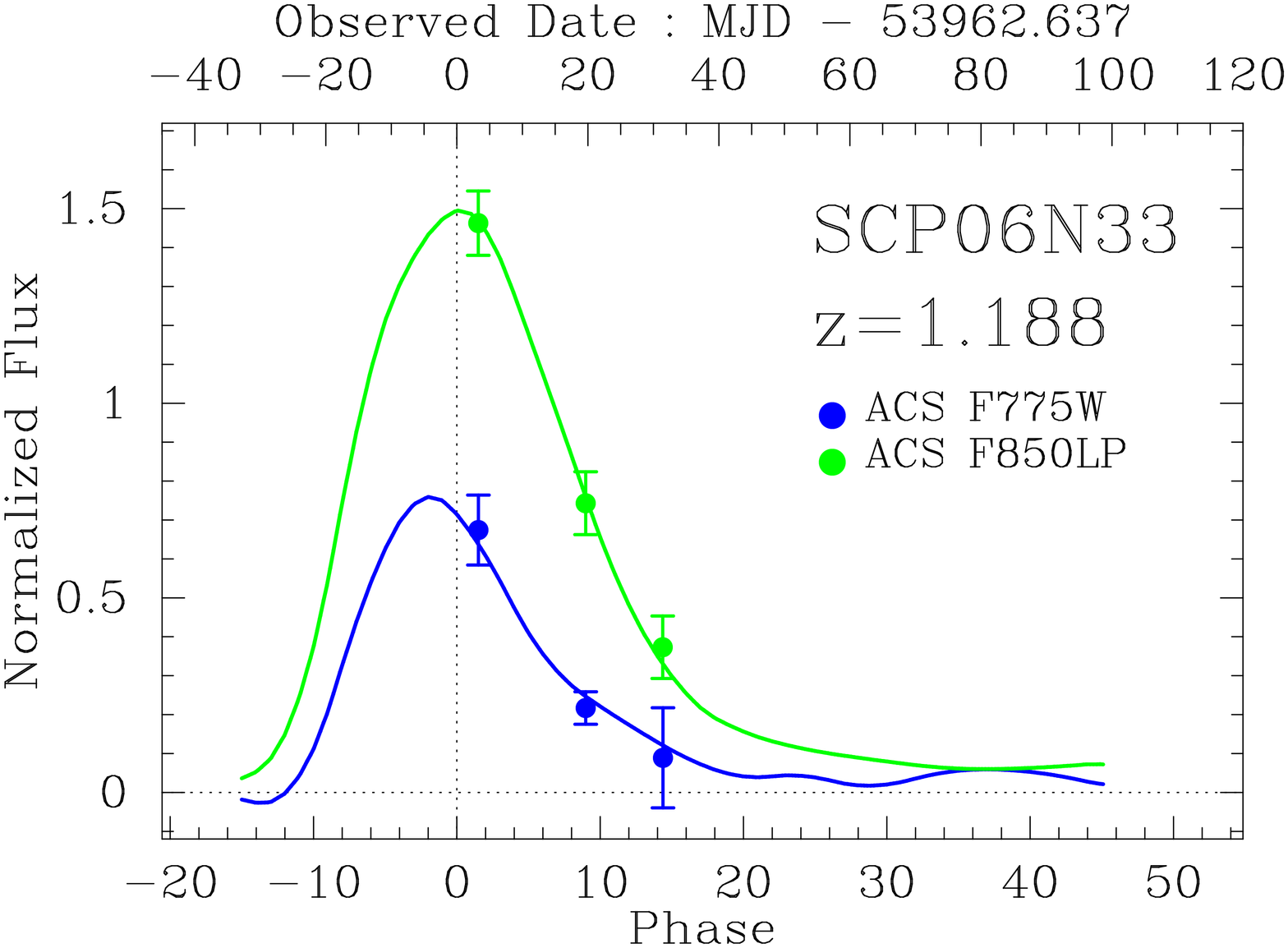}
\includegraphics[angle=270,scale=.33,width=0.68\columnwidth]{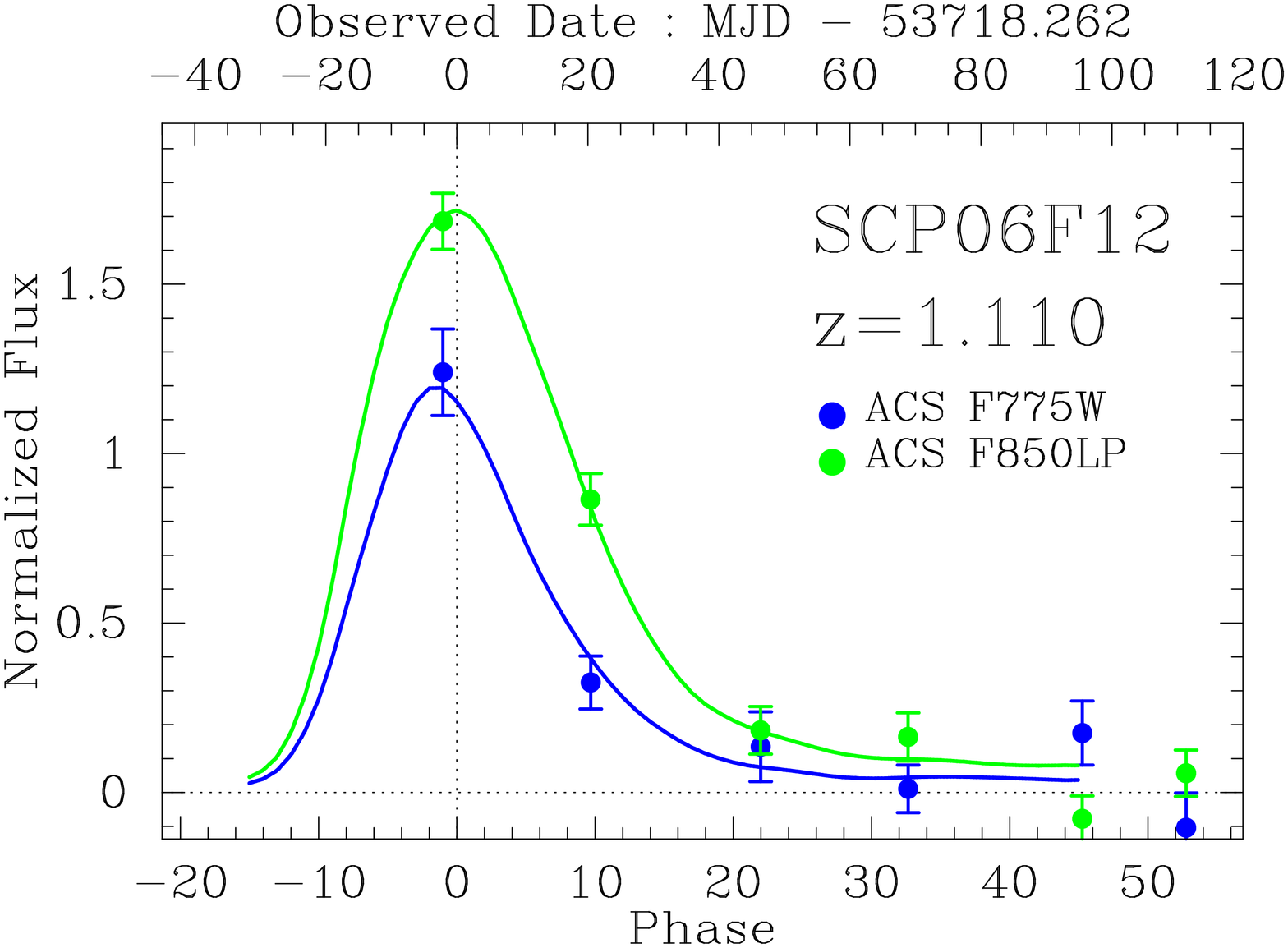}
\includegraphics[angle=270,scale=.33,width=0.68\columnwidth]{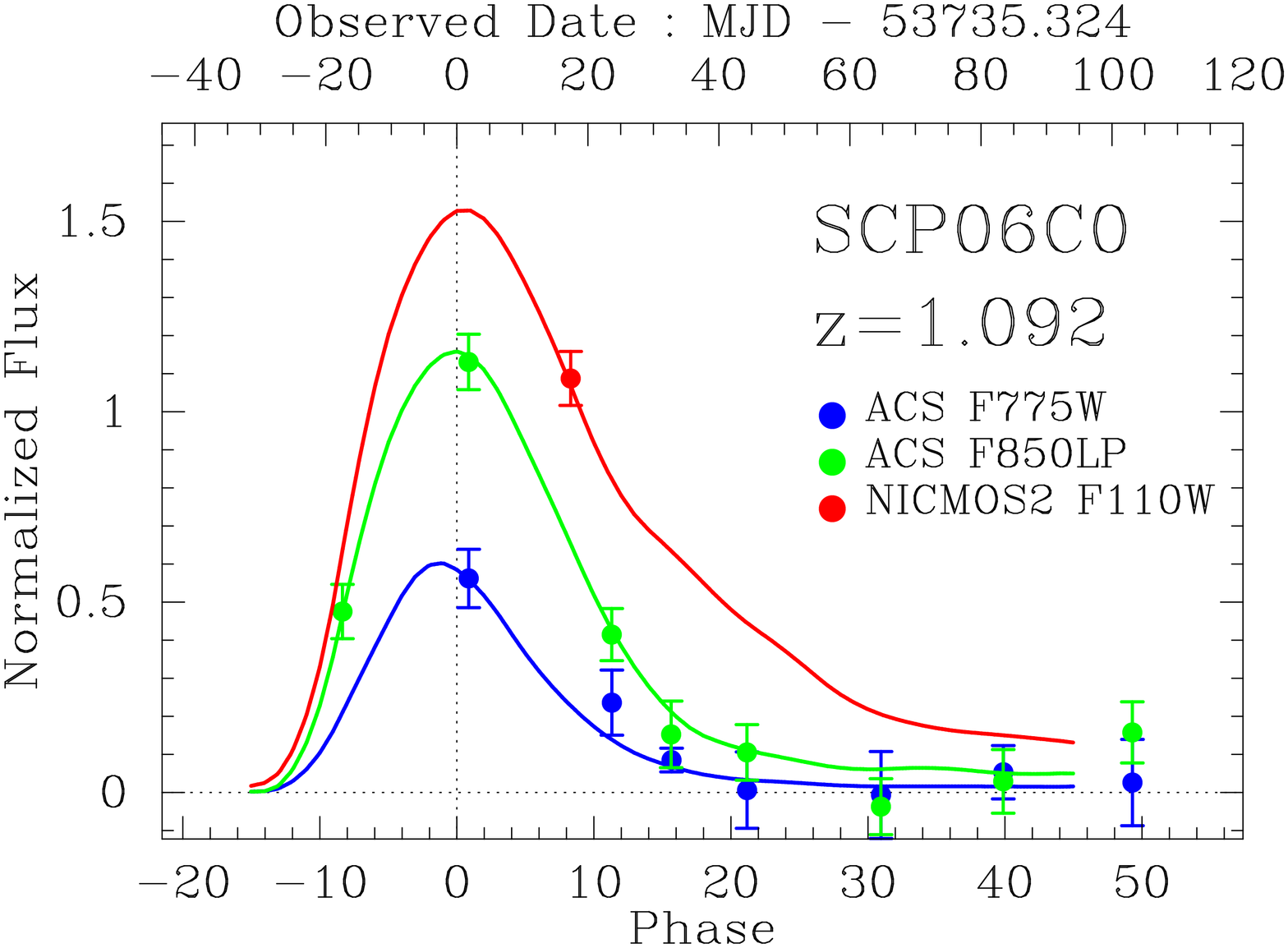}
\vspace{2mm}
\includegraphics[angle=270,scale=.33,width=0.68\columnwidth]{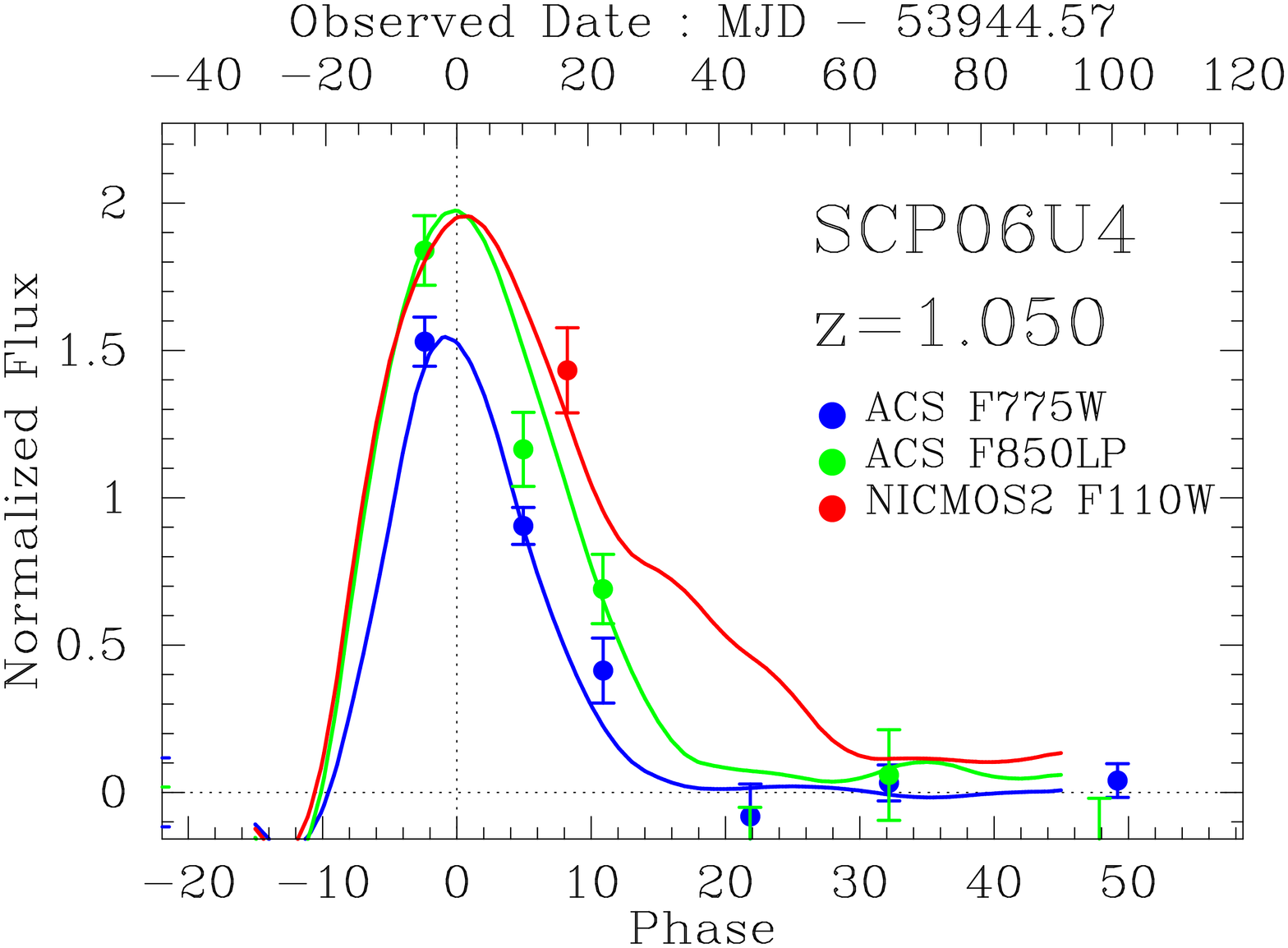}
\includegraphics[angle=270,scale=.33,width=0.68\columnwidth]{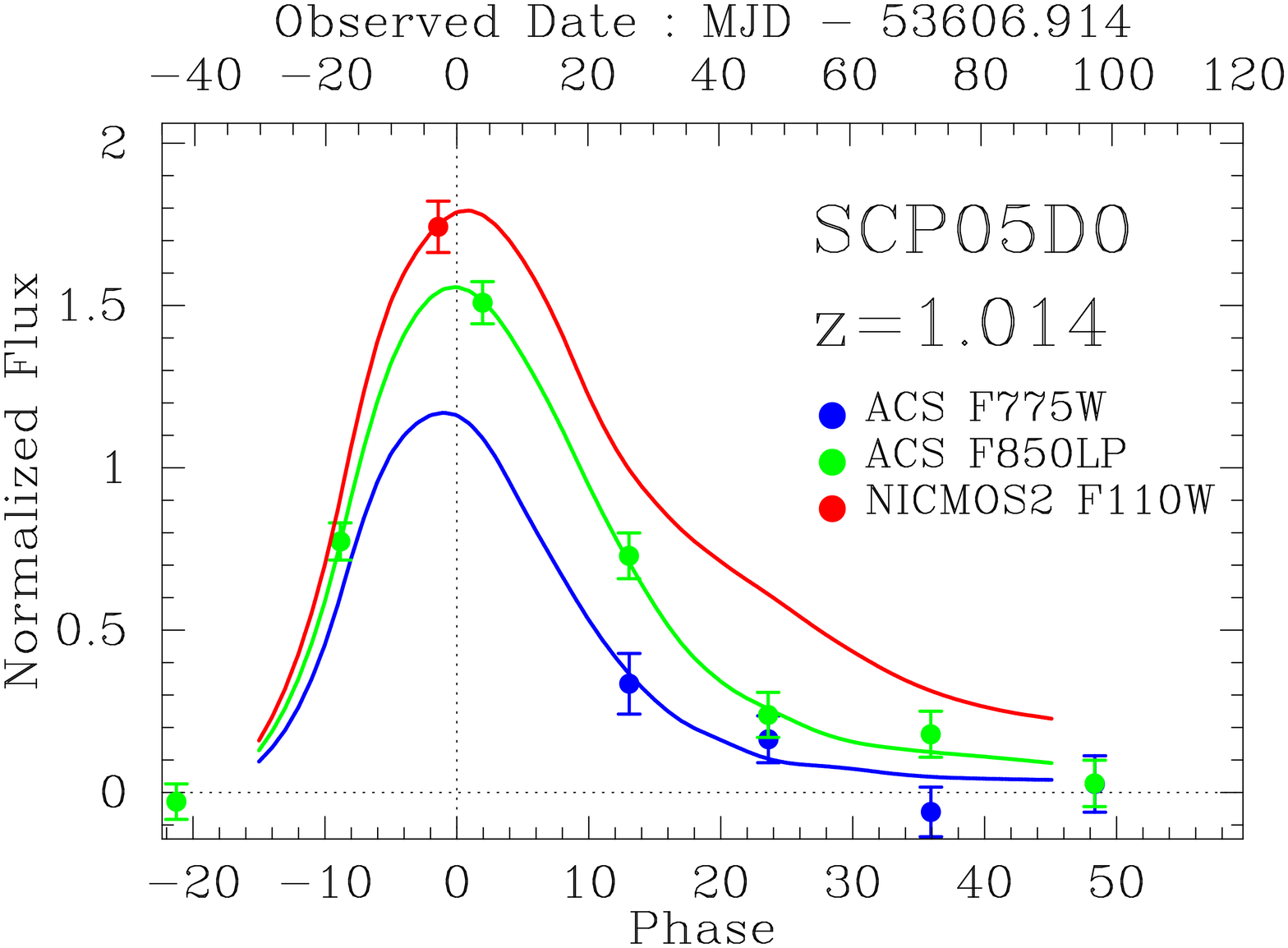}
\includegraphics[angle=270,scale=.33,width=0.68\columnwidth]{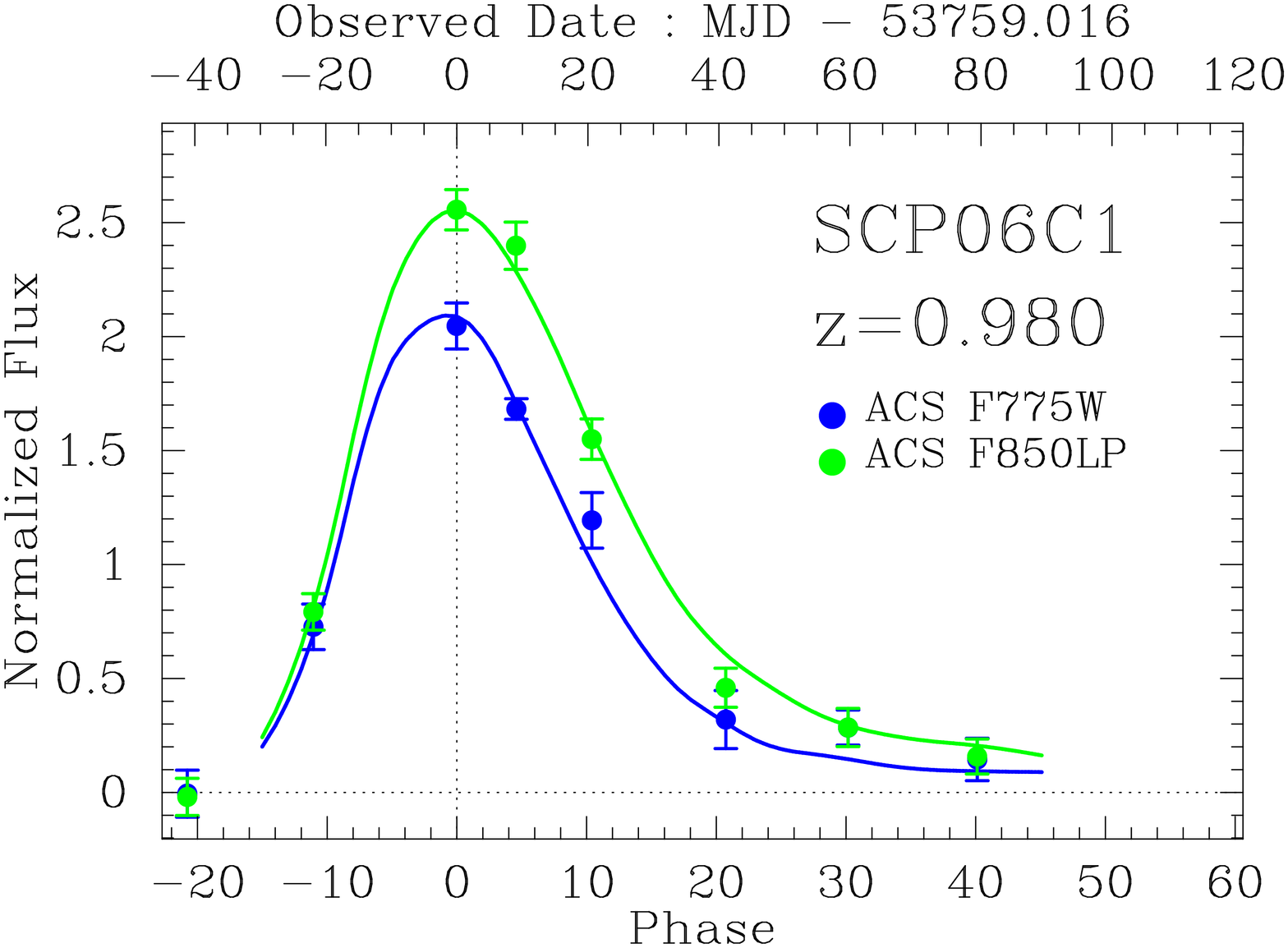}
\vspace{2mm}
\includegraphics[angle=270,scale=.33,width=0.68\columnwidth]{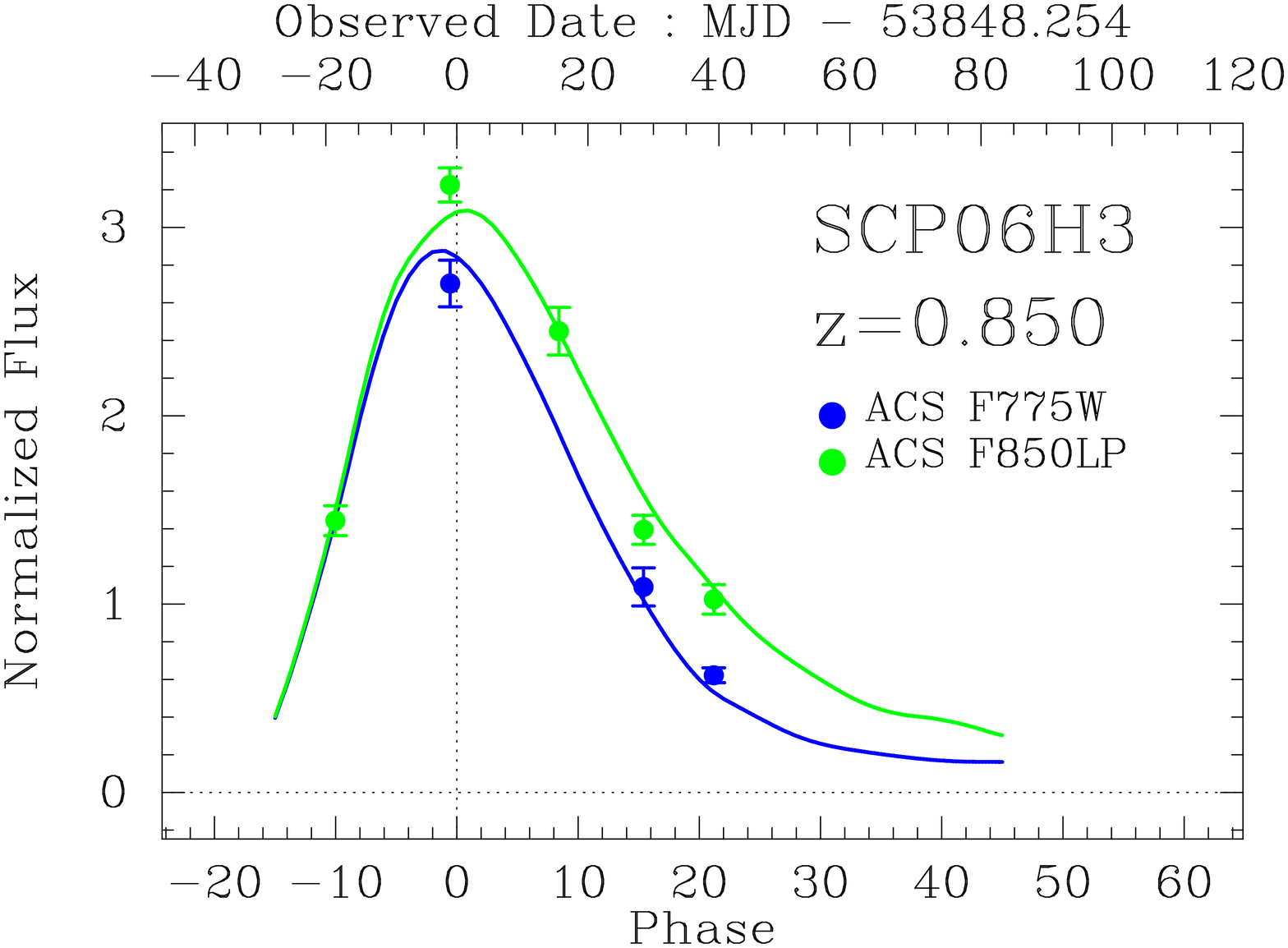}
\includegraphics[angle=270,scale=.33,width=0.68\columnwidth]{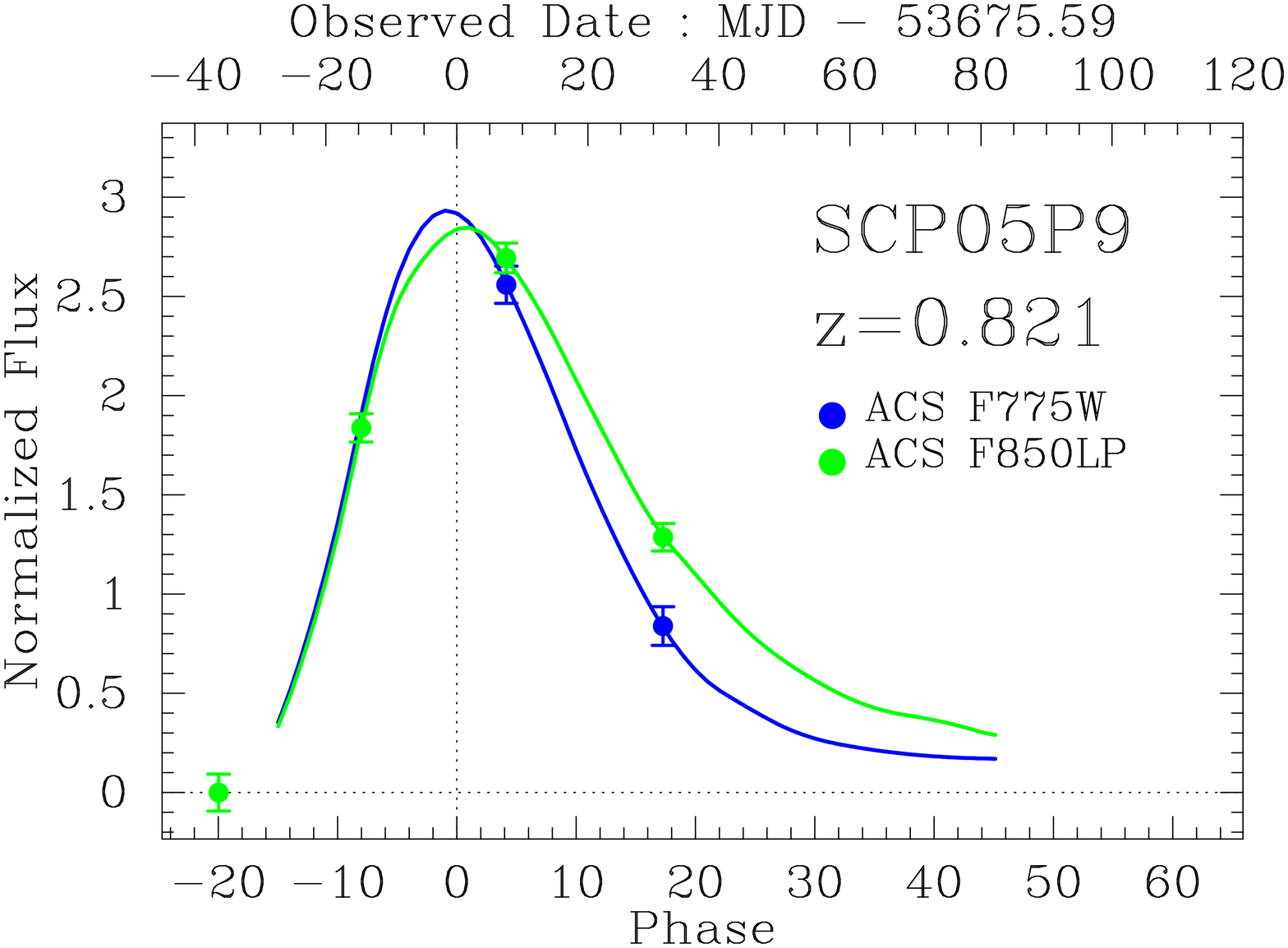}
\includegraphics[angle=270,scale=.33,width=0.68\columnwidth]{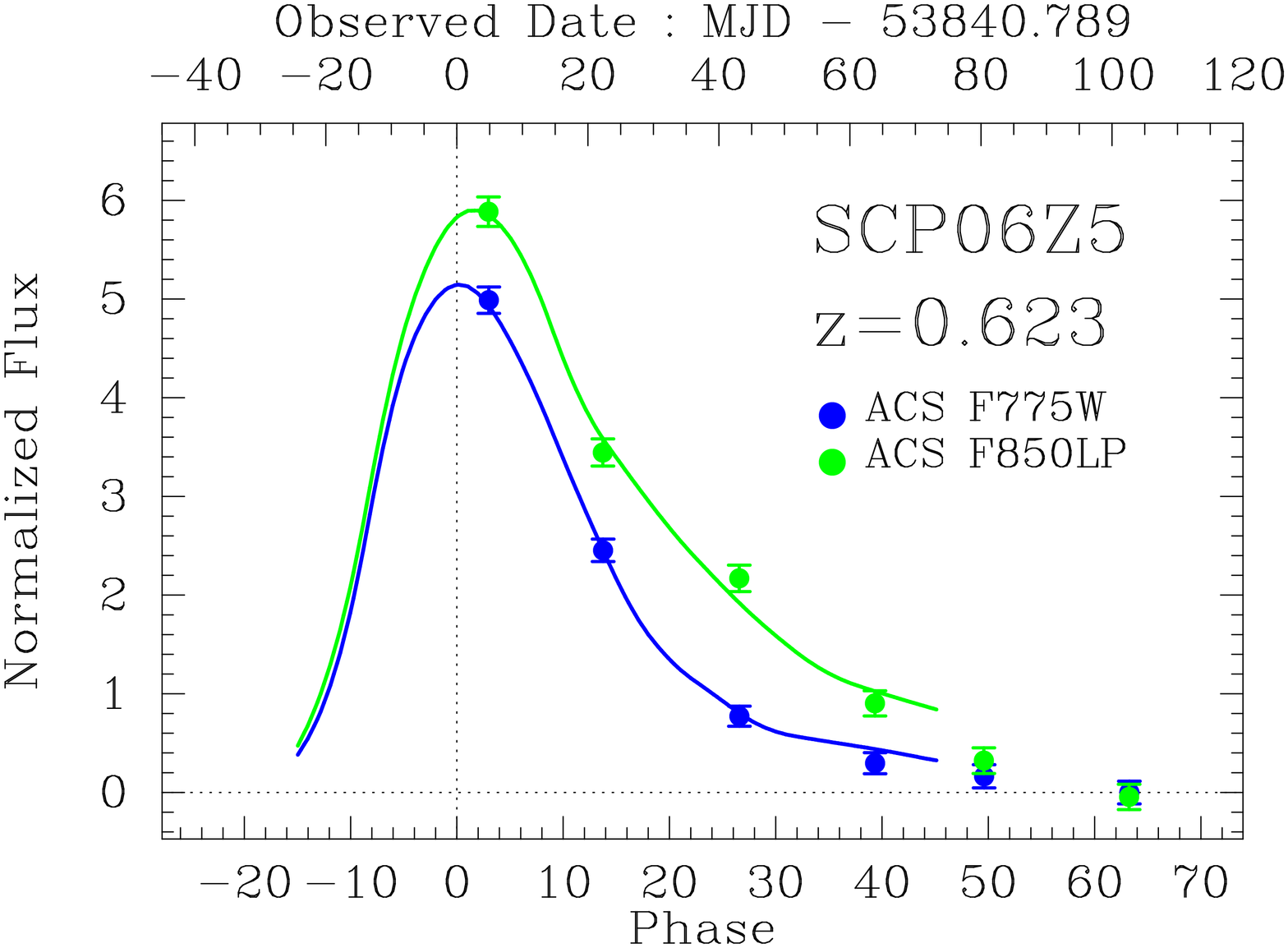}
\end{center}
\caption{\NOTE{fig:b to be revised : v150} 15 \sneia light curve fits by SALT2.
Flux is normalized to the \zacs-band zeropoint magnitude. 
ACS \iacs, ACS \zacs and NICMOS F110W data is color coded in blue, green and
red respectively.  Note that SCP~05D06 (z=1.314) has $H$-band data from Keck AO system (orange)
\citep{melbourne07a} and that this data is consistent with the HST/ACS and HST/NICMOS light curve data.
} \label{fig:b} \label{fig:ltcv}

\ifemapj \end{figure*}
\else  \end{figure}
\fi

\ifemapj 
\begin{deluxetable*}{lrrrrrcc}
\else 
\begin{deluxetable}{lrrrrrcc}
\tabletypesize{\footnotesize}
\fi
\tablewidth{0pt}
\tablecaption{SALT2 Lightcurve Fit Results \label{tbl:c} \label{tbl:ltcvfit}}
\tablehead{
\colhead{SN name} &
\colhead{$z$} &
\colhead{MJD$_{\rm Bmax}$} &
\colhead{$m_{B}$} &
\colhead{$x_{1}$} &
\colhead{$c$} &
\colhead{Galaxy Mass\tablenotemark{a}} &
\colhead{Lens} \\
\colhead{} &
\colhead{} &
\colhead{} &
\colhead{} &
\colhead{} &
\colhead{} &
\colhead{(10$^{11}{\rm M_{\odot}}$)} &
\colhead{Factor\tablenotemark{b}}
}
\startdata
SCP06A4 & 1.192 & $53912.7\pm 1.5$ & $25.497\pm 0.048$ & $-1.45\pm 0.68$ & $0.065\pm 0.084$ & 0.44 & \nodata\\
SCP06C0 & 1.092 & $53735.4\pm 1.0$ & $25.636\pm 0.066$ & $-2.66\pm 0.65$ & $0.257\pm 0.083$ & 1.97 & $1.030^{+0.007}_{-0.005}$\\
SCP06C1 & 0.980 & $53759.0\pm 0.7$ & $24.613\pm 0.028$ & $-0.35\pm 0.33$ & $0.014\pm 0.053$ & \nodata & \nodata\\
SCP06F12 & 1.110 & $53718.4\pm 2.3$ & $25.253\pm 0.068$ & $-2.09\pm 1.29$ & $-0.133\pm 0.142$ & \nodata & \nodata\\
SCP06G4 & 1.350 & $53860.9\pm 1.4$ & $25.424\pm 0.052$ & $0.15\pm 0.64$ & $-0.029\pm 0.052$ & 1.72 & $1.015^{+0.005}_{-0.004}$\\
SCP06H3 & 0.850 & $53848.2\pm 0.6$ & $24.345\pm 0.038$ & $0.58\pm 0.31$ & $0.089\pm 0.067$ & \nodata & \nodata\\
SCP06H5 & 1.231 & $53860.2\pm 1.5$ & $25.389\pm 0.111$ & $-3.12\pm 1.10$ & $-0.103\pm 0.187$ & 3.66 & \nodata\\
SCP06K0 & 1.415 & $53751.3\pm 2.8$ & $25.811\pm 0.087$ & $0.30\pm 0.97$ & $0.147\pm 0.081$ & 2.30 & \nodata\\
SCP06N33 & 1.188 & $53962.6\pm 4.3$ & $25.407\pm 0.132$ & $-2.15\pm 1.32$ & $-0.038\pm 0.175$ & \nodata & $1.066^{+0.017}_{-0.014}$\\
SCP05D0 & 1.014 & $53606.9\pm 0.9$ & $25.201\pm 0.066$ & $-0.61\pm 0.65$ & $0.061\pm 0.085$ & 0.40 & \nodata\\
SCP05D6 & 1.315 & $53658.5\pm 1.3$ & $25.660\pm 0.046$ & $-1.26\pm 0.56$ & $-0.058\pm 0.061$ & 2.61 & $1.021^{+0.012}_{-0.008}$\\
SCP05P9 & 0.821 & $53675.6\pm 0.6$ & $24.367\pm 0.049$ & $0.25\pm 0.50$ & $0.022\pm 0.075$ & \nodata & \nodata\\
SCP06R12 & 1.212 & $53966.6\pm 3.5$ & $25.789\pm 0.114$ & $-2.06\pm 1.50$ & $-0.158\pm 0.198$ & 0.23 & \nodata\\
SCP06U4 & 1.050 & $53944.4\pm 1.1$ & $25.056\pm 0.063$ & $-4.62\pm 1.09$ & $-0.102\pm 0.096$ & 1.11 & \nodata\\
SCP06Z5 & 0.623 & $53840.5\pm 3.0$ & $23.482\pm 0.144$ & $-0.76\pm 0.88$ & $0.070\pm 0.120$ & \nodata & \nodata
\enddata
\tablenotetext{a}{The details of host galaxy identifications, coordinates and 
its stellar mass measurements can be found in \citet{meyers11a}.}
\tablenotetext{b}{Gravitational lensing magnification
  factor (see \S \ref{sec:lens} for details). For cosmological analysis
  we must divide the corrected \sne fluxes by this factor to make use of these supernovae.}
\tablenotetext{c}{SCP06U4 is not included in our current cosmological results, but will likely be included in future compilations
  (see \S \ref{sec:g} for details).}

\ifemapj \end{deluxetable*}
\else  \end{deluxetable}
\fi

\subsection{Union2.1}
\outline{draft}{Union Concept\\}

To the Union2 \snia\ compilation \citep{amanullah10a}, we add
\kitchensinksample \sneia\ from this paper that were classified as
either secure or probable, including six \sneia hosted by 
high-z cluster elliptical galaxies. The four \sneia that were classified as
possible are not used. We also add \contrerasSNe \sneia from the
low-redshift sample of \citet{contreras10a}, \contrerasPrevious of
which were not in Union2 (the others had published data from CfA). As
in Union2, for all \sne we require

\begin{enumerate}

\item that the CMB-centric redshift is greater than 0.015;

\item that there is at least one point between $-15$ and $6$ rest-frame
days from $B$-band maximum light;

\item that there are at least five valid data points;

\item that the entire $68\%$ confidence interval for $x_1$ lies
between $-5$ and $+5$;

\item data from at least two bands with rest-frame central wavelength
  coverage between 2900 \ang\ and 7000 \ang; and
  
\item at least one band redder than rest-frame $U$-band (4000 \ang). This cut
is new to this analysis, but only affects SN~2002fx, a GOODS supernova
which is very poorly measured.

\end{enumerate}

In addition to these quality cuts, we removed any supernova spectroscopically classified as SN~1991bg-like. These \sneia are a distinct subclass which is not modeled well by SALT2. At high redshift, where spectroscopic sub-typing may not be possible, we screen for these supernovae photometrically by searching for any supernovae with red ($c > 0.2$) and narrow-width ($x_1 < -3$) lightcurves, but do not find any. When fit with SALT2, and color-corrected and shape-corrected (as though they were normal \sneia), spectroscopically identified members of this class have an average absolute magnitude only 0.2 magnitudes fainter than normal \sneia; any contamination from the handful of supernovae near this cut will have only a small impact (and one well-accounted for by our contamination systematic, see \citet{amanullah10a}).

From the \kitchensinksample \sneia that were classified as either secure
or probable (see Table \ref{tbl:a}), SN~SCP06U4 and SN~SCP06K18 fail to
pass these cuts. SN~SCP06K18 lacks good enough light-curve coverage
and SN~SCP06U4 fails the $x_1$ cut\footnote{Using an updated version
  (2-18-17) of SALT2 (or using SALT1), SN~SCP06U4 would pass this cut,
  so this supernova may be included in future analyses.}.  This leaves
\goodkitchensinksample \sneia that are used to constrain the cosmology.

\subsection{Fitting the Cosmology}

Following \citet{amanullah10a}, the best-fit cosmology is determined
by minimizing

\begin{equation}
  \chi_{\mathrm{stat}}^2 =
  \sum_\mathrm{\sne}\frac{\left[\mu_B(\alpha, \beta, \delta, M_{\rm B}) - 
       \mu(z;\Omega_{m},\Omega_w,w)\right]^2}
       {\sigma_{\mathrm{lc}}^2 + \sigma_{\rm ext}^2 + \sigsamp^2}.
\label{eq:chisq}
\end{equation}
A detailed discussion of the terms in this equation can be found in
\citet{amanullah10a}.  We only comment on the final term in the
denominator, $\sigsamp^2$, which is computed by setting the
reduced $\chi^{2}$ of each sample to unity. 
This term was referred to as ``$\sigma^2_{\mathrm{systematic}}$'' 
in \citet{kowalski08b,amanullah10a}. We note that
$\sigsamp^2$ includes intrinsic dispersion as well as sample-dependent
effects.
This term effectively further deweights samples with
poorer-quality data that has sources of error which have not been 
accounted for.
As noted in \citet{amanullah10a}, this may occasionally
deweight an otherwise well-measured supernova.

Following \citet{conley06a}, \citet{kowalski08b} and
\citet{amanullah10a}, we hide our cosmology results until the full
analysis approach is settled. As in previous Union analysis, we carry
out an iterative $\chi^{2}$ minimization with outlier rejection. Each
sample is fit for a flat $\Lambda$CDM cosmology independently of the
other samples (but with $\alpha$, $\beta$, and $\delta$ set to their
global values). An $M_B$ is chosen for each sample by minimizing the
absolute variance-weighted sum of deviations, minimizing the effects
of outliers. We then reject any supernova more than $3 \sigma$ from
this fit. All of the \sneia in our new sample pass the outlier
rejection. As each sample is fit independently with its own Hubble
line, systematic errors and the choice of cosmological model are not
relevant in this selection.

\ifemapj \begin{figure*}[]
\else \begin{figure}[]
\fi

\centering
\includegraphics[height=\textwidth,angle=270,clip=true]{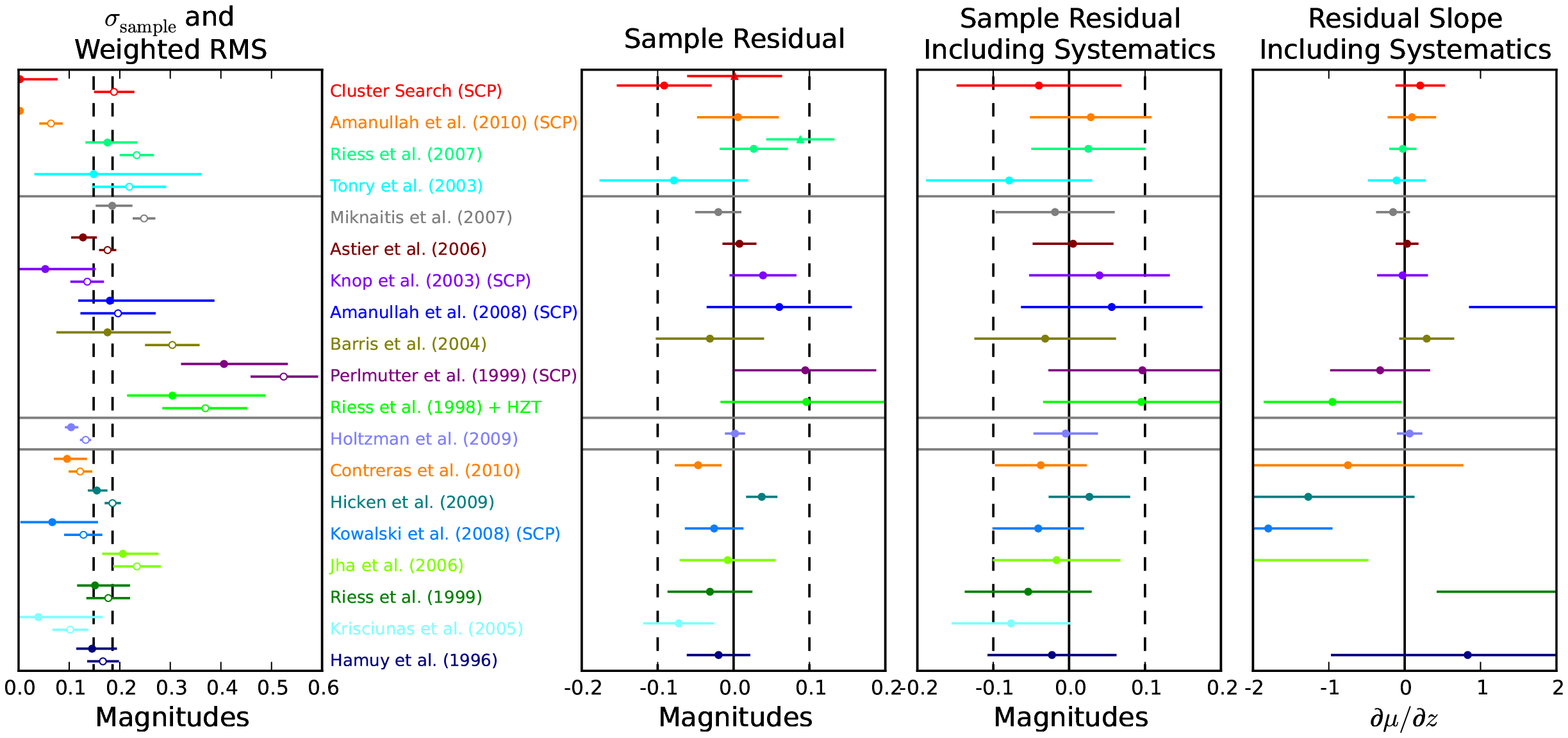}
\caption{\NOTE{fig:i02} Diagnostics plot for the individual data sets.
  From left to right: irreducible sample dispersion (filled circles) and
  variance-weighted RMS about the best-fit model (open circles); the average sample
  residual from the best-fit model ($\mu_{\mathrm{measured}} - \mu_{\mathrm{model}}$) excluding and including systematic
  errors; and the best-fit slope of the Hubble residual (in magnitudes) versus
  redshift --- $\partial \mu_\mathrm{residual}/\partial z$. Note that the errors on the
  sample dispersion include only statistical errors and do not
  include possible systematic errors. The confidence intervals on the weighed RMS are
  obtained with Monte-Carlo simulations. The triangles in the sample residual plot show
  the effect of including the filter shifts discussed in Section \ref{sec:diagnostics}.
  \label{fig:i02}}
\ifemapj \end{figure*}
\else  \end{figure}
\fi

\ifemapj \begin{figure*}[p]
\else \begin{figure}[p]
\fi

  \centering
  \includegraphics[angle=270,width=0.9\textwidth]{%
    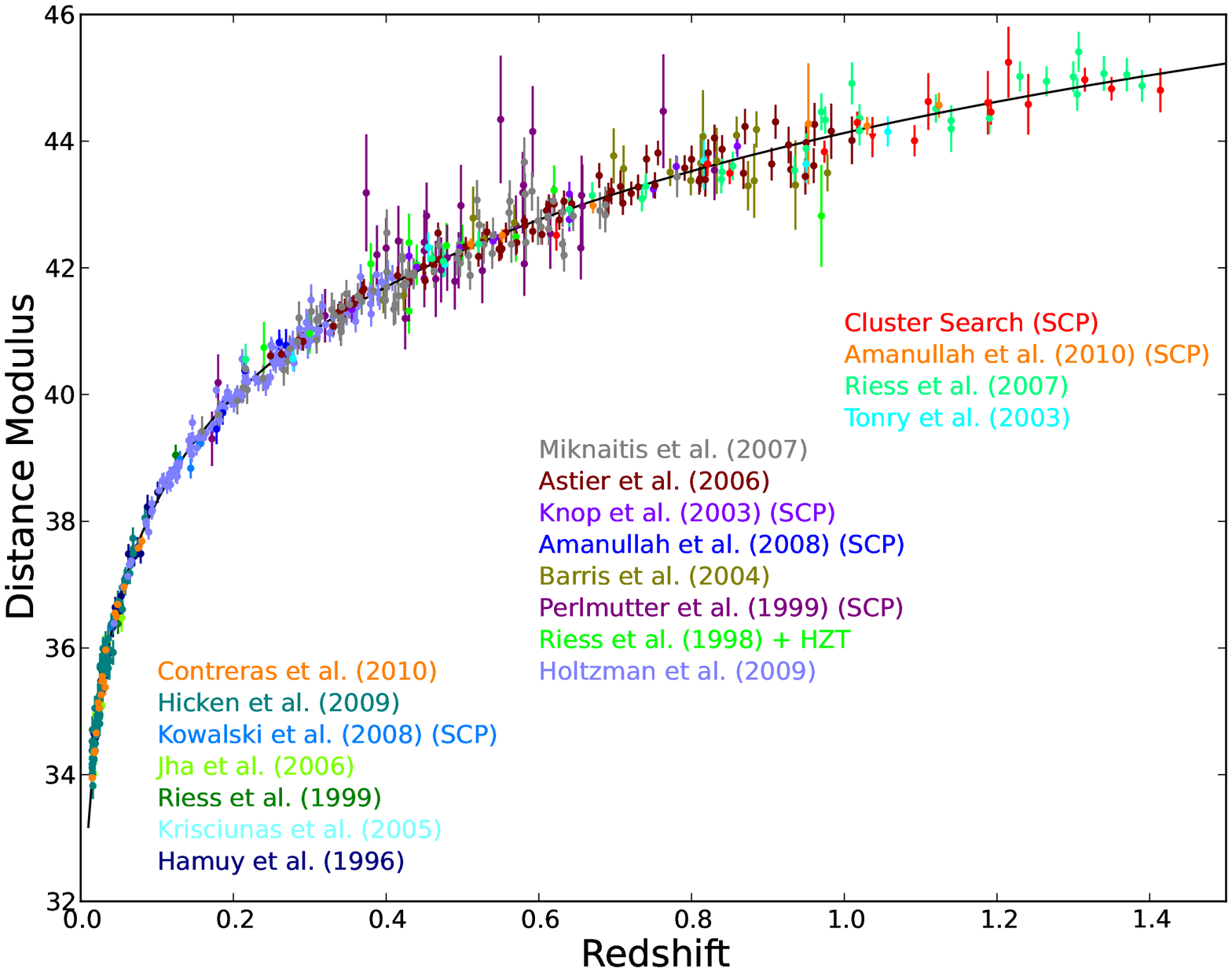}
  \caption{\NOTE{fig:i03} Hubble diagram for the Union2.1
    compilation.  The solid line
    represents the best-fit cosmology for a flat $\Lambda$CDM Universe for supernovae alone. SN~SCP06U4 falls outside the allowed $x_1$ range and is excluded from the current analysis. When fit with a newer version of SALT2, this supernova passes the cut and would be included, so we plot it on the Hubble diagram, but with a red triangle symbol.
    \label{fig:i03}}

\ifemapj \end{figure*}
\else  \end{figure}
\fi

\subsubsection{Diagnostics}\label{sec:diagnostics}
A diagnostic plot, which is used to study possible inconsistencies
between \snia samples, is shown in Figure~\ref{fig:i02}. The median of
$\sigsamp$ can be used as a measure of the intrinsic dispersion
associated with all \sneia. The intrinsic dispersion is a reflection
of how well our empirical models correct for the observed dispersion
in supernova luminosities. The median $\sigsamp$ for this paper
is \mediansigmasys \,mag and is indicated with the leftmost dashed vertical
line in the left panel. 

The variance weighted RMS about the best-fit cosmology gives an indication of
the quality of the photometry. A sample with more accurate photometry
will have a smaller RMS. For \sneia from our survey, the RMS is
\clusterRMS, which is only slightly larger than that measured for the
1st year \snia sample from SNLS, and equal to the median of all
samples (shown as the rightmost dashed line in Figure~\ref{fig:i02},
left panel).

The two middle panels show the tension between data-sets, the first
with statistical errors only, and the second with statistical and
systematic errors (see \S\ref{sec:systematics}). Most samples
land within 1 $\sigma$ of the mean defined by all samples and about
one third lie outside 1 $\sigma$, as expected for a normal
distribution. No sample exceeds 2 $\sigma$. The right hand panel shows
the slope of the residuals, which, for larger data sets, can be used
to reveal Malmquist-like biases or calibration errors.

The supernovae from our sample are \oursamplebrightersigma\ brighter
than the average sample. While the source of the difference may
certainly be a simple statistical fluctuation, part of the difference
might be attributable to errors in the filter responses of the ACS
filters. (The difference is largely driven by the \sneia\ that have
only ACS \iacs\ and \zacs\ data to constrain their light curves.)
Based on photometric observations of spectrophotometric standards,
\citet{bohlin07a} report possible blueward shifts of 94 \ang\ for the
$z_{\mathrm 850}$ filter and 57 \ang\ for the $i_{\mathrm 775}$ filter
(with smaller shifts in bluer filters). The red triangle in the sample
residual panel shows the effect of applying these shifts. The shifts
also affect the GOODS supernovae. The green triangle shows the affect
of applying the filter shifts to those data. \citet{bohlin07a} notes
that more data to confirm the filter shifts are needed, so we do not
apply them in our primary analysis. Instead, we include the
uncertainty in the filter curves as a systematic error, as described
in \S\ref{sec:systematics}.

Part of the difference could also be due to the correction that we
apply for the recently discovered correlation between host galaxy mass
and the luminosity of \sneia\ after the lightcurve width and color
corrections have been applied. Many of the hosts in our sample are
massive early type galaxies. In this analysis, the correction we use
is smaller than the correction that has been noted by others. We add
this difference as a systematic error, as described in
\S\ref{sec:systematics}.

Figure \ref{fig:i03} shows the Hubble Diagram with SNe from the
updated Union2 sample and the best-fit $\Lambda$CDM model. We add
\goodkitchensinksample \sneia\ from this current paper. (As discussed
above, SN~SCP06U4 is likely to be included in future analyses so it is
included on the plot with a different symbol.) Ten (eleven with
SN~SCP06U4) are above a redshift of one, significantly increasing the
number of well-measured supernovae above this redshift.

\subsection{Systematic errors}\label{sec:systematics}

In this paper, we follow the systematics analysis we presented in
\cite{amanullah10a}. Systematic errors that directly affect supernova
distance measurements (calibration, and galactic extinction, for
example) are treated as nuisance parameters to be fit simultaneously
with the cosmology. Minimizing over these nuisance parameters gives
additional terms to add to the distance modulus covariance matrix

\begin{equation}
U_{ij} = \sum_{\epsilon} \frac{d\mu_i(\alpha, \beta)}{d \epsilon}\frac{d\mu_j(\alpha, \beta)}{d \epsilon} \sigma_\epsilon ^2\;,
\end{equation}
where the sum is over each of these distance systematic errors in the analysis.
(Although the distance modulus depends on $\delta$ as well as $\alpha$ and $\beta$, the derivatives with respect to the zeropoints do not.) In this analysis, $\alpha$ and $\beta$ have little interaction with
cosmological parameters. When computing cosmological constraints, we
therefore freeze the covariance matrix in order to avoid multiple
matrix inversions\footnote{As demonstrated in the Union2 appendix,
  these matrix inversions can be simplified at the expense of more
  matrix multiplication; the run-time does not change much.}. 
Only when the $\alpha$ and $\beta$ may vary significantly from the global best-fit (Table \ref{tb:subsets}), do we update $\alpha$ and $\beta$.

Systematic errors that affect sample composition or the color and
shape correction coefficients cannot be parameterized
supernova-by-supernova in this way. These are incorporated by
assigning each dataset its own constant covariance. This is an adequate
treatment, as these systematic errors are subdominant.

There are two systematic errors that were not included in
\citet{amanullah10a}, but are included in this analysis for the first
time: a systematic error on the host-mass correction coefficient, $\delta$
(which might affect $\delta$ at the $\sim 0.05$ level), and uncertainties in
the effective wavelengths of the ACS \iacs\ and \zacs\ filters.

In addition to updating the NICMOS F110W zeropoint and uncertainty, as
described in \S\ref{sec:nicmosZP}, we revise the uncertainty
assigned to the zeropoint for NICMOS F160W to account for the uncertainty
in the count-rate non-linearity at this wavelength \citep{dejong06a}. Table \ref{tb:zperrs} gives
the assumed zeropoint error for each filter.
\ifemapj \begin{deluxetable*}{llll}
\else \begin{deluxetable}{llll}
\fi
  \tablewidth{0pt}
  \tablecaption{Assumed instrumental uncertainties for \sne in
    this paper.\label{tb:zperrs}}
  \tablehead{
    \colhead{Source} & 
    \colhead{Band}   & 
    \colhead{Uncertainty} & 
    \colhead{Reference}}
  \startdata
    HST     & WFPC2         & 0.02 & \citet{heyer04a}\\
            & ACS F850LP & 0.01 & \citet{bohlin07a} \\
            & ACS F775W & 0.01 & \\
            & ACS F606W & 0.01 & \\
            & ACS F850LP & 94 \ang & \citet{bohlin07a} \\
            & ACS F775W & 57 \ang & \\
            & ACS F606W & 27 \ang & \\
            & NICMOS J      & 0.024 & Ripoche et. al. (in prep), Section \ref{sec:nicmosZP} \\
            & NICMOS H      & 0.06 & \citet{dejong06a}\\
    \hline
    SNLS    & $g$, $r$, $i$ & 0.01 & \citet{astier06a}\\
            & $z$           & 0.03 &\\
    \hline
    ESSENCE & $R$, $I$      & 0.014 & \citet{wood-vasey07a}\\
    \hline
    SDSS    & $u$           & 0.014 & \citet{kessler09a}\\
            & $g$, $r$, $i$ & 0.009 &\\
            & $z$           & 0.010 & \\
    \hline
    SCP: \cite{amanullah10a} & $R$, $I$    & 0.03 & \citet{amanullah10a} \\
               & $J$     & 0.02 & \\
    \hline
    Other   & $U$-band      & 0.04 & \citet{hicken09b}\\
            & Other Band    & 0.02 & \citet{hicken09b}
  \enddata
\ifemapj \end{deluxetable*}
\else  \end{deluxetable}
\fi

We note that the nearby supernovae from targeted searches are
sensitive to $\delta$ (relative to the untargeted searches) at the
level of $(\untargetedlow - \targetedlow) \Delta \delta \approx 0.02$
magnitudes, while the covariance weighted mean of the cluster
supernovae varies with $\delta$ as $0.24 \Delta \delta \approx 0.01$
magnitudes. We cannot propagate this systematic on a
supernova-by-supernova basis, as this would be equivalent to fitting
for $\delta$, which we already do. Therefore, we include this error by
adding a covariance of $0.02^2$ to the nearby, targeted supernova
surveys, a covariance of $0.01^2$ to our new data-set, and
$0.02\cdot0.01$ between these data-sets.

Including uncertainties in filter effective wavelength is not as
straightforward as including zeropoint uncertainties. Effective
wavelength is only the first-order method of describing a filter. For
a simple filter shift, as implemented here, $d \mu (\alpha, \beta)/ d
\lambda$ will undergo significant variations as supernova spectral
features shift in and out of the filter. These are likely to be worse
than the actual effect of simply reweighting filter throughput. 
Although in general these variations will get averaged out
with different phases, redshifts, and additional filters, we have
modeled a worst-case in accounting for this systematic (and even then
it only affects the supernovae most dependent on \zacs).

\ifemapj \begin{deluxetable*}{lll}
\else \begin{deluxetable}{lll}
\fi

  \tablecaption{Effect on constant $w$ error bars and area of the $95\%$ $w_0-w_a$ confidence contour (inverse DETF FoM) for each type of systematic error, when \snia\ constraints are combined with constraints from CMB, H$_{0}$, and BAO.
\label{tb:werrsys}}
  \tablehead{
    \colhead{Source} & \colhead{Error on Constant $w$} & \colhead{Inverse DETF FoM}}
  \startdata
Vega               & 0.033 & 0.19\\
All Instrument Calibration &  0.030 & 0.18\\
(ACS Zeropoints) & 0.003 & 0.01 \\
(ACS Filter Shift) & 0.007& 0.04\\
(NICMOS Zeropoints) & 0.007& <0.01\\
Malmquist Bias     & 0.020 & 0.07\\
Color Correction   & 0.020 & 0.07\\
Mass Correction    & 0.016 & 0.08\\
Contamination      & 0.016 & 0.05\\
Intergalactic Extinction           & 0.013 & 0.03\\
Galactic Extinction Normalization  & 0.010 & 0.01 \\
Rest-Frame $U$-Band Calibration      & 0.009 & <0.01\\
Lightcurve Shape                   & 0.006 &  <0.01\\
\hline
\textit{Quadrature Sum of Errors/ Sum of Area (not used)}& \textit{0.061}& \textit{0.68}\\
Summed in Covariance Matrix       & 0.048& 0.42
  \enddata
\ifemapj \end{deluxetable*}
\else  \end{deluxetable}
\fi

Table \ref{tb:werrsys} shows the impact of each type of systematic
error on our cosmological constraints, in combination with BAO, CMB, and H$_0$
data (see \S\ref{sec:cosmology}).
For the purpose of constructing Table \ref{tb:werrsys}, we add, 
for each systematic error in the table, the contribution from 
just that systematic to the statistical-only covariance matrix.
The confidence interval for constant $w$ where the $\chi^2$ is within 1 of the minimum $\chi^2$ (the edges of this confidence interval are hereafter referred to with the notation $\Delta \chi^2 = 1$) is found iteratively; 
the plus and minus errors for constant $w$ are averaged. The statistical-only
constant $w$ error bar is subtracted in quadrature, leaving the effect of each
systematic on constant $w$. We also quote the effect of each systematic error
of the $\Delta \chi^2 = 5.99$ confidence contour in the $(w_0,w_a)$ plane; as this is two-dimensional,
we subtract the area (not in quadrature) of the statistical-only contour.

Since the derived cosmology errors vary with the best-fit cosmology, after a given systematic error has been added, 
the supernova magnitudes are shifted so that the best-fit cosmology including
that systematic matches the best-fit with statistical errors
only. This magnitude adjustment (which is the same adjustment we use
for blinding ourselves to the best-fit cosmology) consists of
repeatedly computing the difference in distance modulus between the
best-fit cosmology and fiducial value and adding it to the supernovae.

As with the Union2 compilation, calibration systematics represent the
largest contribution to the error on constant $w$. Here, we see that they
are also the dominant systematics for $(w_0,w_a)$. As noted by \citet{amanullah10a},
significantly smaller systematic
errors are derived by adding each covariance in the covariance matrix,
rather than adding the cosmological impacts together. This is due to the
different redshift dependence of each
systematic error, as well as some self-calibration that occurs as described in \citet{amanullah10a}.

\ifemapj 
\begin{deluxetable*}{ccccccccc}
\tabletypesize{\footnotesize} 
\else 
\begin{deluxetable}{ccccccccc}
\tabletypesize{\tiny}
\fi
\tablecaption{Constraints on
  standardization and cosmological parameters for subsets. $M_B$ is the $B$-band corrected absolute magnitude; $\alpha$, $\beta$, and $\delta$ are the lightcurve shape, color, and host mass correction coefficients, respectively. The outlier
  rejection is redone each time, so the totals may not add up to the
  whole sample. The constraints are computed including BAO, CMB, and $H_0$
  constraints and supernova systematic errors.
    \label{tb:subsets}}
  \tablehead{
    \colhead{Subset} & \colhead{Number} & \colhead{$M_B (h = 0.7)$} &
    \colhead{$\alpha$} & \colhead{$\beta$}  & \colhead{$\delta$} & \colhead{$\Omega_m$} &
    \colhead{$w$}}
  \startdata
\cutinhead{Whole Sample}
$ z \geq  0.015 $ & 580 & $ -19.321^{+0.030}_{-0.030} $ & $ 0.121^{+0.007}_{-0.007} $ & $ 2.47^{+0.06}_{-0.06} $ & $ -0.032^{+0.031}_{-0.031} $ & $ 0.271^{+0.015}_{-0.014} $ & $ -1.013^{+0.068}_{-0.074} $\\ 
\cutinhead{Correction Coefficients, Split by Redshift}
$ 0.015\leq z \leq  0.10 $ & 175 & $ -19.328^{+0.037}_{-0.038} $ & $ 0.118^{+0.011}_{-0.011} $ & $ 2.57^{+0.08}_{-0.08} $ & $ -0.027^{+0.054}_{-0.054} $ & $ 0.270 $ (fixed)  & $ -1.000 $ (fixed) \\ 
$ 0.100\leq z \leq  0.25 $ & 75 & $ -19.371^{+0.054}_{-0.054} $ & $ 0.146^{+0.019}_{-0.019} $ & $ 2.56^{+0.18}_{-0.17} $ & $ -0.087^{+0.060}_{-0.060} $ & $ 0.270 $ (fixed)  & $ -1.000 $ (fixed) \\ 
$ 0.250\leq z \leq  0.50 $ & 152 & $ -19.317^{+0.046}_{-0.046} $ & $ 0.116^{+0.014}_{-0.013} $ & $ 2.46^{+0.12}_{-0.12} $ & $ -0.042^{+0.066}_{-0.066} $ & $ 0.270 $ (fixed)  & $ -1.000 $ (fixed) \\ 
$ 0.500\leq z \leq  1.00 $ & 137 & $ -19.307^{+0.048}_{-0.049} $ & $ 0.124^{+0.019}_{-0.019} $ & $ 1.46^{+0.19}_{-0.19} $ & $ 0.023^{+0.060}_{-0.060} $ & $ 0.270 $ (fixed)  & $ -1.000 $ (fixed) \\ 
$ z \geq  1.000 $ & 25 & $ -19.289^{+0.217}_{-0.254} $ & $ -0.019^{+0.072}_{-0.076} $ & $ 3.48^{+1.13}_{-0.89} $ & $ -0.151^{+0.384}_{-0.446} $ & $ 0.270 $ (fixed)  & $ -1.000 $ (fixed) \\ 
\cutinhead{Effect of $\delta$ on $w$}
$ z \geq  0.015 $ & 580 & $ -19.340^{+0.026}_{-0.026} $ & $ 0.123^{+0.007}_{-0.007} $ & $ 2.47^{+0.06}_{-0.06} $ & $ -0.080 $ (fixed)  & $ 0.272^{+0.015}_{-0.014} $ & $ -1.004^{+0.067}_{-0.072} $\\ 
$ z \geq  0.015 $ & 580 & $ -19.303^{+0.031}_{-0.031} $ & $ 0.120^{+0.007}_{-0.007} $ & $ 2.47^{+0.06}_{-0.06} $ & $ 0.000 $ (fixed)  & $ 0.271^{+0.015}_{-0.014} $ & $ -1.013^{+0.069}_{-0.075} $\\ 
\cutinhead{Cosmological Results, Split by Lightcurve Color and Shape}
$c \geq 0.05$ & 256 & $ -19.387^{+0.037}_{-0.038} $ & $ 0.118^{+0.011}_{-0.011} $ & $ 2.77^{+0.09}_{-0.09} $ & $ -0.057^{+0.052}_{-0.052} $ & $ 0.269^{+0.015}_{-0.014} $ & $ -1.028^{+0.077}_{-0.084} $\\ 
$c \leq 0.05$ & 321 & $ -19.323^{+0.030}_{-0.030} $ & $ 0.125^{+0.011}_{-0.010} $ & $ 1.29^{+0.32}_{-0.33} $ & $ -0.057^{+0.038}_{-0.038} $ & $ 0.275^{+0.015}_{-0.014} $ & $ -0.982^{+0.069}_{-0.075} $\\ 
$x_1 \geq -0.25$ & 311 & $ -19.366^{+0.041}_{-0.041} $ & $ 0.020^{+0.026}_{-0.025} $ & $ 2.58^{+0.10}_{-0.10} $ & $ -0.004^{+0.047}_{-0.047} $ & $ 0.269^{+0.015}_{-0.014} $ & $ -1.037^{+0.077}_{-0.085} $\\ 
$x_1 \leq -0.25$ & 269 & $ -19.386^{+0.044}_{-0.045} $ & $ 0.152^{+0.021}_{-0.020} $ & $ 2.43^{+0.08}_{-0.08} $ & $ -0.087^{+0.050}_{-0.050} $ & $ 0.267^{+0.015}_{-0.014} $ & $ -1.045^{+0.077}_{-0.084} $\\ 
\cutinhead{Correction Coefficients and $M_B$ for the Large Datasets}
Hicken et al. (2009) & 94 & $ -19.314^{+0.055}_{-0.055} $ & $ 0.115^{+0.015}_{-0.015} $ & $ 2.74^{+0.11}_{-0.11} $ & $ -0.053^{+0.098}_{-0.099} $ & $ 0.270 $ (fixed)  & $ -1.000 $ (fixed) \\ 
Holtzman et al. (2009) & 129 & $ -19.336^{+0.051}_{-0.051} $ & $ 0.149^{+0.014}_{-0.013} $ & $ 2.40^{+0.15}_{-0.14} $ & $ -0.061^{+0.050}_{-0.050} $ & $ 0.270 $ (fixed)  & $ -1.000 $ (fixed) \\ 
Miknaitis et al. (2007) & 74 & $ -19.325^{+0.078}_{-0.080} $ & $ 0.113^{+0.037}_{-0.035} $ & $ 2.49^{+0.17}_{-0.16} $ & $ 0.000 $ (fixed)  & $ 0.270 $ (fixed)  & $ -1.000 $ (fixed) \\ 
Astier et al. (2006) & 71 & $ -19.292^{+0.047}_{-0.048} $ & $ 0.145^{+0.019}_{-0.018} $ & $ 1.70^{+0.18}_{-0.18} $ & $ -0.023^{+0.040}_{-0.040} $ & $ 0.270 $ (fixed)  & $ -1.000 $ (fixed) \\ 
\cutinhead{$z > 0.9$, Split by Galaxy Host}
Early Type $z > 0.9$ & 13 & $ -19.388^{+0.139}_{-0.186} $ & $ 0.112^{+0.139}_{-0.151} $ & $ 3.16^{+1.84}_{-1.26} $ & $ 0.000 $ (fixed)  & $ 0.270 $ (fixed)  & $ -1.000 $ (fixed) \\ 
Late Type $z > 0.9$ & 15 & $ -19.141^{+0.067}_{-0.067} $ & $ 0.094^{+0.049}_{-0.041} $ & $ 0.49^{+0.85}_{-0.69} $ & $ 0.000 $ (fixed)  & $ 0.270 $ (fixed)  & $ -1.000 $ (fixed) \\ 
\enddata
\ifemapj \end{deluxetable*}
\else  \end{deluxetable}
\fi

Some potential systematic errors can be investigated by dividing the
whole dataset into subsets. Table \ref{tb:subsets} shows many of these
divisions. All of the numbers are computed including supernova
systematics; the cosmological constraints are computed including BAO
and CMB data. In short, we do not see any evidence of unknown
systematic errors, requiring the cosmological impact to be smaller
than the current errors.

The first subsets are subsets in redshift. These can be used to study
possible evolution of correction coefficients for shape, color, and
host mass. The redshift range 0.5 to 1 seems to show $\beta$ and
$\delta$ smaller in magnitude, but the revised SNLS sample
\citep{guy10a} which uses a newer version of the calibration and
lightcurve fitting (as well as many more supernovae), shows no signs
of this. As we have already budgeted these systematic
uncertainties, these updates will be within our error bars.

The next rows show the effect of changing $\delta$ from 0 to $-0.08$
(the size of the correction in \citet{sullivan10b}). Because a large
error on $\delta$ is already included in the systematic error
covariance matrix, this has less than a 0.01 effect on $w$, about ten
times smaller than it would have if we did not include this
systematic.

Next, we consider systematics caused by potentially different
populations of supernovae. We perform a cut on the best-fit true $x_1$
or $c$ of each supernova (see \citet{amanullah10a} for details). The
cosmology in each case is compatible with the cosmology derived from
the whole sample.

We now look at each of the four largest datasets for evidence of
tension. The only tension found is in the first-year SNLS sample 
\citep{astier06a}. 
Here, $\beta$ and $\delta$ are both at odds with the whole
sample, but as noted above, we do not believe this is a cause for
concern.

The final two rows show the high-redshift sample split by host type;
this is discussed in \S \ref{sec:splitbyhost}.

\section{Constraints on Dark Energy}\label{sec:cosmology}\label{sec:h}
\outline{draft}{Overview\\}

Following \citet{amanullah10a}, we constrain the properties of dark
energy first using \sneia\ alone (with and without systematics),
and then by combining the constraints derived from \sneia with those
derived from the 7-year WMAP data of the CMB \citep{komatsu11a}, the
position of the BAO peak from the combined analysis of the SDSS DR7
and 2dFGRS data \citep{percival10a}, and the measurement of the Hubble
constant ($H_{0}$) from Cepheids \citep{riess11a}.

The rate of expansion at redshift $z$, $H(z)$, is described by the
Friedman equation:
\begin{eqnarray}
\frac{H^{2}(z)}{H^{2}_{0}}&=&\Omega_{m}(1+z)^{3} + \Omega_{k}(1+z)^{2}\\
& & + \ \Omega_{\mathrm{DE}} 
     \exp{\left[ \int 3(1+w(z)) {\rm dln} (1+z) \right ]} 
     ,\nonumber
\label{eqn:hz}
\end{eqnarray}
where $H_{0}$ is the rate of expansion today, \omegam\ and $\Omega_{\mathrm{DE}}$ are 
the matter and dark energy density with respect to the critical density
today, $w(z)$, is the dark energy equation-of-state parameter, and $\Omega_k = 1 - \Omega_m - \Omega_{\mathrm{DE}}$ is 
the spatial curvature density of the universe. Distances, such as the 
luminosity distance, depend on the integral of $1/H(z)$ over redshift.

In this section, we consider the following models for dark energy:

\begin{itemize}

\item[-] $\Lambda$CDM: A cosmological constant in a flat universe.

\item[-] $w$CDM: A constant equation-of-state parameter in a flat universe.

\item[-] $ow$CDM: A constant equation-of-state parameter in a curved universe.

\item[-] $w_{z}$CDM models: A time-varying equation-of-state parameter in universes with and without curvature.

\end{itemize}

The results for each of the models are listed in Table~\ref{tbl:d} and
discussed in turn in the following sub-sections. Unless stated
otherwise, the uncertainties represent the 68\% confidence limits
($\Delta \chi^2 = 1$) and include both statistical uncertainties and
systematic errors.

\subsection{{\boldmath$\Lambda$}CDM}
\ifemapj \begin{figure*}[h]
\else \begin{figure}[]
\fi

\includegraphics[width=0.42\textwidth,clip=true]{%
  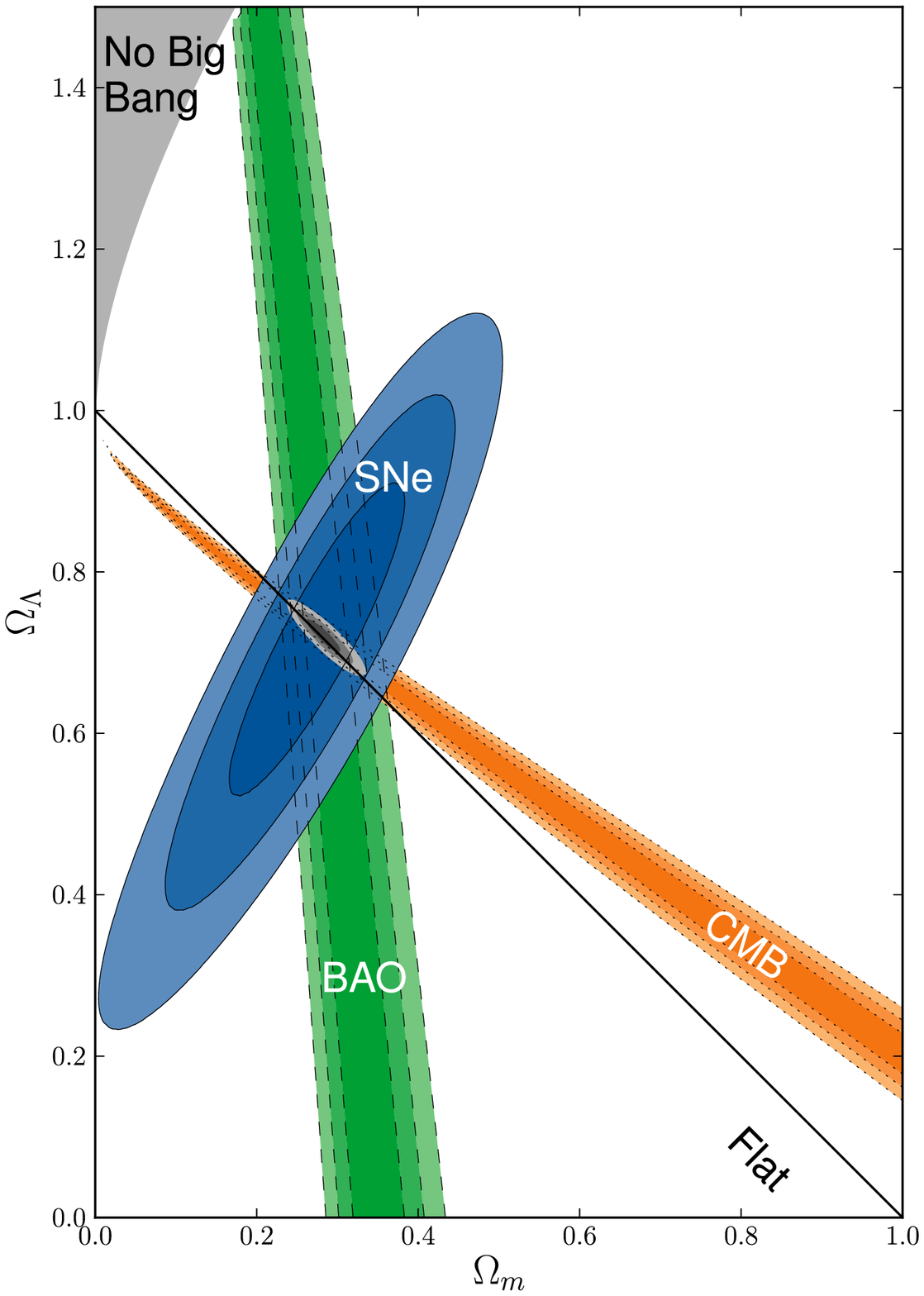}%
\hspace{0.06\textwidth}%
\includegraphics[width=0.42\textwidth,clip=true]{%
  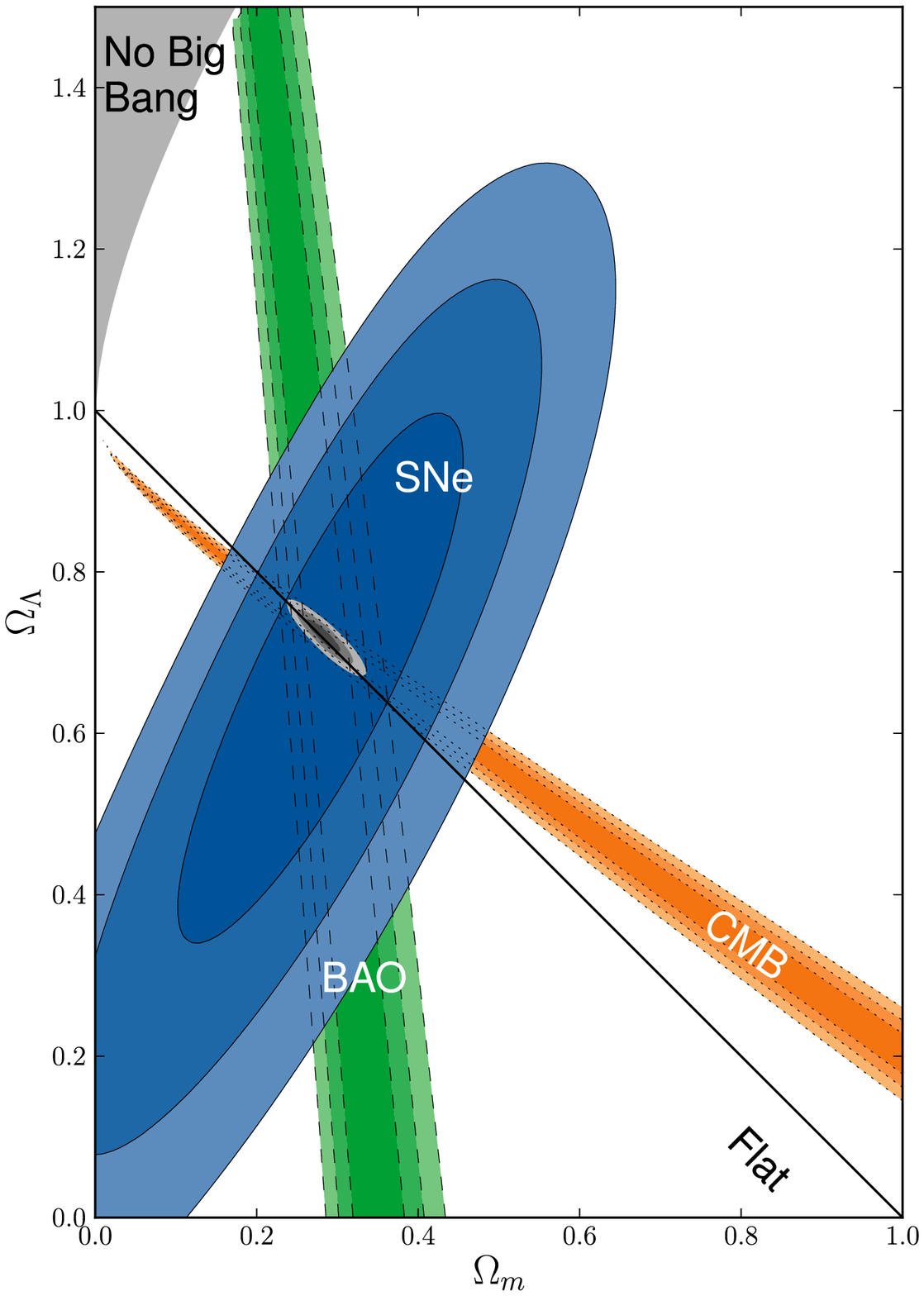}%
\caption{\NOTE{fig:i04} $\Lambda$CDM model: 
  $68.3\%$, $95.4\%$, and $99.7\%$ confidence regions of the
  $(\omegam,\omegal)$ plane from \sneia\ combined with the constraints 
  from BAO and CMB.
  The left panel shows the \snia\ confidence region only including statistical
  errors while the right panel shows the \snia\ confidence region
  with both statistical and systematic errors.
  \label{fig:omol}}

\ifemapj \end{figure*}
\else  \end{figure}
\fi

\outline{draft}{LCDM\\}
In the $\Lambda$CDM model, the equation-of-state parameter is exactly $-1$ and
does not vary with time. In a flat Universe, \sneia\ alone constrain the
dark-energy density, $\Omega _{\Lambda}$, to be
$\Omega _{\Lambda} = \omegalambdasnealone$ including systematics.
Adding the constraints from CMB, BAO and $H_0$ reduces the
uncertainty. Under the assumption of a flat Universe, the four
probes yield
$$\flatLCDMomegal
 {\rm \hspace{3mm} (\Lambda CDM : SN+CMB+BAO+H_{0})} $$

In this $\Lambda$CDM model, the expansion of the universe switched from deceleration
to acceleration at $z=0.752\pm0.041$, which corresponds to a look back time of $6.62\pm0.22$~Gyr, 
about the half of the age of the universe.
Equality between the energy density of dark energy and matter occurred later, at
$z=0.391\pm0.033$ or the look back time of $4.21\pm0.27$~Gyr.

In Figure \ref{fig:omol}, we show the confidence intervals on \omegam\
and \omegal\ from SNe, CMB and BAO. Both the individual
constraints and the combined constraint are shown (the BAO constraints
are computed with an \omegam$h^{2}$ prior from the CMB). The
SN constraint is almost orthogonal to that of the CMB. If we remove the flatness
prior, the best-fit \omegam\ and \omegal\ change by a fraction of their errors with \omegakoLCDM.

\subsection{{\boldmath$w$}CDM : Constant Equation of State with {\boldmath $w \neq -1$}}
\outline{draft}{w in a flat universe\\} 
In $w$CDM models, $w$ is constant but is allowed to be different from $-1$.
While few dark energy theories give $w\ne-1$ and yet constant \citep{copeland06a}, 
the constant $w$ model is still useful to constrain as it
contains fewer parameters than the dynamical dark energy models
 considered in the next section, and a value different from $-1$ would
rule out the cosmological constant.

In a flat universe (\omegak$=0$), \sneia\ alone give
\snealoneconstw\ (including systematics).
Adding the constraints from the other three probes tightens the constraint on $w$
considerably, as the constraints from \sneia\ in the \omegam--\,$w$
parameter plane are almost orthogonal to those provided by BAO
and the CMB (Figure \ref{fig:omw}). 

\ifemapj \begin{figure*}[h]
\else \begin{figure}[h]
\fi

\includegraphics[width=0.45\textwidth,clip=true]{%
  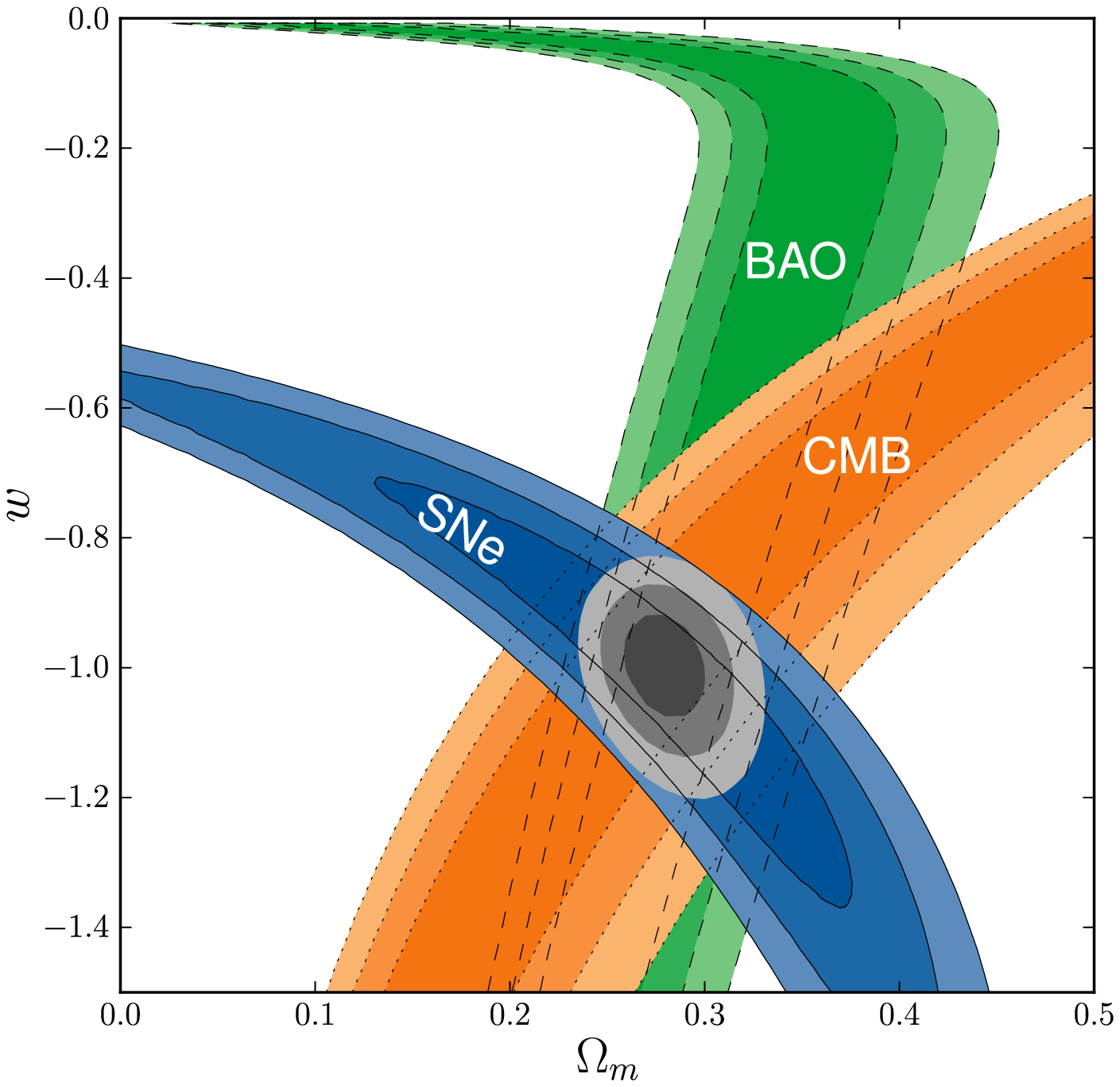}%
\hspace{0.06\textwidth}%
\includegraphics[width=0.45\textwidth,clip=true]{%
  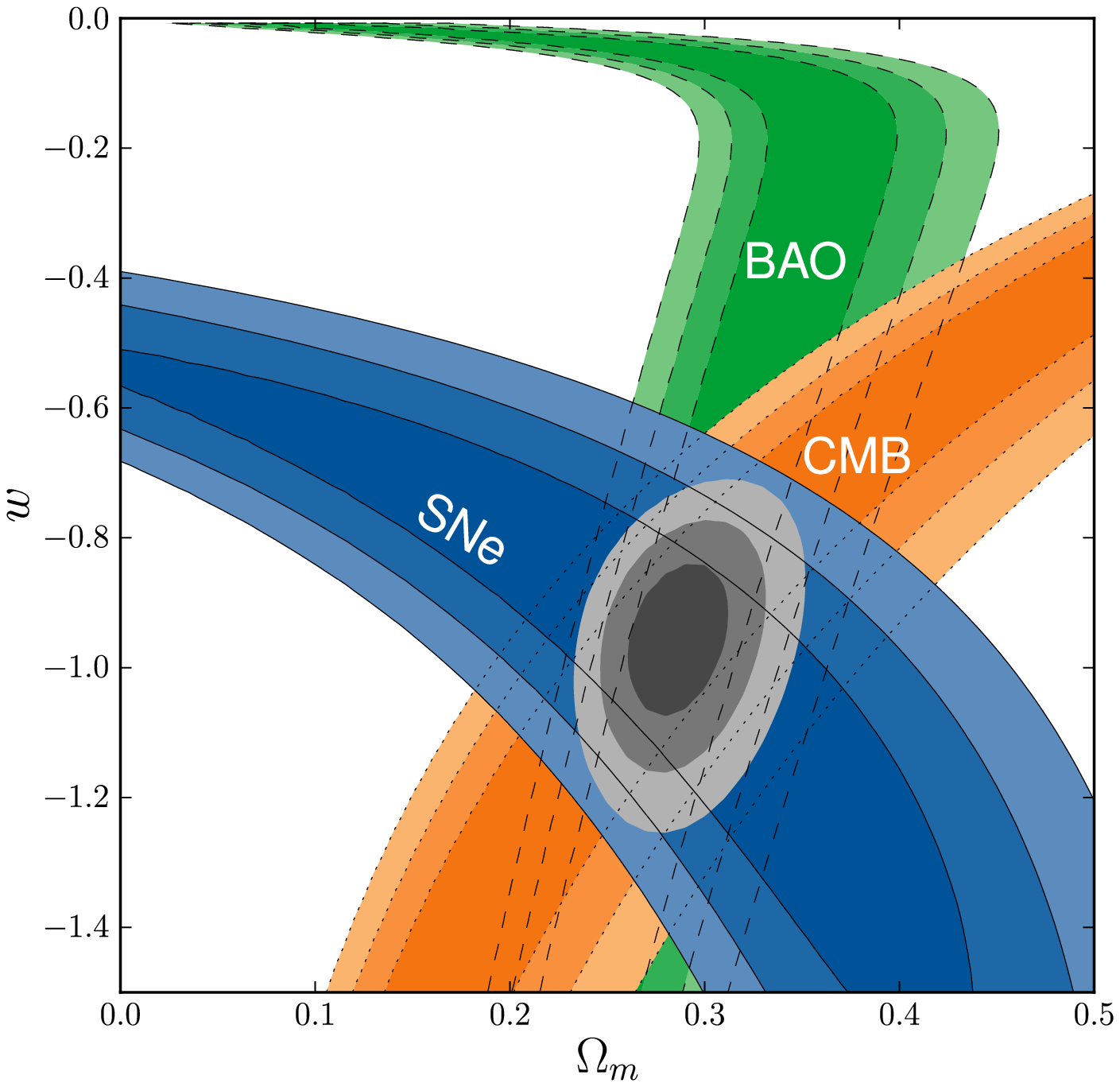}%
\caption{\NOTE{fig:i05}$w$CDM model: $68.3\%$, $95.4\%$, and $99.7\%$ 
confidence regions in the $(\omegam,w)$ plane from \sneia\, BAO and CMB.
The left panel shows the \snia\ confidence region for statistical 
uncertainties only, while the right panel shows the confidence region including both 
statistical and systematic uncertainties.
We note that CMB and \snia\ constraints are orthogonal, making this combination
of cosmological probes very powerful for investigating the nature of dark
energy.
  \label{fig:i05} \label{fig:omw}}

\ifemapj \end{figure*}
\else  \end{figure}
\fi

In principle, a constraint on $H_{0}$ helps to break the degeneracy
between \omegam\ and $h$ for CMB, which measures \om
$h^{2}$ \citep{spergel03a}. However, in this case adding
supernova data helps more, as narrowing the degeneracy between
\omegam\ and $w$ allows the CMB itself to constrain $H_0$.
By combining all four probes, we find
$\flatwfit$.
 As seen in Table~\ref{tbl:d}, neither BAO nor $H_0$ currently make much of a
 difference in the error bars for this model.

\subsection{{\boldmath$ow$}CDM : Constant Equation of State in a Curved Universe}

\outline{draft}{w in a curved universe\\} 

Inflation models generally predict that the curvature of the Universe, \omegak, is
$\sim 10^{-5}$ \citep{guth81a,liddle00a}. In curved universes, \sneia\
play the critical role in constraining $w$, while CMB+BAO constrain \omegak\ and \omegam.  
By combining all four probes, we find
$\owCDMomegak$ and $\wcurvedconstant$.
Even with the additional freedom for non-zero curvature, 
a flat universe is supported from observations. 
Among many cosmological parameters, the curvature of
the universe is the most well-determined parameter.

We note CMB alone does not place a tight constraint on curvature\footnote{\url{http://lambda.gsfc.nasa.gov/product/map/dr4/parameters.cfm}},
$\omegak=-0.102^{+0.085}_{-0.097}$ 
\citep{komatsu11a}, and it is the combination of \snia\ and BAO which improves the 
constraint by a factor of ten.  
Including $H_0$ improves the curvature constraints but the current
measurements from \sneia\ and BAO have more impact.

For the equation-of-state parameter $w$, 
as seen in Table \ref{tbl:d}, BAO constraints are now needed to
 constrain \omegam, while $H_0$ again has a small impact on the $w$
 measurement.

\ifemapj 
\begin{deluxetable*}{ccccccccc}
\tabletypesize{\footnotesize}
\else 
\begin{deluxetable}{ccccccccc}
\tabletypesize{\tiny}
\fi
  \tablecaption{Fit results on cosmological parameters $\om$, $w_0$, $w_a$ and
    $\Omega_k$. The parameter values are followed by their statistical
    (first column)
    and statistical and systematic (second column) $1 \sigma$ ($\Delta \chi^2$ = 1) uncertainties. For the fits including curvature and time-varying $w$, the confidence intervals can be quite non-gaussian and we also show $\Delta \chi^2 = 4$ confidence intervals (with and without systematics) for comparison.\label{tbl:d}}
    
  \tablehead{
    \colhead{Fit} & \colhead{$\Omega_m$} & \colhead{$\Omega_m$ w/ Sys}
    & \colhead{$\Omega_k$}  & \colhead{$\Omega_k$ w/ Sys} &
    \colhead{$w_0$} & \colhead{$w_0$ w/ Sys} & \colhead{$w_a$} & \colhead{$w_a$ w/ Sys}}
  \startdata
\cutinhead{$\Lambda$CDM}
SNe & $ 0.277^{+0.022}_{-0.021} $ & $ 0.295^{+0.043}_{-0.040} $ & $0$ (fixed) & $0$ (fixed) & $-1$ (fixed) & $-1$ (fixed)  & $0$ (fixed) & $0$ (fixed) \\
SNe+BAO+CMB & $ 0.278^{+0.014}_{-0.013} $ & $ 0.282^{+0.017}_{-0.016} $ & $0$ (fixed) & $0$ (fixed) & $-1$ (fixed) & $-1$ (fixed)  & $0$ (fixed) & $0$ (fixed) \\
SNe+BAO+CMB+$H_0$ & $ 0.271^{+0.012}_{-0.012} $ & $ 0.271^{+0.014}_{-0.014} $ & $0$ (fixed) & $0$ (fixed) & $-1$ (fixed) & $-1$ (fixed)  & $0$ (fixed) & $0$ (fixed) \\
\cutinhead{$o\Lambda$CDM}
SNe+BAO+CMB & $ 0.282^{+0.015}_{-0.014} $ & $ 0.286^{+0.018}_{-0.017} $ & $ -0.004^{+0.006}_{-0.006} $ & $ -0.004^{+0.006}_{-0.007} $ & $-1$ (fixed) & $-1$ (fixed)  & $0$ (fixed) & $0$ (fixed) \\
SNe+BAO+CMB+$H_0$ & $ 0.271^{+0.013}_{-0.012} $ & $ 0.272^{+0.014}_{-0.014} $ & $ 0.002^{+0.005}_{-0.005} $ & $ 0.002^{+0.005}_{-0.005} $ & $-1$ (fixed) & $-1$ (fixed)  & $0$ (fixed) & $0$ (fixed) \\
\cutinhead{$w$CDM}
SNe & $ 0.281^{+0.067}_{-0.092} $ & $ 0.296^{+0.102}_{-0.180} $ & $0$ (fixed) & $0$ (fixed) & $ -1.011^{+0.208}_{-0.231} $ & $ -1.001^{+0.348}_{-0.398} $ & $0$ (fixed) & $0$ (fixed) \\
SNe+BAO+$H_0$ & $ 0.309^{+0.029}_{-0.028} $ & $ 0.320^{+0.035}_{-0.033} $ & $0$ (fixed) & $0$ (fixed) & $ -1.097^{+0.091}_{-0.106} $ & $ -1.076^{+0.117}_{-0.133} $ & $0$ (fixed) & $0$ (fixed) \\
SNe+CMB & $ 0.271^{+0.018}_{-0.017} $ & $ 0.279^{+0.025}_{-0.023} $ & $0$ (fixed) & $0$ (fixed) & $ -0.983^{+0.051}_{-0.056} $ & $ -0.955^{+0.075}_{-0.079} $ & $0$ (fixed) & $0$ (fixed) \\
SNe+BAO+CMB & $ 0.278^{+0.014}_{-0.014} $ & $ 0.285^{+0.018}_{-0.017} $ & $0$ (fixed) & $0$ (fixed) & $ -0.993^{+0.052}_{-0.055} $ & $ -0.951^{+0.075}_{-0.081} $ & $0$ (fixed) & $0$ (fixed) \\
SNe+BAO+CMB+$H_0$ & $ 0.272^{+0.013}_{-0.013} $ & $ 0.271^{+0.014}_{-0.014} $ & $0$ (fixed) & $0$ (fixed) & $ -1.008^{+0.050}_{-0.054} $ & $ -1.013^{+0.068}_{-0.073} $ & $0$ (fixed) & $0$ (fixed) \\
\cutinhead{$ow$CDM}
SNe+CMB & $ 0.281^{+0.069}_{-0.087} $ & $ 0.295^{+0.109}_{-0.161} $ & $ -0.003^{+0.034}_{-0.027} $ & $ -0.005^{+0.067}_{-0.041} $ & $ -1.007^{+0.179}_{-0.194} $ & $ -0.993^{+0.299}_{-0.331} $ & $0$ (fixed) & $0$ (fixed) \\
SNe+BAO+CMB & $ 0.283^{+0.016}_{-0.015} $ & $ 0.287^{+0.018}_{-0.017} $ & $ -0.004^{+0.007}_{-0.007} $ & $ -0.002^{+0.008}_{-0.008} $ & $ -1.012^{+0.058}_{-0.062} $ & $ -0.975^{+0.094}_{-0.098} $ & $0$ (fixed) & $0$ (fixed) \\
SNe+BAO+CMB+$H_0$ & $ 0.272^{+0.013}_{-0.013} $ & $ 0.272^{+0.015}_{-0.014} $ & $ 0.002^{+0.006}_{-0.006} $ & $ 0.002^{+0.007}_{-0.007} $ & $ -1.006^{+0.056}_{-0.060} $ & $ -1.003^{+0.091}_{-0.095} $ & $0$ (fixed) & $0$ (fixed) \\
\cutinhead{$w_{z}$CDM}
SNe+CMB & $ 0.273^{+0.022}_{-0.020} $ & $ 0.281^{+0.043}_{-0.028} $ & $0$ (fixed) & $0$ (fixed) & $ -1.006^{+0.165}_{-0.182} $ & $ -0.993^{+0.263}_{-0.307} $ & $ 0.11^{+0.75}_{-0.77} $ & $ 0.17^{+1.08}_{-1.19} $\\
SNe+BAO+CMB & $ 0.278^{+0.014}_{-0.014} $ & $ 0.284^{+0.018}_{-0.017} $ & $0$ (fixed) & $0$ (fixed) & $ -1.052^{+0.126}_{-0.120} $ & $ -1.013^{+0.183}_{-0.173} $ & $ 0.30^{+0.48}_{-0.62} $ & $ 0.26^{+0.57}_{-0.74} $\\
SNe+BAO+CMB+$H_0$ & $ 0.271^{+0.013}_{-0.013} $ & $ 0.270^{+0.015}_{-0.014} $ & $0$ (fixed) & $0$ (fixed) & $ -1.021^{+0.123}_{-0.117} $ & $ -1.046^{+0.179}_{-0.170} $ & $ 0.07^{+0.49}_{-0.60} $ & $ 0.14^{+0.60}_{-0.76} $\\
\cutinhead{$ow_{z}$CDM}
SNe+CMB & $ 0.177^{+0.086}_{-0.093} $ & $ 0.190^{+0.208}_{-0.154} $ & $ 0.075^{+0.065}_{-0.128} $ & $ 0.073^{+0.115}_{-0.141} $ & $ -0.988^{+0.176}_{-0.202} $ & $ -0.969^{+0.284}_{-0.345} $ & $ 0.90^{+0.26}_{-3.88} $ & $ 0.89^{+0.43}_{-5.25} $\\
SNe+BAO+CMB & $ 0.283^{+0.019}_{-0.017} $ & $ 0.286^{+0.022}_{-0.023} $ & $ -0.004^{+0.017}_{-0.010} $ & $ -0.001^{+0.037}_{-0.013} $ & $ -1.010^{+0.169}_{-0.178} $ & $ -0.997^{+0.266}_{-0.293} $ & $ -0.01^{+1.04}_{-1.05} $ & $ 0.13^{+1.16}_{-1.57} $\\
SNe+BAO+CMB+$H_0$ & $ 0.270^{+0.014}_{-0.013} $ & $ 0.274^{+0.016}_{-0.015} $ & $ 0.025^{+0.008}_{-0.008} $ & $ 0.027^{+0.012}_{-0.011} $ & $ -1.218^{+0.069}_{-0.072} $ & $ -1.198^{+0.100}_{-0.112} $ & $ 1.21^{+0.10}_{-1.14} $ & $ 1.19^{+0.13}_{-0.13} $\\
SNe+BAO+CMB+$H_0$ ($\Delta \chi^2 = 4.0$) & $ 0.270^{+0.029}_{-0.026} $ & $ 0.274^{+0.032}_{-0.029} $ & $ 0.025^{+0.016}_{-0.035} $ & $ 0.027^{+0.026}_{-0.036} $ & $ -1.218^{+0.425}_{-0.147} $ & $ -1.198^{+0.293}_{-0.227} $ & $ 1.21^{+0.19}_{-2.49} $ & $ 1.19^{+0.27}_{-2.40} $
  \enddata

\ifemapj \end{deluxetable*}
\else  \end{deluxetable}
\fi

\subsection{Time Dependent Equation of State}

We next examine models in which dark energy changes with time.  For a
wide range of dark energy models, it can be shown \citep{linder03b}
that, to good approximation, the dark energy equation-of-state can be
parametrized by
\begin{equation}
w(a)=w_{0}+w_{a}(1-a)
\end{equation}
where $a=1/(1+z)$ is a scale factor. The $\Lambda$CDM model is recovered when
$w_{0}=-1$ and $w_a=0$.  The constraints on $w_{0}$ and $w_{a}$ are
shown in Figure \ref{fig:i07} and Table \ref{tbl:d}.

The Dark Energy Task Force \citep{albrecht06a} proposed a figure of
merit (FoM) for cosmological measurements equal to the inverse of the
area of the $95\%$ confidence contour in the $w_0-w_a$ plane. When we
make this measurement, using the $\Delta \chi^2 = 5.99$ contour, we
find a FoM of \FoMstat\ (statistical-only) and \FoMsys\ (including
systematics). Frequently, the FoM is also defined in terms of the $1\sigma$ errors ($\Delta \chi^2 = 1$); this FoM is \Fomonesigstat\
(statistical-only) and \FoMonesigsys\ (including systematics).
Surprisingly, even with $w_a$ floating, we still find an
interesting constraint on $\Omega_k$ of $\sim~0.02$.

\ifemapj \begin{figure*}[h]
\else \begin{figure}[h]
\fi

\centering
\includegraphics[width=0.45\textwidth,clip=true]{%
  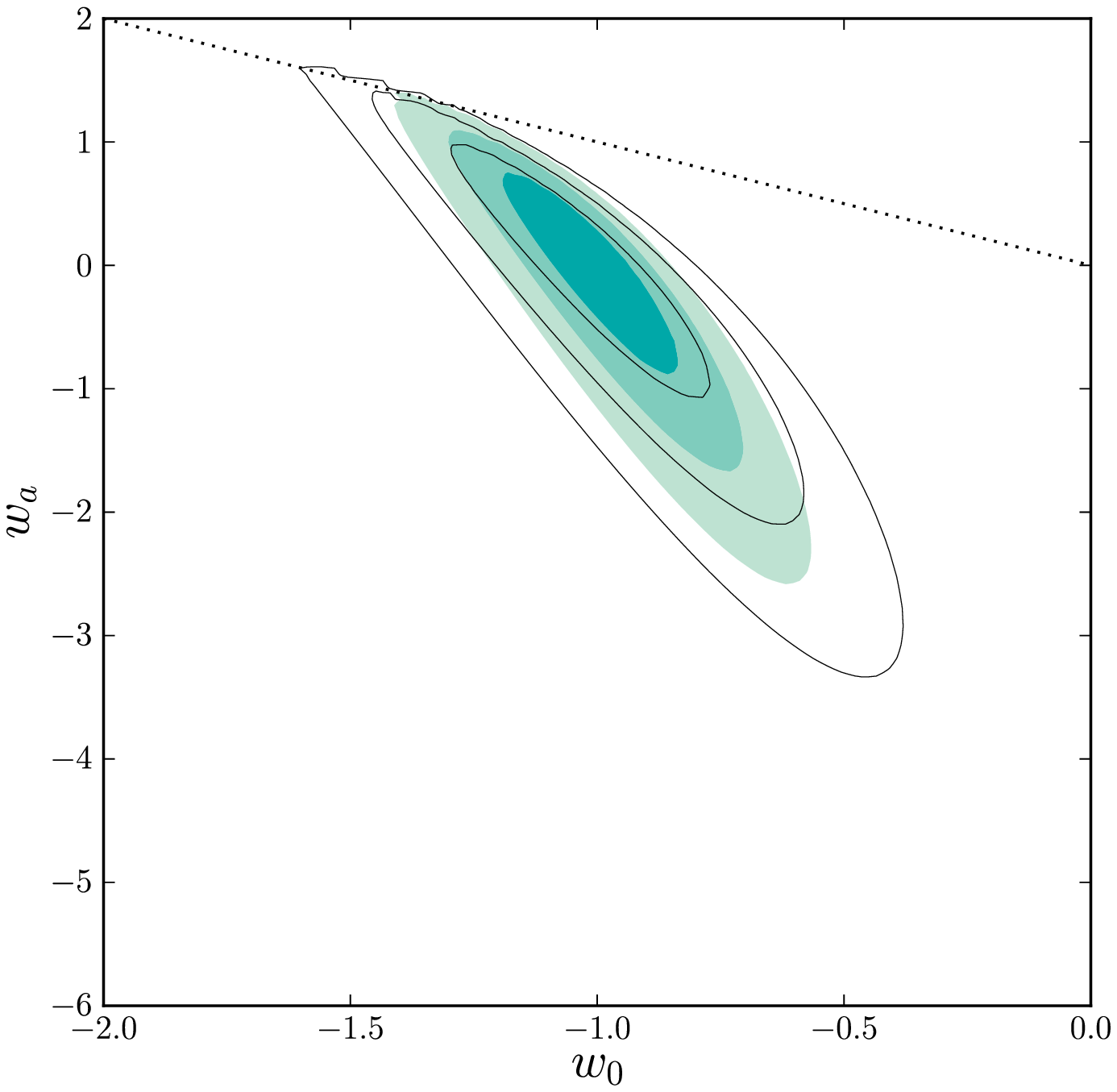}
\caption{\NOTE{fig:i07}$68.3\%$, $95.4\%$, and $99.7\%$ confidence regions of the
  $(w_0,w_a)$ plane from \sne combined with the constraints from BAO,
  CMB, and $H_0$, both with (solid contours) and without (shaded contours)
  systematic errors. Zero curvature has been assumed. Points above the
  dotted line ($w_0 + w_a > 0$) violate early matter domination and
  are disfavored by the data.
  \label{fig:i07} \label{fig:waw0}}

\ifemapj \end{figure*}
\else  \end{figure}
\fi

We next consider a model in which the dark-energy equation-of-state parameter is
constant inside fixed redshift bins. This model has more parameters
(and thus more freedom) than $w_0-w_a$. The results are shown in
Figure~\ref{fig:i08} and Table \ref{tbl:d}. We adopt the redshift bins used in
\citet{amanullah10a}, so that a direct comparison can be made.

In the left panel with broad bins, we show a reasonably good measurement of the
equation-of-state parameter from redshift 0 to 0.5. From redshift 0.5 to 1,
there is no real constraint. For example, any scalar field model ($|w| < 1$)
is reasonably compatible with the data. Above redshift 1, the
constraints are weaker. $w \gtrsim 0$ is ruled out, as this
violates early matter domination. 

We separate the supernova and early universe constraints by defining a
bin at redshift 1.6, as shown in the middle panel. This shifts the confidence
interval for $w(1.0 < z < 1.6)$ towards higher $w$.  Eliminating this division, and instead adding more bins
up to redshift 0.5 (right panel), gives three constraints of moderate
quality with a possible crossing of $w = -1$. No matter the binning,
we will need more data extending above redshift 1 to investigate the
dark energy equation-of-state parameter where the uncertainty is still very
large.

To examine constraints on the existence of dark energy at different epochs, 
we study $\rho(z)$, which is the density of the dark energy 
and allowed to have different values in fixed redshift bins. 
Within each bin, $\rho$ is constant. (Note that the
discontinuities in $\rho(z)$ at the bin boundaries introduce
discontinuities in $H(z)$.) We choose the same binning as above, but
note that binned $\rho$ and binned $w$ models give different expansion
histories. Our results are shown in Figure~\ref{fig:i09} and Table
\ref{tbl:g}.

Although there is no real constraint on the
equation-of-state parameter at redshift 0.5 to 1, dark energy is seen at high
significance in both panels. There is weak evidence for the existence of dark energy
above redshift 1, as can be seen in the left panel. However, if we again separate the supernova data and 
early universe constraints (right panel) we see neither probe has any 
constraint on the existence of dark energy above redshift 1.

\NOTE{NS: figures will be rearranged properly}

\ifemapj \begin{figure*}[h]
\else \begin{figure}[h]
\fi
\includegraphics[width=0.3\textwidth]{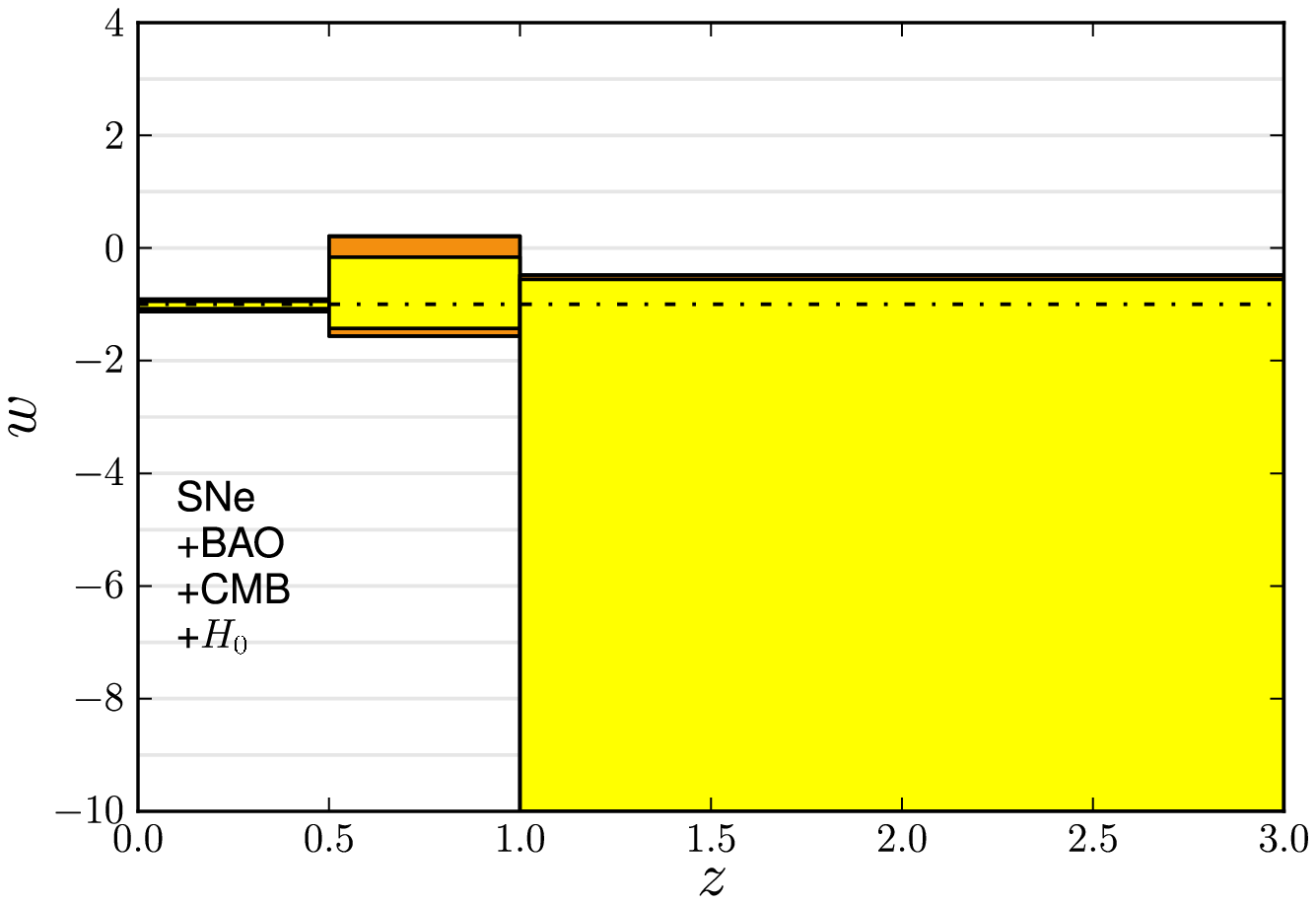}
\hspace{0.02\textwidth}
\includegraphics[width=0.3\textwidth]{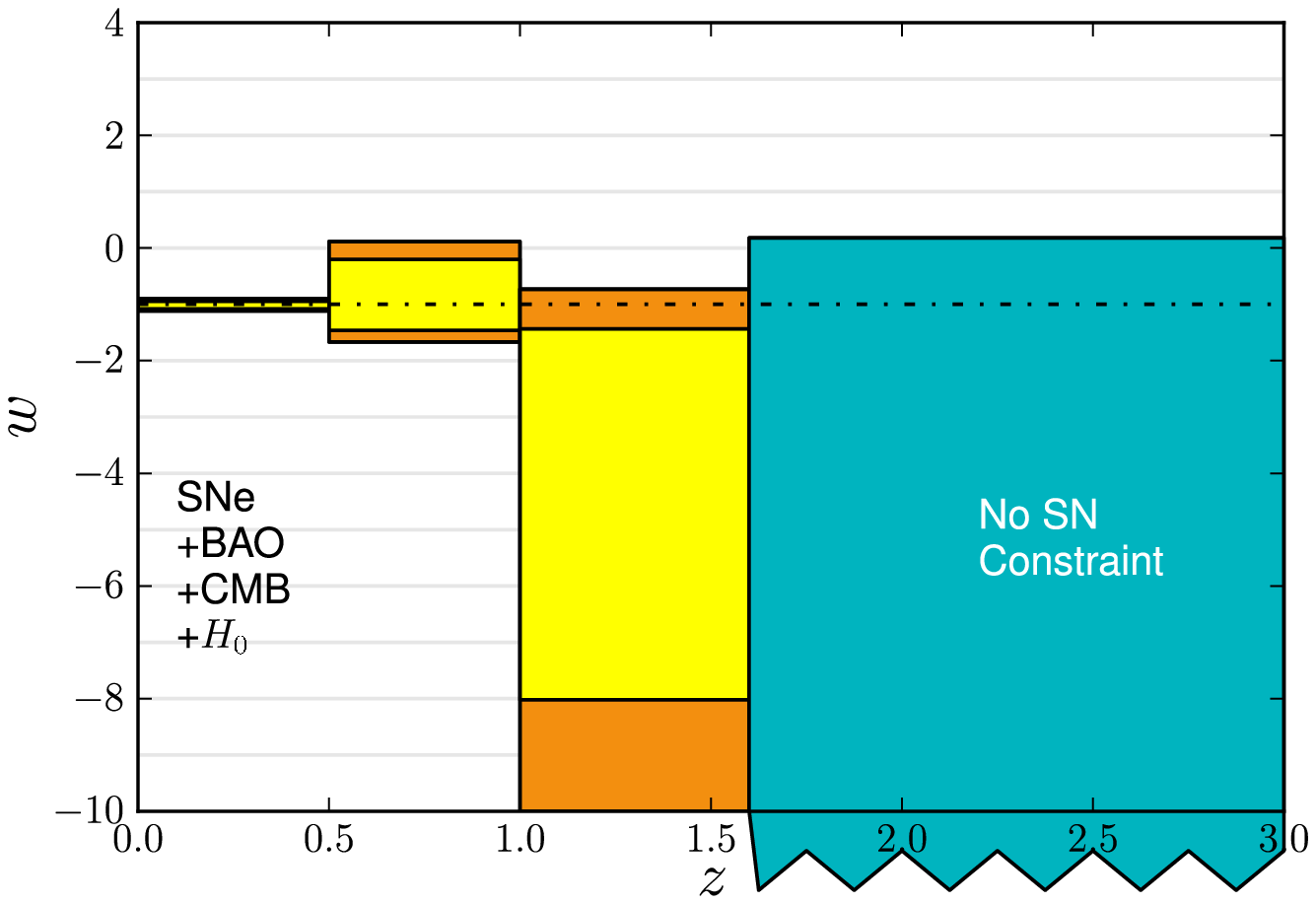}
\hspace{0.02\textwidth}
\includegraphics[width=0.3\textwidth]{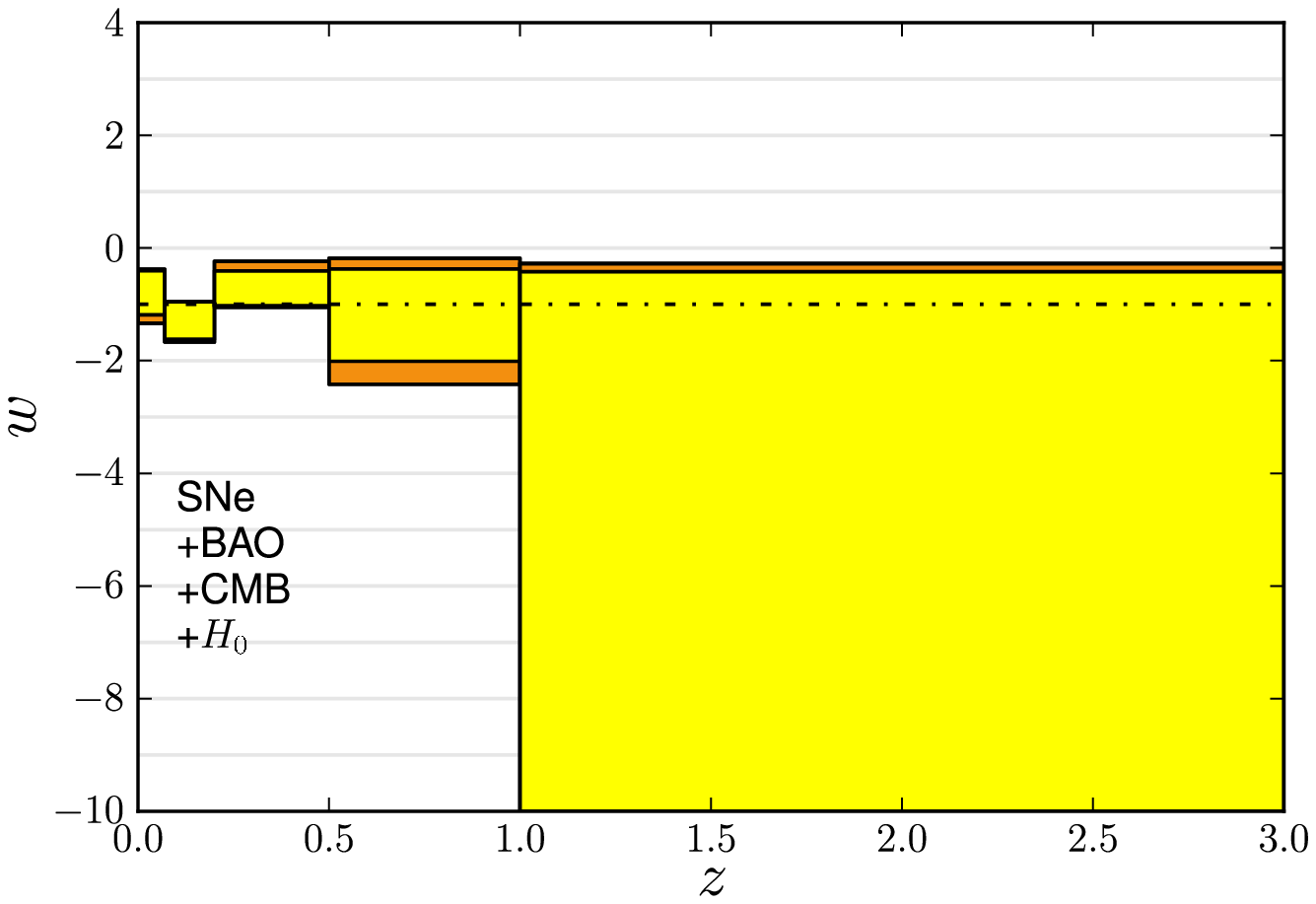}
\caption{\NOTE{fig:i08} \label{fig:binnedw}
    Constraints on $w(z)$, where $w(z)$ is assumed to be constant in
    each redshift bin, are plotted at the 68\% probability level ($\Delta \chi^2 = 1$). 
    Each panel shows different redshift binning.
    The results were obtained assuming a flat universe for the joint data
    set of SNe, BAO, CMB, and $H_0$, with (dark/orange) and
    without (light/yellow) SN systematics.
    The middle panel takes a closer look at the $z>1$ constraints, while the right
    panel shows the effects of $w$ binning at low redshift. In this panel the best
    fit values of $w$ cross $w = -1$ twice at low redshift, an unusual feature 
    in dark energy models.
    We note that the $\Lambda$CDM model is consistent with our $w(z)$ constraints for each of these binnings.
%
%
    \label{fig:i08}}
\ifemapj \end{figure*}
\else  \end{figure}
\fi

\ifemapj \begin{figure*}[h]
\else \begin{figure}[h]
\fi
\includegraphics[width=0.3\textwidth]{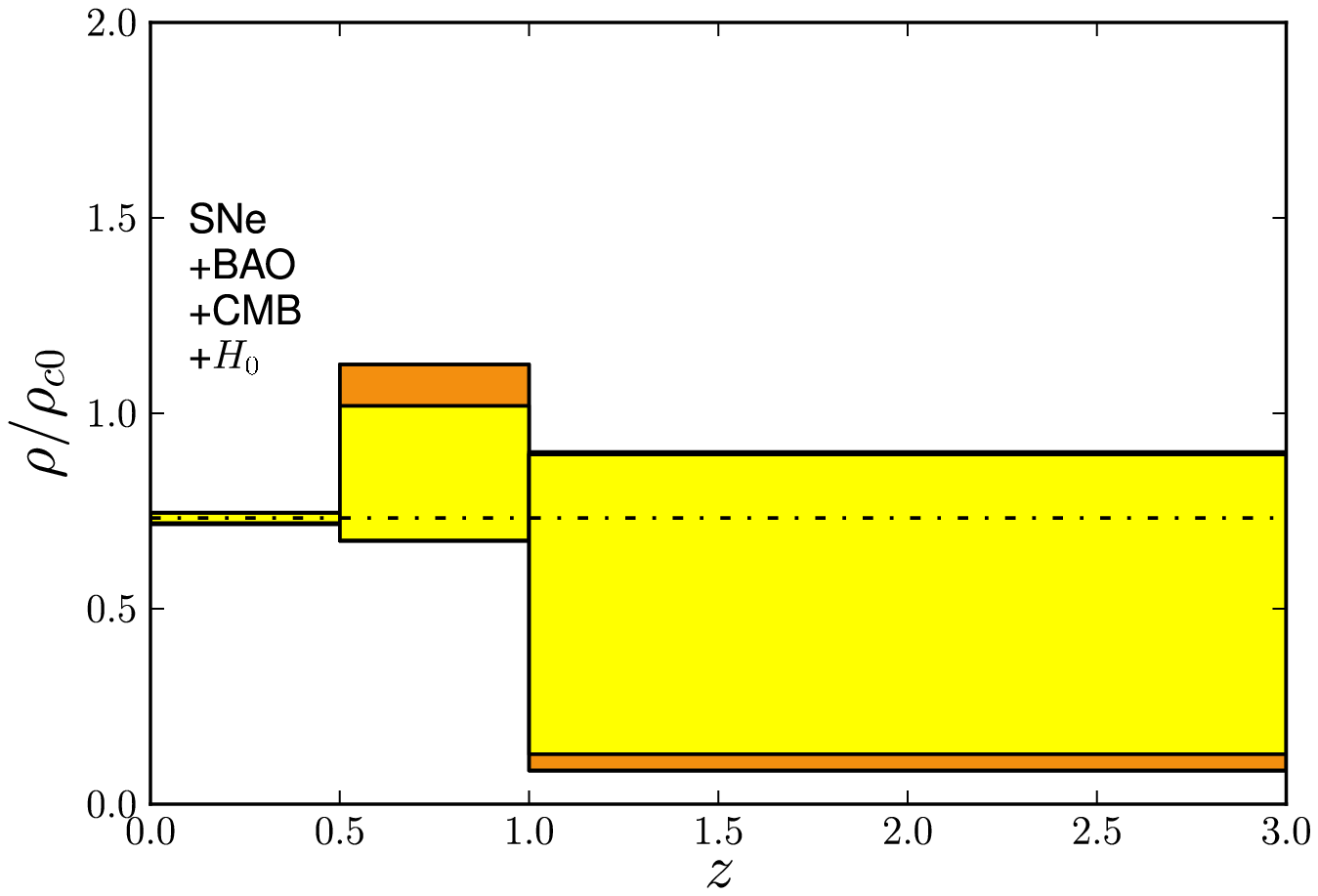}
\hspace{0.02\textwidth}
\includegraphics[width=0.3\textwidth]{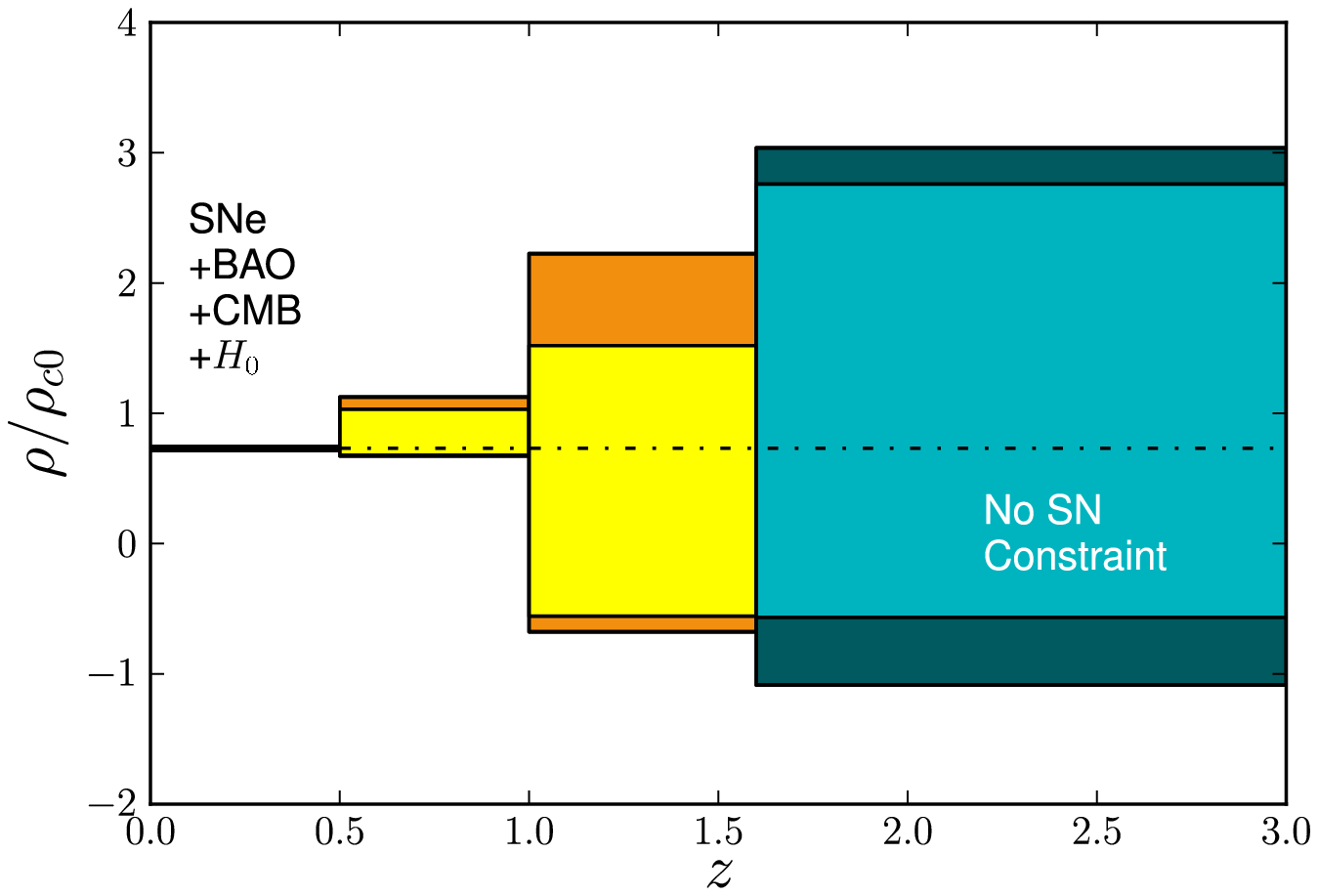}
\caption{\NOTE{fig:i08} \label{fig:binnedrho}
    Redshift evolution of dark energy density: 
    Constraints on $\rho(z)$ are shown as a function of redshift, 
    where $\rho(z)$ is the density of the dark energy at a given redshift bin 
    and assumed to be constant within the redshift bin.
    $\rho(z)$ is normalized by the critical density today ($\rho_{c0}$) and 
    is plotted at the 68\% probability level ($\Delta \chi^2 = 1$). 
    The results were obtained assuming a flat Universe for the joint data
    set of \sneia, BAO, CMB, and $H_0$, with (dark/orange) and
    without (light/yellow) SN systematics. 
    The two panels demonstrate different redshifts binning and have different scales.
    \label{fig:i09}}
\ifemapj \end{figure*}
\else  \end{figure}
\fi

\ifemapj \begin{deluxetable*}{ll|rrrl}
\else \begin{deluxetable}{ll|rrrl}
\fi

  \tablecaption{ Constraints on redshift binned equation of state $w$ and 
   density $\rho$ (normalized by the current critical density). 
   The constraints are computed including SNe, BAO, $H_0$, and CMB data. \label{tbl:g} \label{tbl:binnedrho}}
      \tablehead{
      \colhead{} & 
      \colhead{} & 
      \colhead{z $<$ 0.5} &
      \colhead{0.5 $<$ z $<$ 1.0} &
      \colhead{1.0 $<$ z $<$ 1.6} &
      \colhead{z   $>$ 1.6 \tablenotemark{a} } }
  \startdata
  $w$(z)               & Stat Only: & $ -1.013^{+0.067}_{-0.069} $& $ -0.78^{+0.58}_{-0.68} $& $ -3.7^{+2.2}_{-4.4} $& $ < 0.18 $\\ 
                       &  w/ Sys: & $ -1.006^{+0.110}_{-0.113} $& $ -0.69^{+0.80}_{-0.98} $& $ -3.9^{+3.2}_{-8.2} $& $ < 0.24 $\\             
                                    
  \hline
  $\rho_{\mathrm{DE}}(z) / \rho _{c0}$ & Stat Only: & $ 0.732^{+0.013}_{-0.014} $& $ 0.85^{+0.18}_{-0.17} $& $ 0.23^{+1.29}_{-0.79} $& $ 0.9^{+1.9}_{-1.5} $\\ 
                &   w/ Sys: & $0.731^{+0.014}_{-0.015} $& $ 0.88^{+0.24}_{-0.21} $& $ 0.33^{+1.90}_{-1.00} $& $ 0.7^{+2.4}_{-1.8} $
  \enddata
\tablecomments{Constraints on binned $\rho_{\mathrm{DE}}(z)/\rho_{c0}$ and $w(z)$. 
This redshift binning corresponds to the middle panel of Figure \ref{fig:binnedw} and the right panel of Figure \ref{fig:binnedrho}}
\tablenotetext{a}{We note that the weak constraints in these bins come mostly from the CMB (which tells us that the early universe was matter-dominated) and are only indirectly constrained by supernovae.}
\ifemapj \end{deluxetable*}
\else  \end{deluxetable}
\fi

\section{Discussion}
\outline{draft}{Discussion on the questions and points we made }

\subsection{Improving the Constraints on Time-Varying $w$ by Efficiently Adding $z>1$ Supernovae}

Beyond $z=1$, we add 10 new well-measured \sneia to the Hubble
diagram. The variance-weighted RMS scatter of the new sample is
\clusterzgtoneWRMS mag. As a comparison, the \numbergoodszgtone\ $z>1$ \sneia from the
GOODS survey that pass our Union2 selection cuts have a
variance-weighted RMS scatter of
\GOODSzgtoneWRMS\,mag. The new sample almost doubles the weight of
HST-discovered \sneia beyond $z=1$. The increase provides improvements
on the most difficult-to-measure parameters, those that describe the
time-varying properties of dark energy: $\rho(z)$ and $w(z)$ at the
higher redshifts. In particular, the supernovae from this search
improve the constraint on $\rho(z)$ at redshifts $1.0 z < 1.6$ by
$28\%$ (statistical errors only) and 18\% (including supernova systematics)
after adding the constraints from the CMB, BAO and H$_{0}$ (using the binning
illustrated in the right panel
of Figure \ref{fig:binnedrho}). (It is more difficult to compare binned $w$ results,
as the constraints are much less gaussian and more sensitive to the location of the best fit.)

The new sample is also obtained with greater observing
efficiency with HST. Considering the number of $z>1$ \sneia that make the
Union2 selection cuts, the yield of \sneia\ increases from a rate of
one \snia per 43 HST orbits in the GOODS survey to one \snia per 22
HST orbits in this survey.

\subsection{Splitting the sample according to host galaxy type} \label{sec:splitbyhost}

\sneia\ are well-standardized with a small dispersion in magnitudes
across the whole class. Any clues to heterogeneous characteristics
therefore offer exciting possibilities to {\it further} improve
standardization, enhancing the use of SNe as a cosmological
probe. There is now evidence from studies of large samples of \sneia\
at both low and intermediate redshifts ($0 < z \lesssim 0.8$) that
\snia properties are related to the properties of the host. The
clearest of these is the relation between light curve width and the
specific star formation rate. \sneia\ in passive galaxies tend to have
narrower light curves than \sneia\ that are in galaxies that are
actively forming stars.

More than two-thirds of our new \sneia beyond $z=0.9$ are hosted by
early-type galaxies (Meyers et al. (2011)). In field
surveys, such as the GOODS survey, this ratio is inverted. By
combining \sneia\ from our HST Cluster SN Survey and GOODS, together
with our $z>0.9$ \sneia\ in \citet{amanullah10a}, which have HST images
of the host, we can create a sample of \sneia\ that has roughly equal
numbers when split according to host type. When split this way, we
find that $z>1$ \sneia\ in early galaxies rise and fall more quickly
than \sneia\ in later host types, thus extending the redshift interval
over which the effect is now detected. Finding that low and high
redshift \sneia\ follow similar trends gives us confidence that we can
use very distant events to constrain cosmological parameters. This finding is
reported in more detail in \citet{meyers11a}.

There is also evidence from \sneia\ at low and intermediate redshifts
for other correlations with host type. \citet{sullivan10b} find that
both $\beta$ and the RMS scatter about the Hubble diagram are smallest
for \sneia\ in passive galaxies. These trends suggest that dust plays a
greater role in reddening and dimming \sneia\ in late-type galaxies. We
examined our $z>0.9$ sample for evidence of similar correlations using
 our host classification from Tables 3 and 4 of Meyers et al (2011).

After correcting \snia luminosities for lightcurve shape, \snia\ color
and host galaxy mass (with the global values of these correction coefficients), we measure a sample dispersion of
\earlyhizintdisp mag for \sneia\ in early-type galaxies and
\latehizintdisp mag for \sneia\ in late-type galaxies. In terms of the
RMS, we find \earlyhizWRMS mag and \latehizWRMS mag for early and
late-type samples, respectively. The uncertainties are currently too
large to distinguish between the two samples. Similarly for $\beta$,
the errors are larger than the difference between the two samples, as
seen in Table \ref{tb:subsets}. Clearly, higher quality data of a
larger number of $z>0.9$ \sneia\ in both early and late-type galaxies
are required before the trends that are seen at low redshift can be
detected in high redshift samples.

We also examined the error-weighted difference in the brightness of
\sneia\ in the two samples after correcting for lightcurve shape and color, but without correcting for the host-mass
luminosity relation (setting $\delta=0$) and find that \sneia\ in
early-type galaxies are \earlyhizbrighter mag brighter. Since
early-type galaxies are typically more massive than late-type
galaxies, this $2\sigma$ difference, if confirmed with larger
statistics, may be related to host galaxy mass.

\subsection{Future directions with current instrumentation.}\label{sec:future}

Due to the much improved sensitivity of the WFC3 IR detector, it will
be feasible to measure $z > 1$ SNe with much better precision. The
color measurement errors ($\sim~0.03$ in $B-V$) can be made comparable
to the color measurement errors in the SDSS supernova survey \citep{smith02a,holtzman08a}. 
Assuming that the intrinsic dispersion of \snia luminosities does not change
with redshift, 
the variance weighted RMS of the WFC3 sample should be similar to that 
measured for the SDSS, i.e. $\sim~0.14$ mag.  A well-observed \snia\ 
with WFC3 should have a statistical weight of two to three \sneia\ from the
Cluster and GOODS surveys.

With a sufficient number of well-measured $z>1$ \sneia\ with WFC3, it
should be possible to search for the correlations between the properties
of \sneia\ and their hosts that are seen at lower redshifts. As
discussed above, current samples at $z>1$ are too small to detect
these differences. With the improved WFC3 photometry, only 40
\sneia, split evenly between early and late-type hosts, would be
needed to constrain a difference in $\beta$ to an uncertainty of 0.4,
which is about half the difference found for lower redshift \sneia\
\citep{sullivan10b}. These samples would be just enough to see
evidence of the lower RMS for passive hosts seen by
\citet{sullivan10b}.

Current WFC3 \snia\ surveys target empty fields, which means that there
will be few \sneia\ in passive host galaxies. A WFC3 \snia\ survey that
spends part of its time targeting $z \gtrsim 1$ clusters would ensure a better
balance between host types while increasing the overall yield.

In order to investigate the figure of merit constraints possible with
WFC3, we simulate a sample of 40 supernovae at redshift 1.2 and add
this sample into the current compilation. As there is a hard wall at
$w_0 + w_a = 0$ when including BAO and CMB data, we simply fix
$\Omega_m$, rather than including BAO and CMB data (the alternative
would be to adjust the supernova magnitudes to a cosmology model
far away from the wall). 
When adding these supernovae, the statistical figure of merit improves by
$39\%$. By the same metric, the current cluster sample improves the
figure of merit by $10\%$.

\subsection{Reducing the Systematic Errors for Future Surveys}

As has been stressed by several authors, systematic errors are now
larger than statistical errors. To fully utilize the potential of
current and future \snia\ surveys to constrain cosmology, it will be
necessary to reduce these errors significantly.

The largest current source of systematic uncertainty is calibration.
Calibration uncertainties can be split into uncertainties related to
the primary standard, and uncertainties in instrumental zeropoints and
band passes. In principle, all of these uncertainties can be reduced
by establishing a network of well-calibrated standard stars and
monitoring telescope system throughputs \citep{regnault09a}. The Sloan
Digital Sky Survey demonstrated that a 1\% relative photometric
calibration is possible with the current standard star network and 
system throughput monitoring \citep{doi10a}.

The ongoing Nearby Supernova Factory (SNf) project \citep{aldering02a}
is aiming to provide the network of standard stars.
SNf will also address the systematic uncertainty due to host-mass
correction since the range of host masses would become comparable 
to that of high redshift for the first time. 
Additionally, the comprehensive \snia spectral time series from the SNf 
will allow one to tackle systematic uncertainties related to modeling of 
the lightcurves.

In the future, recently approved experiments such as ACCESS \citep[Absolute Color 
Calibration Experiment for Standard Stars][]{kaiser10a} and the proposed NIST STARS project
\citep[National Institute for Standards and Technology][]{mcgraw10a,zimmer10a} 
are aiming to achieve sub-percent absolute flux calibration for the network 
of stars in the wavelength range of visible to NIR.
With this network of stars and with new techniques for monitoring throughput of the telescopes 
\citep{stubbs07a}, we will be able to cross-calibrate systems and
reduce the systematic errors below the statistical errors.

\section{Summary and Conclusions}
\outline{draft}{Summary and Acknowledgment}

We present HST ACS, HST NICMOS and Keck AO-assisted photometry of
\fullsample \sneia\ in the redshift range $0.63 < z < 1.42$. The \sneia\
were discovered in the HST Cluster Supernova Survey, a survey run by
the Supernova Cosmology Project to search for \sneia\ in fields
centered on 25 distant galaxy clusters \citep{dawson09a}.

We implement new techniques to improve the accuracy of HST
photometry. In particular, for data taken with NICMOS, which samples
the rest-frame $B$ and $V$-bands of $z>1$ \sneia, we use a more direct,
more accurate measure of the NICMOS zeropoint (Ripoche et al. 2011),
and we remove the residual background that persists after standard
processing of NICMOS data with the {\tt CALNICA} pipeline (Hsiao et
al. 2010). For data taken with ACS WFC in the \zacs filter, we
incorporate a SED-dependent aperture correction (see Appendix A).

Following the procedures outlined in \citet{kowalski08b} and
\citet{amanullah10a}, we add our \sneia\ to the Union2 compilation. 
Fourteen of the 20 \sneia\ of our supernovae pass the Union2 selection cuts. 
Ten of them are at $z>1$. 
The strategy of targeting high-redshift galaxy clusters results 
in factor of two improvement in the yield per HST orbit of 
well-measured \sneia beyond $z=1$ and a factor of three to five improvement
for \sne hosted by early-type galaxies. 
For WFC3, with its smaller field of view, the advantage of a cluster 
search is even greater.

We use the new Union2.1 sample to constrain the properties of dark
energy. \sneia\ alone constrains the existence of dark energy to very
high significance. After adding constraints from the CMB, BAO, and
$H_0$ measurements, we provide the tightest limits yet on the
evolution of dark energy with time: \waflatwithsys. Our sample
improves the constraints on binned $\rho$ by $18\%$ in the
difficult-to-measure high redshift bin, $1.0 < z < 1.6$. Even with a
time-varying $w_{0}$-$w_a$ model, the universe is constrained to be flat
with an accuracy of 2\% in \omegak.

The results from this new cluster-hosted supernova sample point the way 
to the next steps that are now possible with the WFC3 on HST, 
an instrument that can obtain high signal-to-noise, multifilter SN~Ia 
lightcurves at $z > 1$. 
The cluster approach, used in this paper, would make it feasible to 
build a significantly larger sample at these highest redshifts, evenly 
balanced between early and late-type hosts.  
With such a sample, we can mitigate the effects of dust and evolution 
that may ultimately limit constraints on time-varying $w$.

\vspace{4mm}
Financial support for this work was provided by NASA through program
GO-10496 from the Space Telescope Science Institute, which is operated
by AURA, Inc., under NASA contract NAS 5-26555.  This work was also
supported in part by the Director, Office of Science, Office of High
Energy and Nuclear Physics, of the U.S. Department of Energy under
Contract No. AC02-05CH11231, as well as a JSPS core-to-core program
``International Research Network for Dark Energy'' and by JSPS
research grant 20040003.  Support for MB was provided by the
W. M. Keck Foundation.  The work of SAS was performed under the
auspices of the U.S. Department of Energy by Lawrence Livermore
National Laboratory in part under Contract W-7405-Eng-48 and in part
under Contract DE-AC52-07NA27344.  The work of PE, JR, and DS was
carried out at the Jet Propulsion Laboratory, California Institute of
Technology, under a contract with NASA. 
TM and YI have been financially supported by the Japan Society for 
the Promotion of Science through its Research Fellowship.
HH acknowledges support from a VIDI grant from the Netherlands Organization
for Scientific Research (NWO) and a Marie Curie International Reintegration Grant.
NS, CL and SP wish to thank the support and hospitality of the Aspen 
Center for Physics, where much of this paper was written.
We would like to thank Jay Anderson, L. E. Bergeron, 
Ralph Bohlin, Roelof de Jong, Anton Koekemoer, Jennifer Mack, 
Bahram Mobasher, Adam Riess, Kenneth Sembach, and 
ACS and NICMOS team at Space Telescope Institute
for their advice on the HST data calibration.
We also thank Alex Conley for calibration discussions. Finally,
we would like to thank our referee, who carefully read our paper and gave valuable feedback.

~


\appendix
\section{Appendix : ACS color dependent aperture correction} 
\label{sec:k}
\outline{draft}{Red Halo Scattering, Fig : wavelength vs APC\\}

The scattering of long wavelength photons ($> 8000$ \ang ) within the
ACS CCDs causes the point spread function (PSF) of images taken in the
ACS \zacs-band to depend on the spectral energy distribution (SED) of
the source \citep{sirianni05a,jee07b}. The SED-dependent PSF means that
aperture corrections also depend on the SED. In this section, we
describe how we derive aperture corrections for our observations taken
with the ACS \zacs-band.

\outline{draft}{Color Dependence Measurements\\} 

We use two stars (GRW70 and GD72 from ACS programs PID9020 and
PID10720) to measure the aperture correction (AC) as a function of
wavelength. The stars were observed with 15 narrow-band filters
between 7660\ang and 10360\ang. We processed the data in the same
way as the \snia\ images. To derive the aperture correction, we
compare the flux in two apertures, one with a 3-pixel radius and the
other with a 110-pixel radius (the radius used for defining the
zeropoint). Errors from removing the sky are the dominant source of
the uncertainty of this measurement.

\outline{draft}{Wavelength vs Aperture Correction Factor\\}

The measured encircled energy and its best-fit curve for a 3-pixel radius are 
shown as a function of wavelength in Figure ~\ref{fig:k01}. 
We apply an aperture correction to the observed flux with this best-fit curve.
Figure \ref{fig:k02} demonstrates that the \iacs - \zacs color is not
accurate enough for \sneia\ to infer the accurate aperture correction.  
Therefore, we need to introduce an SED-weighted aperture correction.

\ifemapj \begin{figure*}[h]
\else \begin{figure}[h]
\fi

\begin{center}
\includegraphics[angle=90,scale=.75,width=0.5\textwidth]{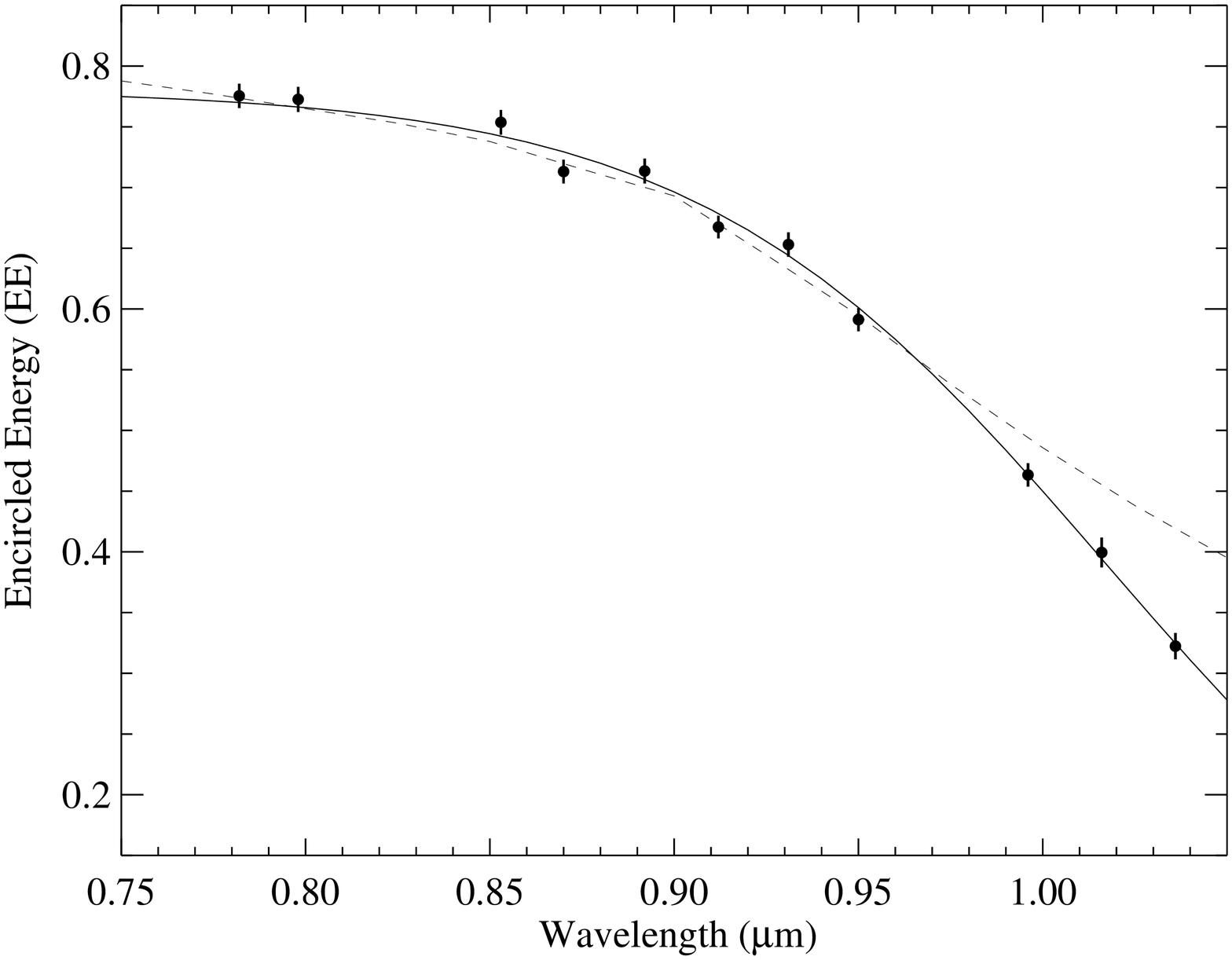}
\end{center}
\caption{\NOTE{fig:k01 to be revised : place holder} 
Wavelength vs. Encircled Energy (EE) Fraction for a 3-pixel aperture. The solid line
is a fit (Eq.~\ref{eqn:ac}) to the measured data points. 
The dotted line is the relation from \citet{sirianni05a}}. \label{fig:k01}
\ifemapj \end{figure*}
\else  \end{figure}
\fi

\ifemapj \begin{figure*}[h]
\else \begin{figure}[h]
\fi

\begin{center}
\includegraphics[angle=90,scale=.75,width=0.5\textwidth]{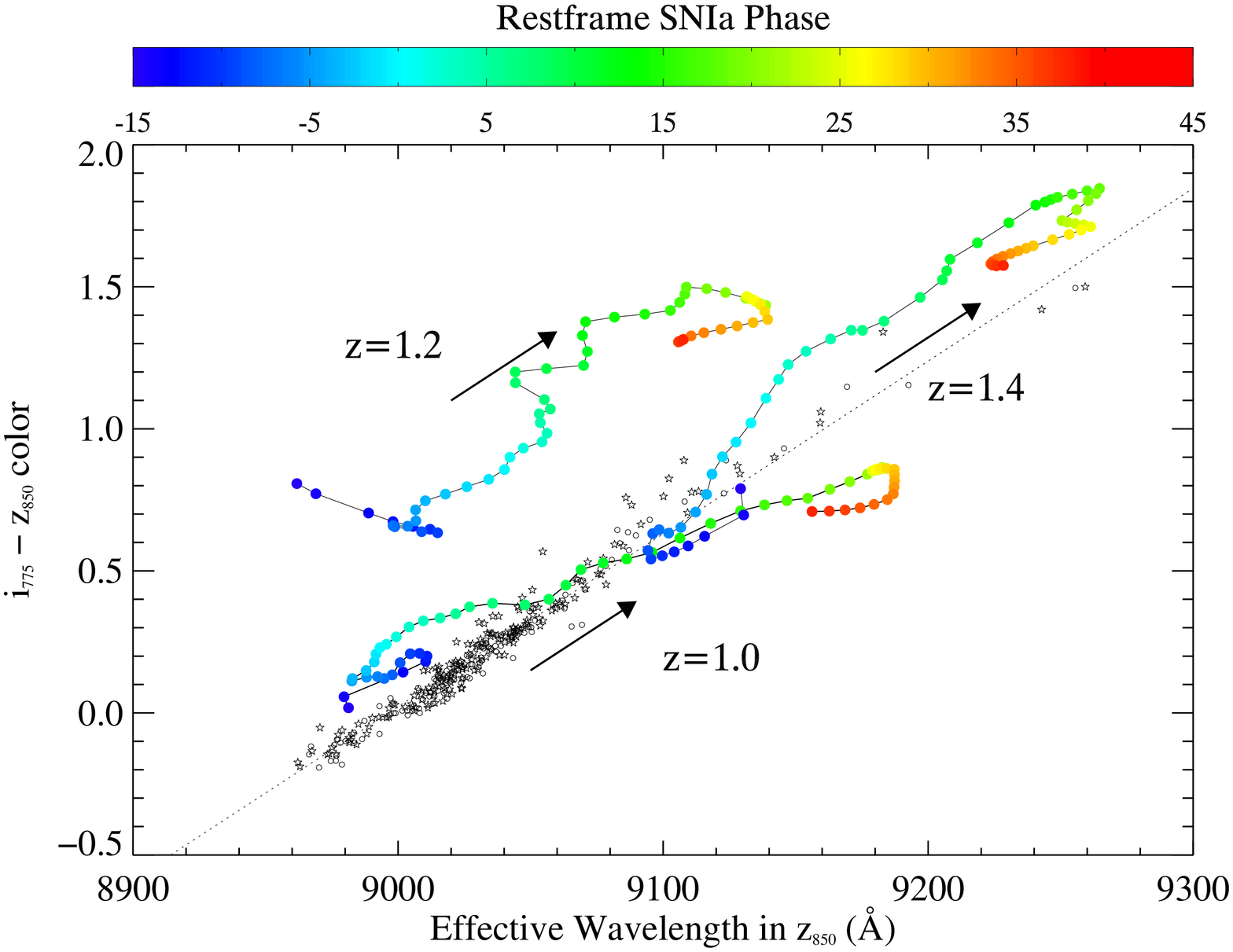}
\caption{\NOTE{fig:k02 to be revised : place holder} Wavelength
  vs. \iacs$-$\zacs color: solid asterisks are stars from
  Gunn-Stryker Catalog \citep{gunn83a}; open asterisks are stars
  from Pickles catalog \citep{pickles98a}. The dotted line is the
  best-fit line to the stellar locus. SN~Ia loci \citep{hsiao07a} are
  plot for three redshifts with color-coded phases and arrows which
  show the direction of the phase evolution. The SN~Ia loci
  deviate from the stellar locus as the SED is different.
  Therefore, \iacs$-$\zacs colors are not accurate enough to perform
  aperture corrections, necessitating the introduction of an SED-weighted aperture
  correction.} \label{fig:k02}
\end{center}

\ifemapj \end{figure*}
\else  \end{figure}
\fi

We find that the best-fit encircled energy curve can be described with

\begin{equation}
{\rm EE}(\lambda) =\frac{1}{1.28+1.45 \times 10^{-8} \times {\rm exp} (17.99 \times \lambda)}
\label{eqn:ac}
\end{equation}
where $\lambda$ is wavelength in $\mu m$. The formula is valid in wavelength for virtually all the F850LP filter. 

\outline{draft}{EE for Stars/Galaxies\\}

Since the SED of a \snia\ is a function of redshift and phase, so is
the aperture correction. We demonstrate two methods to account for
this dependence. 
The first method, ``Method I,'' calculates the aperture correction 
iteratively using the SED produced by SALT2. The advantage of this 
method is that the final aperture-corrected fluxes can be used with 
the standard throughputs and zeropoints.
The second method, ``Method II,'' computes a new zeropoint and F850LP 
throughput for direct use with the flux measurements that have not been 
aperture-corrected. 
We note here that an apparent color difference between SNe with only ACS 
data and those that also had NICMOS observations was seen 
unblinded {\em before} the development of these methods, so the 
relative colors of the ACS-only and NICMOS-included subsamples 
should not be considered blinded.

Both methods give the same answer, although the error in the 
color from ``Method II'' is slightly larger than that of ``Method I'',
because the effective wavelength of the \zacs shifts towards the \iacs 
filter, thus shortening the wavelength separation of the two filters. 
For this paper, we adopt ``Method II'' and report all 
results using this method.

\subsection{Method I: Iterative approach}

The magnitude of an object in the Vega magnitude system is defined \citep{fukugita95a,sirianni05a} as 
\begin{equation}
m=-2.5 {\mathrm log_{10}} \left ( \frac{\int \lambda R f_{\lambda} d\lambda}{\int \lambda R f_{\lambda {\rm Vega}} d\lambda} \right ),
\label{eqn:mag}
\end{equation}
where $\lambda$ is wavelength, $R$ is the system response, 
$f_\lambda$ is the SED of the object (e.g. \snia), and
$f_{\lambda {\rm Vega}}$ is the SED of Vega. We define the Vega magnitude
zeropoint (Zpt) as
\begin{equation}
{\rm Zpt_{Vega}}= 2.5 {\mathrm log_{10}} \left (\int \lambda R f_{\lambda {\rm Vega}} d\lambda  \right ).
\label{eqn:zpt}
\end{equation}
Eq.~\ref{eqn:mag} can then be rewritten as:

\begin{equation}
m=-2.5 {\mathrm log_{10}} \left ( {\int \lambda R f_{\lambda} d\lambda} \right ) + {\rm Zpt_{Vega}}.
\label{eqn:mag2}
\end{equation}
The observed magnitude, $m_{obs}$, within a given aperture is
\begin{equation}
m_{\rm obs}=-2.5 {\mathrm log_{10}} \left( {\int \lambda \, R \, \mbox{EE}(\lambda) \, f_{\lambda} d\lambda} \right ) + {\rm Zpt_{Vega}},
\label{eqn:magobs}
\end{equation}
where EE($\lambda$) is the wavelength dependent encircled energy (EE), which can be derived from
Eq.~\ref{eqn:ac}.

We define the SED dependent aperture correction ($\Delta m_{\rm corr}$) as

\begin{equation}
\Delta m_{\rm corr}= 
-2.5 {\mathrm log_{10}} \left ( \frac{\int \lambda \, R \, \mbox{EE}(\lambda) \, f_{\lambda} d\lambda}
                    {\int \lambda R f_{\lambda} d\lambda} \right )
\end{equation}

In practice, we do not know $f_{\lambda}$ in advance, so we cannot
compute the aperture correction directly. Instead, we derive it
iteratively using the SED derived from fitting the \snia light curve
with SALT2 as input to the next iteration.  With this method, we use the
STScI filter response function and zeropoints.

\subsection{Method II: Modified Filter with Zeropoint}

We can rewrite Eq.~\ref{eqn:magobs} as:

\begin{equation}
m_{\rm obs}=-2.5 {\mathrm log_{10}} \left ( \frac{\int \lambda \, R \, \mbox{EE}(\lambda) f_{\lambda} d\lambda}
                            {\int \lambda \, R \, \mbox{EE}(\lambda) \, f_{\lambda {\rm Vega}} d\lambda} \right )
        + {\rm Zpt_{Vega}}
        -2.5 {\mathrm log_{10}} \left ( {\int \lambda \, R \, \mbox{EE}(\lambda) \, f_{\lambda {\rm Vega}} d\lambda} \right ). 
\label{eqn:magobs2}
\end{equation}

Effectively, the last term serves as a zeropoint offset for a given aperture radius.

We rewrite Eq.~\ref{eqn:magobs2} as:

\begin{equation}
m_{\rm obs}=-2.5 {\mathrm log_{10}} \left ( \frac{\int \lambda \, R \, \mbox{EE}(\lambda) \, f_{\lambda} d\lambda}
                            {\int \lambda \, R \, \mbox{EE}(\lambda) \, f_{\lambda {\rm Vega}} d\lambda} \right )
        + {\rm Zpt_{Vega}} - \Delta {\rm Zpt_{Vega}}, 
\label{eqn:magobs3}
\end{equation}

$\Delta {\rm Zpt_{Vega}}=0.438$ for a 3-pixel radius aperture and the \zacs filter. 
We interpret Eq.~\ref{eqn:magobs3} as a
magnitude measurement that uses a modified filter response, $R\,\mbox{EE}(\lambda)$, and an adjusted
zeropoint, ${\rm Zpt_{Vega}} - \Delta {\rm Zpt_{Vega}}$.
We then run SALT2 using the counts in a fixed aperture, the modified filter response 
and the adjusted zeropoint.





\end{document}